\documentclass[final,3p,times,twocolumn,authoryear]{elsarticle}

\usepackage{graphicx}
\usepackage{wrapfig}
\usepackage{lipsum} 
\usepackage{amssymb}
\usepackage{amsmath}
\journal{Neurocomputing}

\usepackage{subcaption}
\usepackage{orcidlink}
\usepackage{tabularx}
\usepackage[ruled]{algorithm2e}
\usepackage{gensymb}
\usepackage{latexsym}
\usepackage[table,xcdraw]{xcolor}
\usepackage{multirow}
\usepackage{multicol}
\usepackage{float}
\usepackage{makecell}
\usepackage{booktabs}
\usepackage{threeparttable}
\usepackage{enumitem}
\usepackage{orcidlink}
\usepackage{ulem}
\usepackage{amsthm}
\theoremstyle{plain}
\newtheorem{lemma}{Lemma}
\newtheorem{theorem}{Theorem}
\usepackage{balance}
\usepackage{amsmath,amsfonts}
\usepackage{algorithmic}
\usepackage{array}
\usepackage{textcomp}
\usepackage{graphicx}
\usepackage{url}
\usepackage{verbatim}
\usepackage{stfloats}
\newcolumntype{"}{!{\hskip\tabcolsep\vrule width 1.5pt\hskip\tabcolsep}}
\usepackage {soul}

\soulregister{\cite}7
\soulregister{\sout}7
\soulregister{\ref}7
\soulregister{\it}7
\soulregister{\myeq}7
\usepackage{hyperref}
\hypersetup{
    colorlinks=true,
    linkcolor=blue,
    filecolor=blue,      
    urlcolor=blue,
    anchorcolor=blue,
    citecolor=blue,
}
\usepackage{color}

\newcommand{\myeq}[1]{Eq.~(\ref{#1})}
\renewcommand{\cite}{\citep}

\begin{document}
\begin{frontmatter}

\title{CaPGNN: Optimizing Parallel Graph Neural Network Training with Joint Caching and Resource-Aware Graph Partitioning}

\author{Xianfeng Song}
\ead{misongxf@mail.scut.edu.cn}
\author{Yi Zou\corref{cor1}}
\ead{zouyi@scut.edu.cn}
\author{Zheng Shi}
\ead{mieesz@mail.scut.edu.cn}

\cortext[cor1]{Corresponding author.}
\affiliation{organization={School of Microelectronics, South China University of Technology}, city={Guangdong}, country={China}}

\begin{abstract}
Graph-structured data is ubiquitous in the real world, and Graph Neural Networks (GNNs) have become increasingly popular in various fields due to their ability to process such irregular data directly. However, as data scale, GNNs become inefficient. Although parallel training offers performance improvements, increased communication costs often offset these advantages. To address this, this paper introduces CaPGNN, a novel parallel full-batch GNN training framework on single-server with multi-GPU. Firstly, considering the fact that the number of remote vertices in a partition is often greater than or equal to the number of local vertices and there may exist many duplicate vertices, we propose a joint adaptive caching algorithm that leverages both CPU and GPU memory, integrating lightweight cache update and prefetch techniques to effectively reduce redundant communication costs. Furthermore, taking into account the varying computational and communication capabilities among GPUs, we propose a communication- and computation-aware heuristic graph partitioning algorithm inspired by graph sparsification. Additionally, we implement a pipeline to overlap computation and communication. Extensive experiments show that CaPGNN improves training efficiency by up to 18.98x and reduces communication costs by up to 99\%, with minimal accuracy loss or even accuracy improvement in some cases. Finally, we extend CaPGNN to multi-machine multi-GPU environments. The code is available at \href{https://github.com/songxf1024/CaPGNN}{https://github.com/songxf1024/CaPGNN}.

\end{abstract}



\begin{keyword}
Graph Neural Network\sep Distributed System\sep Graph Partition
\end{keyword}
\end{frontmatter}

\section{Introduction}
\label{sec:introduction}
Unlike structured data such as images and text, graphs represent versatile data structures consisting of vertices and edges, which can effectively capture the complex relationships in irregular data. Graph Neural Networks (GNNs) are deep learning models specifically designed for graph data, capable of directly handling graph-structured information. Consequently, GNNs have seen widespread application in areas such as social networks~\cite{sharma2024survey}, recommendation systems~\cite{wu2022graph}, knowledge graphs~\cite{ye2022comprehensive}, image processing~\cite{sarlin2020superglue}, and electronic design automation~\cite{mirhoseini2021graph}. In addition, GNNs can also integrate with other technologies such as reinforcement learning~\cite{li2023task} and Transformers~\cite{yang2021graphformers}, to further expand their application scope and development potential.

With the rapid growth in data scale, traditional GNN training methods face scalability challenges. Although some methods attempt to improve computational efficiency through operator-level optimizations~\cite{10738209, 9355302}, the computation and memory demands of large-scale graphs often exceed the capacity of a single device~\cite{yang2022gnnlab}. Distributed GNNs overcome these limitations by leveraging multiple devices, partitioning large graphs into smaller subgraphs for parallel processing, thus improving training efficiency. Depending on different computational resources, distributed GNN training systems can be categorized into four types~\cite{shao2024distributed}: (a) single-machine multi-GPU systems, (b) GPU cluster systems, (c) CPU cluster systems, and (d) serverless and edge-device systems. Among these, single-machine multi-GPU systems are widely used in practice due to their high performance, simplicity, and cost-efficiency.

Distributed GNN training methods mainly include full-batch~\cite{gong2022graphite} and mini-batch~\cite{grattarola2021graph}. Mini-batch training uses graph sampling to reduce memory consumption and scale to large graphs. However, it introduces computational overhead, risks losing global information, and lacks theoretical convergence guarantees~\cite{huang2020ge}. In contrast, full-batch training preserves global graph information, offers stable gradients, and achieves higher accuracy~\cite{CDFGNN}, albeit at the expense of increased communication and synchronization overhead.

However, the inherent dependencies in graph structures introduce additional challenges for efficient distributed GNN training~\cite{shao2024distributed}. For example, under vertex-centric partitioning, a halo vertex may appear in multiple partitions, which requires inter-partition communication to exchange its features and leads to redundant communication due to repeated access. In addition, the large number of boundary vertices and high-dimensional vertex features exacerbate the communication overhead. Additionally, load imbalance arises from uneven vertex distribution, which existing graph partitioning methods fail to address, as they focus on balancing vertex count without considering GPU performance disparities.

To address these challenges, we propose \textbf{CaPGNN}, an efficient {\it \textbf{C}aching \textbf{a}nd \textbf{P}artitioning} framework for parallel full-batch GNN training on single-server with multi-GPU. We introduce a novel joint caching strategy to reduce communication redundancy associated with boundary vertices and propose a resource-aware graph partitioning algorithm that accounts for GPU performance differences, aiming to reduce communication and synchronization overhead. We summarize the key contributions as follows:
\begin{itemize}
    \item \textbf{To provide a holistic view of the problem of large scale GNN training using both CPU and GPU, we conduct systematical studies.} We share our analytical findings and thoughts, particularly on the impact of boundary vertices, i.e. {\it halo vertices}, on parallel GNN training. In addition, we summarize our findings in the form of three major research observations that motivate this work.

    \item \textbf{To reduce communication overhead from boundary vertices, we propose a \textit{Joint Adaptive Caching Algorithm} (JACA).} Specifically, we leverage CPU and GPU memory as a two-level cache to minimize redundant vertex communication and prioritize caching high-importance vertices evaluated by the analytical model. Performance is further optimized through pinned memory scheduling, lightweight cache updates, prefetching, asynchronous queues, and staleness-tolerant pipelines.
    
    \item \textbf{To minimize communication overhead and synchronization costs, we introduce a \textit{Resource-Aware Partitioning Algorithm} (RAPA).} It incorporates intuitive metrics to model GPU communication and computational capabilities and considers memory availability. Inspired by graph sparsification, it adjusts the scale of halo vertices during graph partitioning to reduce communication overhead while preserving accuracy.

    \item \textbf{To thoroughly evaluate the proposed CaPGNN, we perform comprehensive experiments.} Both theoretical analysis and experimental results confirm the convergence of CaPGNN. The results demonstrate that CaPGNN substantially improves training efficiency successfully. We further extend CaPGNN to multi-machine multi-GPU environments, demonstrating its applicability to distributed settings.
\end{itemize}

This paper is organized as follows. We analyze the related work in Section \ref{sec:related_work}. Section \ref{sec:motivation} explains our motivation. In Section \ref{sec:system_design}, we describe the proposed joint caching strategy and graph partitioning algorithm in detail. We describe the experimental setup and comparative studies in Section \ref{sec:experiments}. We also share our thoughts on the limitations of the current work and areas for future exploration in Section \ref{sec:discussion}. Finally, we conclude this paper in Section \ref{sec:conclusion}.

\section{Related Work}
\label{sec:related_work}
\subsection{Parallel and Distributed GNN Training}
With the exponential growth of data scale, traditional GNNs encounter efficiency bottlenecks in processing capacity. Distributed GNNs, which leverage multi-machine computation and storage resources efficiently, have gradually become a research hotspot. ROC~\cite{jia2020improving} improves system performance and scalability through optimizations in graph partitioning and memory management, although it also increases system complexity. SANCUS~\cite{peng2022sancus} introduces historical embeddings in full-graph training to reduce communication overhead; while effective in improving efficiency, this approach imposes significant computational burdens. NeutronStar~\cite{wang2022neutronstar} proposes a distributed GNN training system combining the Dependencies Cached approach and Dependencies Communicated approach, using a cost model to estimate communication overhead. DistDGL~\cite{zheng2020distdgl} is the distributed version of DGL; however, it is designed for CPU clusters. Similarly, AliGraph~\cite{yang2019aligraph} supports only CPU clusters as well. PipeGCN~\cite{wan2022pipegcn} optimizes DistDGL by incorporating stale embeddings and the pipeline to improve training efficiency. AdaQP~\cite{wan2023adaptive} reduces inter-worker communication by random quantization of messages but does not adequately account for the impact of heterogeneous GPUs and halo vertices.

Distributed GNNs are primarily designed to address scenarios in which graphs and features cannot fit in the memory of a single machine. However, their high economic costs and strict network environment requirements limit their applicability. In contrast, single-machine multi-GPU parallel training is simpler and more efficient. NeuGraph~\cite{ma2019neugraph} is a pioneering work that seamlessly integrates deep learning systems with graph processing systems. PaGraph~\cite{lin2020pagraph} supports sample-based GNN training on multiple GPUs. Legion~\cite{sun2023legion} optimizes multi-GPU training by leveraging the NVLink connectivity architecture. GNNLab~\cite{yang2022gnnlab} improves data parallelism and load balancing by dividing the GNN training process into a Sampler and a Trainer. NeutronOrch~\cite{ai2023neutronorch} optimizes data transfer between CPU and GPU using a cost model. However, these methods are focused primarily on sampling-based training scenarios.

\subsection{Hybrid CPU–GPU systems for GNN training.}
Beyond sampling-based training, recent work has also explored hybrid CPU–GPU architectures to scale full-graph GNN training. HongTu~\cite{wang2023hongtu} stores vertex data in CPU memory and offloads full-graph training to multiple GPUs, combining partition-based training with a recomputation–caching-hybrid management of intermediate data, deduplicated host–GPU communication, and cost-model guided graph reorganization to reduce communication overhead. G3~\cite{wan2023scalable} targets large-scale full-graph GNN training on GPU clusters, proposing a distributed training design with fine-grained peer-to-peer sharing of intermediate results, locality-aware iterative partitioning, and multi-level pipeline scheduling to balance workloads and overlap computation with communication. However, their designs primarily target multi-GPU servers evaluated under homogeneous high-end GPUs or distributed GPU clusters, lacking explicit modeling of GPU heterogeneity. In contrast, CaPGNN focuses on a common single-server multi-GPU environment where GPUs can be heterogeneous, making load imbalance and redundant cross-partition feature exchanges more pronounced.

\subsection{Caching Strategies in GNNs}
Caching mechanisms effectively reduce communication overhead. PaGraph~\cite{lin2020pagraph} introduces a caching strategy for sampling-based GNN training, but its static caching approach suffers from efficiency bottlenecks when dealing with large-scale graphs. BGL~\cite{liu2023bgl} employs a FIFO strategy combined with proximity-aware ordering to enhance cache hit rates and reduce overhead effectively. GNNLab~\cite{yang2022gnnlab} adopts a pre-sampling caching strategy, achieving efficient and robust training results. Although these caching strategies demonstrate notable performance improvements in specific scenarios, there are still many areas for further exploration. For instance, the design of multi-level caching could better accommodate storage requirements at different tiers, optimizing data transmission efficiency could further reduce communication latency, and leveraging hardware characteristics may unlock greater performance potential.

\subsection{Graph Partitioning in GNNs}
Balanced graph partitioning is a key factor in achieving efficient parallel and distributed graph neural network systems. Pregel~\cite{malewicz2010pregel} adopts a vertex-based partitioning approach. PowerGraph~\cite{gonzalez2012powergraph} proposes an edge-based partitioning strategy that evenly distributes edges across machines to mitigate the effects of power-law graphs. PowerLyra~\cite{chen2019powerlyra} introduces a hybrid partitioning strategy that selects the partitioning method based on a vertex degree threshold. The classic METIS~\cite{karypis1998fast} method generates subgraphs of similar sizes through three stages: Coarsening, Initial Partitioning, and Uncoarsening. For massive-scale graphs, streaming partitioners such as Fennel~\cite{tsourakakis2014fennel} process the graph in a single pass and optimize a locality--balance objective, providing an efficient alternative to offline partitioning while still aiming to reduce cross-partition dependencies. Cluster-GCN~\cite{chiang2019cluster} uses METIS to partition the graph into multiple clusters and performs random sampling on these clusters to construct subgraphs. DGCL~\cite{cai2021dgcl} calculates the number of vertices in subgraphs based on the replication factor. However, most of these methods assume equal partition sizes, neglecting variations in device performance and resource constraints, which can lead to load imbalance in practice.

Moreover, graph partition introduces halo vertices and remote neighbor accesses across partitions, often becoming a dominant source of communication overhead, which has been recognized in prior GNN systems. HongTu~\cite{wang2023hongtu} reduces redundant host-GPU communication by exploiting reuse of duplicated neighbor accesses among partitions and reorganizing subgraphs to improve reuse while offloading vertex data to CPU memory. BNS-GCN~\cite{wan2022bns} identifies the explosion of boundary vertices as a key bottleneck in partition-parallel full-graph training and adopts random boundary-vertex sampling to reduce communication and memory costs in a lossy manner. Sancus~\cite{peng2022sancus} reduces cross-partition communication by caching historical embeddings and adaptively skipping broadcasts with bounded staleness in decentralized training. Different from these works, CaPGNN elevates halo vertices as a first-class abstraction in a single-server multi-GPU full-batch setting, and co-designs resource-aware partitioning with a halo-centric two-level cache to reduce redundant halo replicas.

\section{Background and Motivation}
\label{sec:motivation}
In this section, we introduce the background related to parallel full-batch GNN training and present a series of key observations that motivate CaPGNN. The key symbols and notations within the paper are summarized in Table~\ref{tab:notations}.

\subsection{Graph Neural Networks}
\begin{figure}[t]
    \centering
    \setlength{\abovecaptionskip}{0pt}
    \setlength{\belowcaptionskip}{0pt}
    \includegraphics[width=\linewidth]{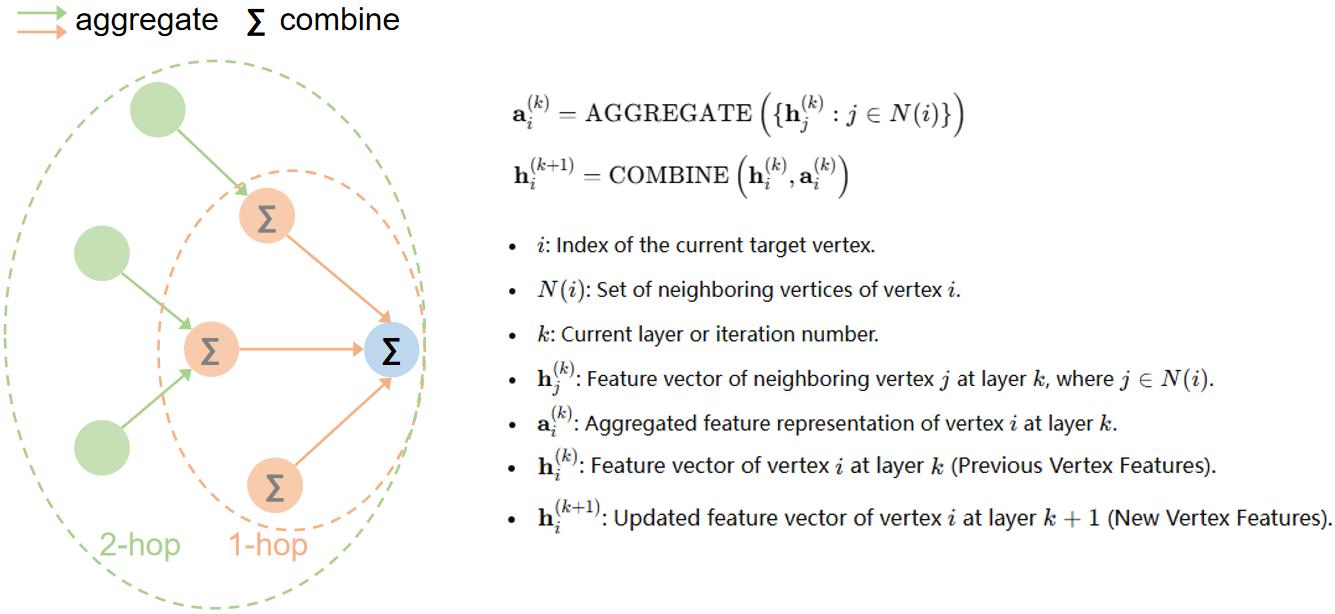}
    \caption{Message passing in a 2-layer GNN. The blue vertex as the target vertex updated via aggregation and combination.}
    \label{fig:gnn_processing}
    \vspace{-1em}
\end{figure}

A graph $G=(V,E)$ is a data structure that represents pairwise relationships between objects, consisting of a set of vertices $V$ and edges $E$. GNNs are deep learning models that learn directly from graph-structured data \cite{schlichtkrull2018modeling}. A basic GNN involves two steps: \textbf{Aggregate} and \textbf{Combine}. In the aggregation step, each vertex gathers information from its neighboring vertices to form a unified representation, capturing the dependencies between vertices in the graph. In the combination step, the vertex integrates the aggregated information with its own features to generate a new vertex representation. Fig.~\ref{fig:gnn_processing} illustrates this process. Through multiple steps of message passing, a vertex progressively acquires information from increasingly distant vertices. This process is represented as:
\begin{equation}
    \label{eq:gnn_formula}
	\begin{aligned}
		\resizebox{0.89\hsize}{!}{
			   $h_i^{(k+1)} = \text{COMBINE}\left(h_i^{(k)}, \text{AGGREGATE}\left(\{h_j^{(k)} : j \in N(i)\}\right)\right)$,
		}
	 \end{aligned}
\end{equation}
where $h_i^{(k)}$ is the embedding of the vertex $v_i$ at the $k$-th iteration, and $N(i)$ denotes the neighbors of the vertex $v_i$. The \textit{AGGREGATE} function combines the embeddings of these neighbors, while the \textit{COMBINE} function generates new embeddings by combining the previous embeddings of the vertex with the aggregated neighborhood information. It is evident that GNN models are highly sensitive to dependencies between vertices.

\subsection{Graph Partition}
In distributed and parallel GNN training, graph partitioning is typically required, where each worker such as a GPU is responsible for training on a partitioned subgraph. Balanced graph partitioning is a key factor in achieving efficient distributed and parallel GNN systems. Depending on the partitioning algorithm, graph partitioning can be classified into vertex-centric (a.k.a. edge-cut) and edge-centric (a.k.a. vertex-cut). The former partitions the graph based on vertices and focuses on minimizing the number of edges crossing partitions, while the latter partitions the graph based on edges, aiming to improve load balancing. Edge-centric partitioning is preferred for large-scale power-law graphs~\cite{windgp}.

\begin{figure}[htbp]
    \centering
    \setlength{\abovecaptionskip}{0pt}
    \setlength{\belowcaptionskip}{0pt}
    \includegraphics[width=\linewidth]{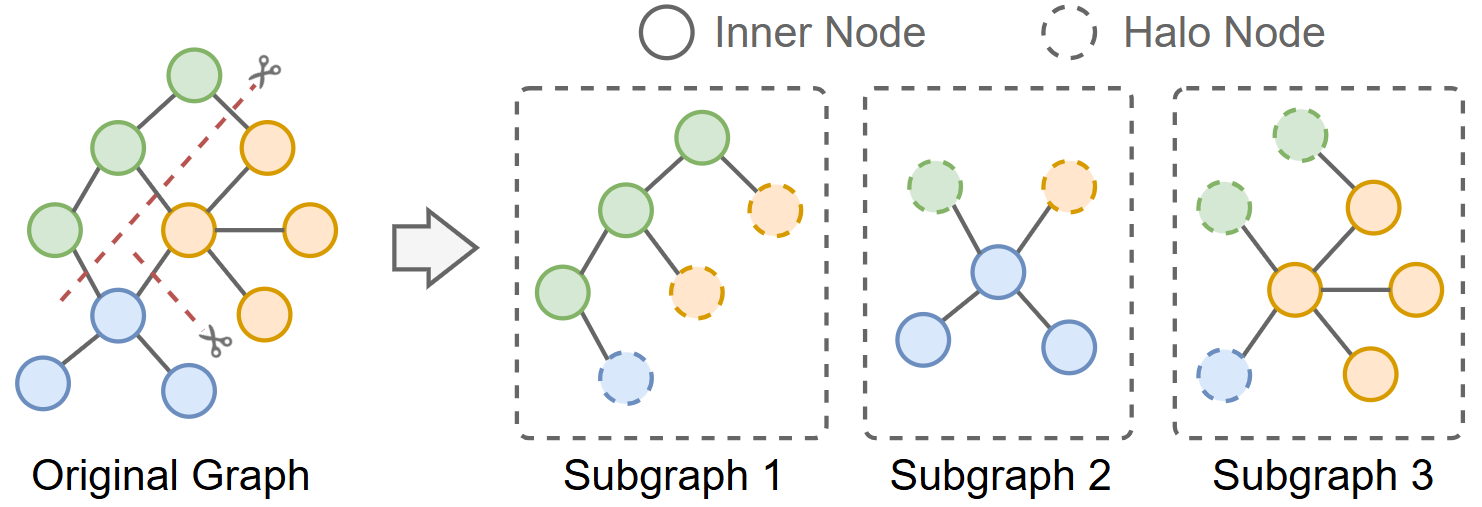}
    \caption{Vertex-centric graph partition. The original graph is partitioned into two subgraphs, with solid circles as inner vertices and dashed circles as halo vertices.}
    \label{fig:graph_partition}
    \vspace{-1em}
\end{figure}

Some partitioning algorithms~\cite{karypis1998metis} typically retain halo vertices on each subgraph. These vertices do not belong to the current partition but represent the connections between the current partition and others. In other words, halo vertices are the endpoints of edges that span across partitions, necessitating inter-partition communication during computation. Fig.~\ref{fig:graph_partition} shows an example of vertex-centric graph partitioning. For GNNs, the communication costs of vertex features and embeddings are typically higher than gradients~\cite{wan2023adaqp}. Consequently, communication overhead is significantly influenced by the number of halo vertices and feature dimensionality~\cite{verma2017an}.

\subsection{Parallel Full-Batch GNN Training}
\begin{figure}[htbp]
    \centering
    \includegraphics[width=\linewidth]{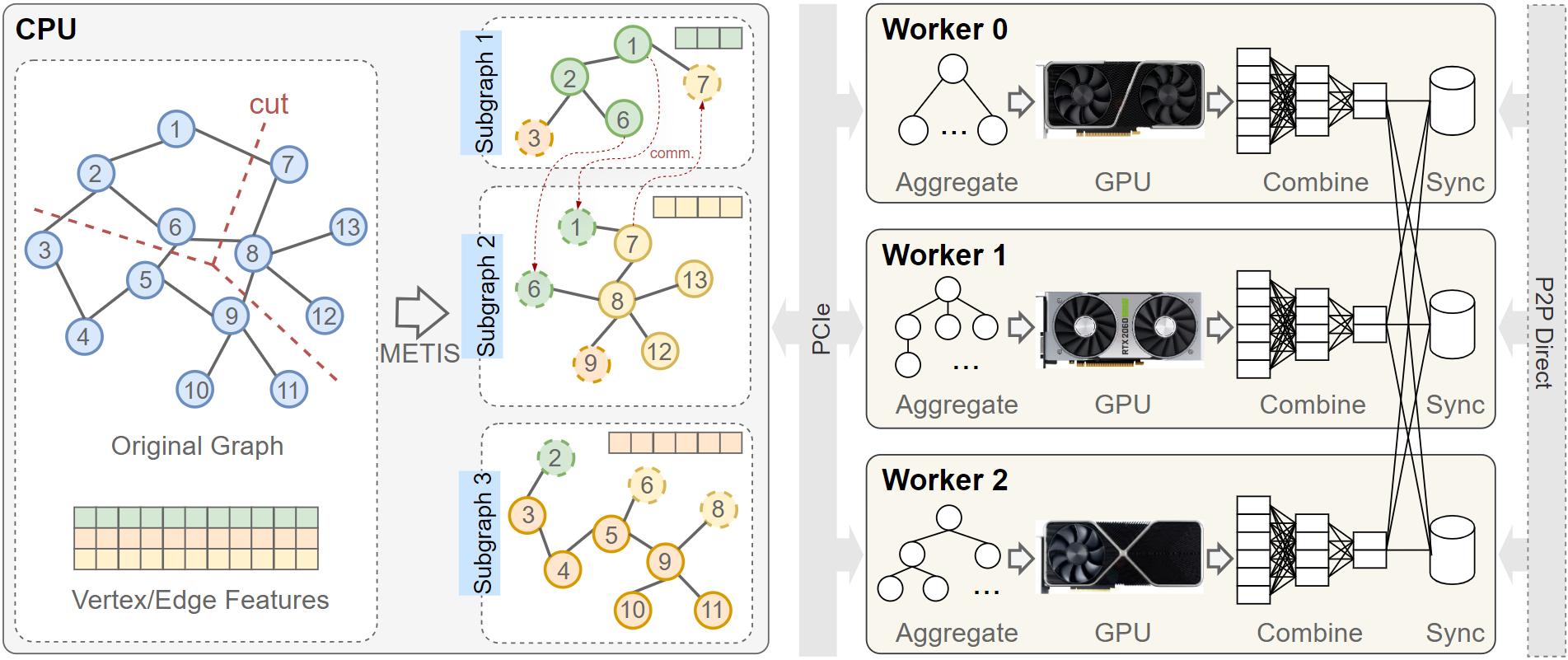}
    \caption{The workflow of a typical parallel GNN process. The original graph is partitioned into subgraphs using METIS. Then, subgraphs and vertex/edge features are distributed to workers. Each worker performs localized computation on its assigned subgraph. Boundary informations are exchanged between workers through PCIe, or via P2P links if available.}
    \label{fig:parallel_gnn_workflow}
\end{figure}

Parallel GNN training can be broadly categorized into data parallelism and model parallelism, with data parallelism being more common due to the typically small size of GNN models. As shown in Fig.~\ref{fig:parallel_gnn_workflow}, in full-batch training, the entire graph is used within a single iteration. Compared to sampling-based mini-batch training, full-batch training generally achieves higher accuracy and faster convergence. Therefore, parallel full-batch GNN training involves partitioning the graph into multiple subgraphs, distributing these subgraphs across different workers (e.g., GPUs) and performing localized training on each subgraph. During this process, boundary vertex features are exchanged between workers. At the end of each backpropagation step, all workers synchronize gradients to update global model parameters.

Unlike traditional deep learning methods, GNN computations are inherently dependent on vertices. Therefore, in large-scale distributed and parallel GNN processing, the bottleneck arises from load imbalance caused by irregular memory access patterns and computational issues~\cite{9761891Huang}. This also indicates that achieving a balance between worker performance and partition sizes in graph partitioning is crucial for enabling efficient parallel GNN systems.

\subsection{Motivation}
\label{subsec:motivation}
Our motivation is primarily based on the following empirical observations, which are consistent with prior studies~\cite{wang2023hongtu,wan2022bns}, and we further quantify these phenomena.

\begin{figure*}[htbp]
    \centering
    \setlength{\abovecaptionskip}{0pt}
    \setlength{\belowcaptionskip}{0pt}
    \begin{subfigure}[t]{0.24\linewidth}
        \centering
        \setlength{\abovecaptionskip}{0pt}
        \setlength{\belowcaptionskip}{0pt}
        \includegraphics[width=\linewidth]{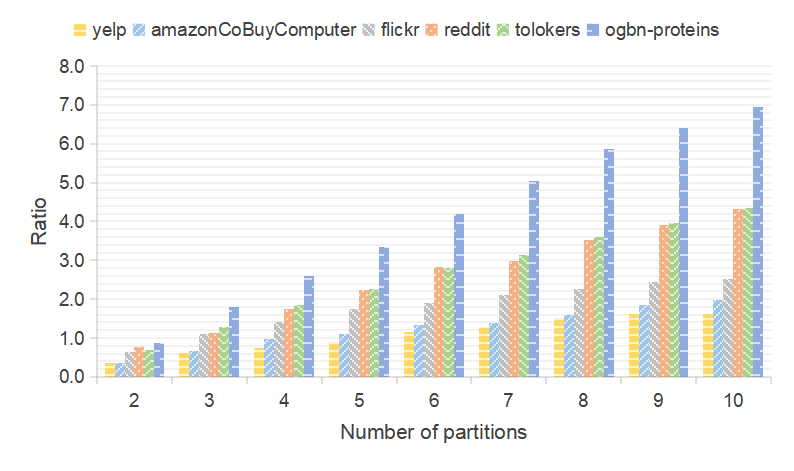}
        \caption{METIS, 1-hop neighbors.}
        \label{fig:ratio_halo2inner_hop1_metis}
    \end{subfigure}
    \hfill
    \begin{subfigure}[t]{0.24\linewidth}
        \centering
        \setlength{\abovecaptionskip}{0pt}
        \setlength{\belowcaptionskip}{0pt}
        \includegraphics[width=\linewidth]{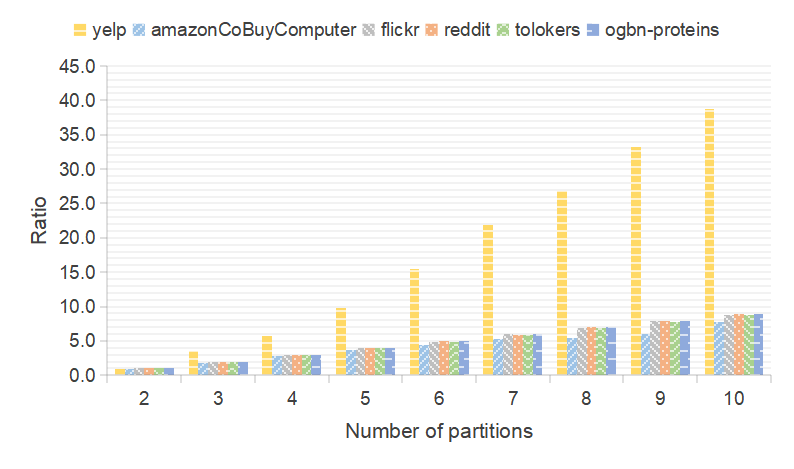}
        \caption{METIS, 2-hop neighbors.}
        \label{fig:ratio_halo2inner_hop2_metis}
    \end{subfigure}
    \hfill
    \begin{subfigure}[t]{0.24\linewidth}
        \centering
        \setlength{\abovecaptionskip}{0pt}
        \setlength{\belowcaptionskip}{0pt}
        \includegraphics[width=\linewidth]{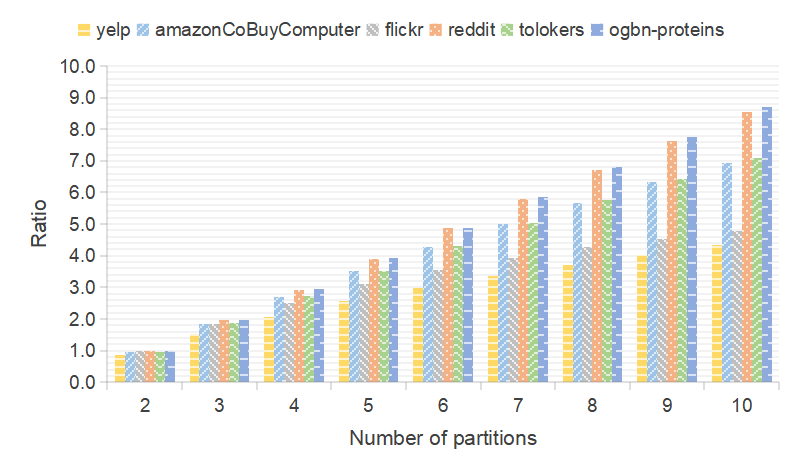}
        \caption{Random, 1-hop neighbors.}
        \label{fig:ratio_halo2inner_hop1_random}
    \end{subfigure}
    \hfill
    \begin{subfigure}[t]{0.24\linewidth}
        \centering
        \setlength{\abovecaptionskip}{0pt}
        \setlength{\belowcaptionskip}{0pt}
        \includegraphics[width=\linewidth]{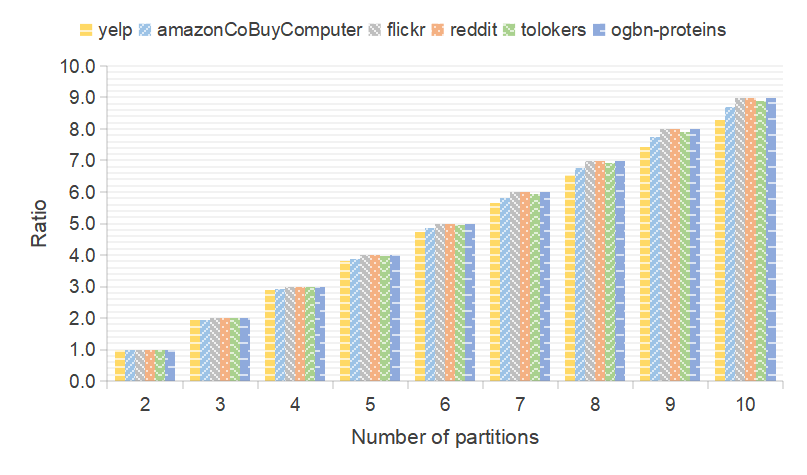}
        \caption{Random, 2-hop neighbors.}
        \label{fig:ratio_halo2inner_hop2_random}
    \end{subfigure}
    \caption{Ratio of halo vertices to inner vertices for different numbers of partitions, hops, datasets, and partition methods.}
    \label{fig:ratio_halo2inner}
\end{figure*}

\textbf{\textit{Observation 1}: The number of total halo vertices can be greater than or equal to the number of inner vertices.}
Graph partitioning methods often retain a certain number of halo vertices for each subgraph, their quantity influenced by factors such as the number of partitions, hops, and partitioning methods. To further analyze this phenomenon, we conduct experiments on several graph datasets using two classical partitioning methods: METIS and Random. These experiments evaluate the relationship between halo vertices and inner vertices under various conditions. Fig.~\ref{fig:ratio_halo2inner} indicates a clear trend: the number and proportion of total halo vertices increase significantly with the number of partitions and hops. This phenomenon directly results in a substantial increase in storage and communication resource demands, particularly in large-scale distributed graph training. Storing all halo vertices significantly increases memory usage and exacerbates communication costs. These issues can significantly constrain the overall system performance. Therefore, optimizing halo vertices is critical.

\begin{figure*}[!htbp]
    \centering
    \setlength{\abovecaptionskip}{0pt}
    \setlength{\belowcaptionskip}{0pt}
    \begin{subfigure}[t]{0.19\linewidth}
        \centering
        \includegraphics[width=\linewidth]{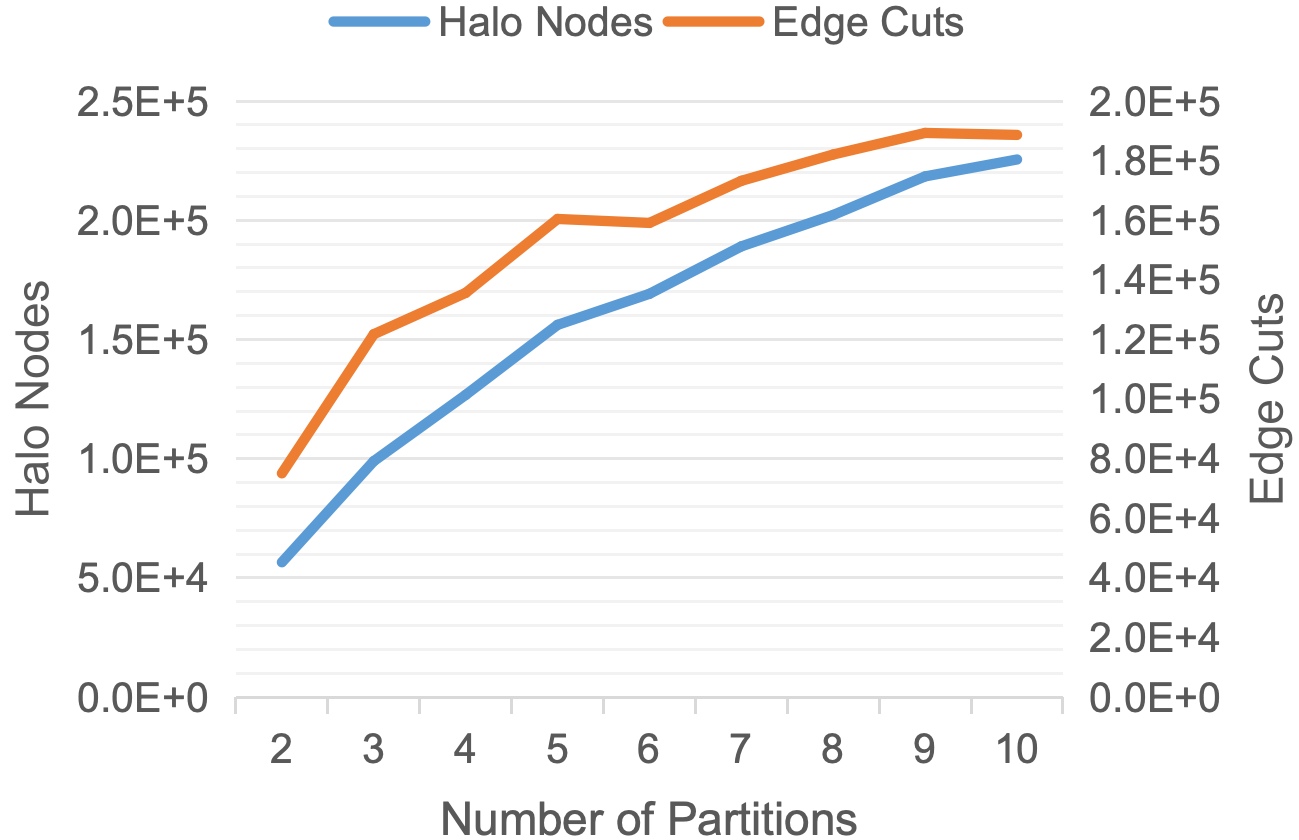}
        \caption{Flickr.}
        \label{fig:ec_hl_flickr}
    \end{subfigure}
    \hfill
    \begin{subfigure}[t]{0.19\linewidth}
        \centering
        \includegraphics[width=\linewidth]{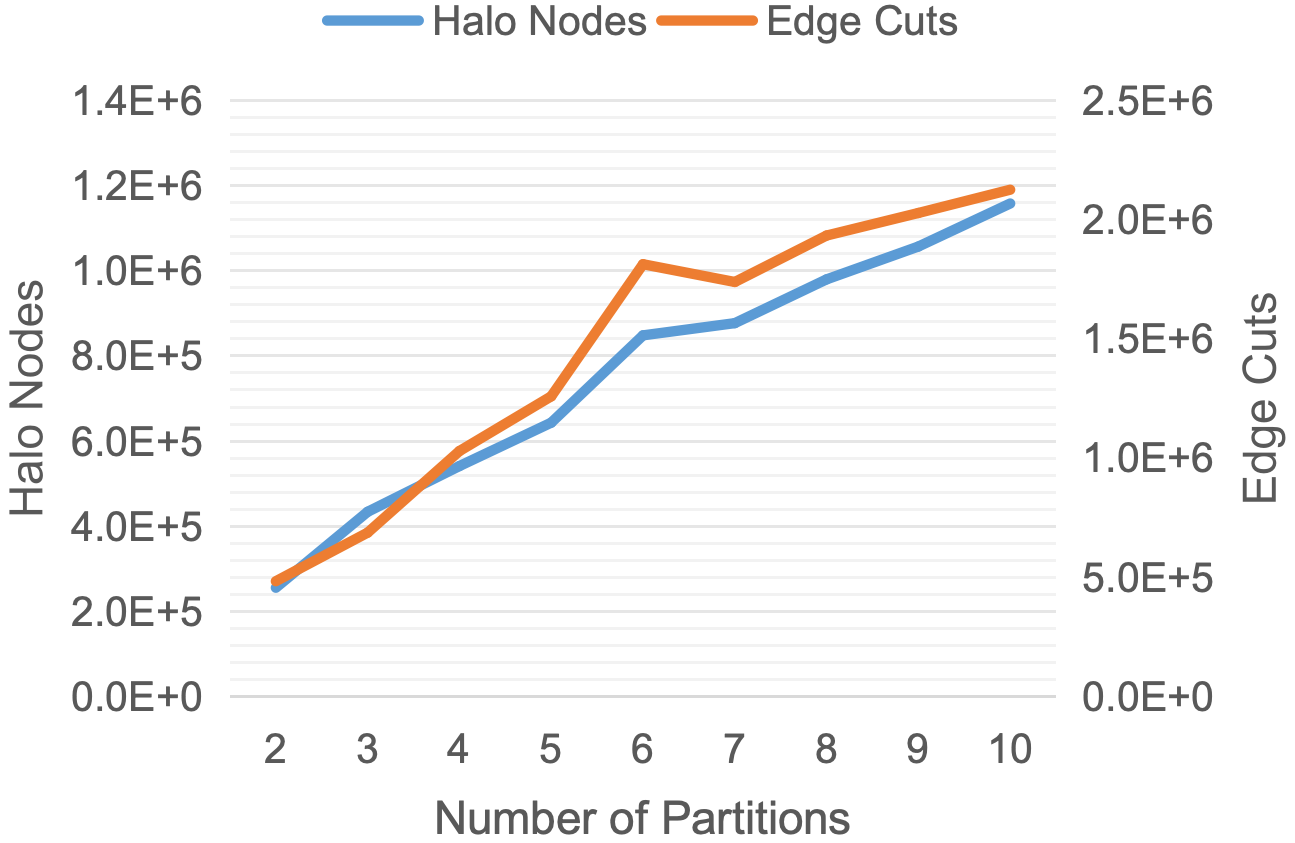}
        \caption{Yelp.}
        \label{fig:ec_hl_yelp}
    \end{subfigure}
    \hfill
    \begin{subfigure}[t]{0.19\linewidth}
        \centering
        \includegraphics[width=\linewidth]{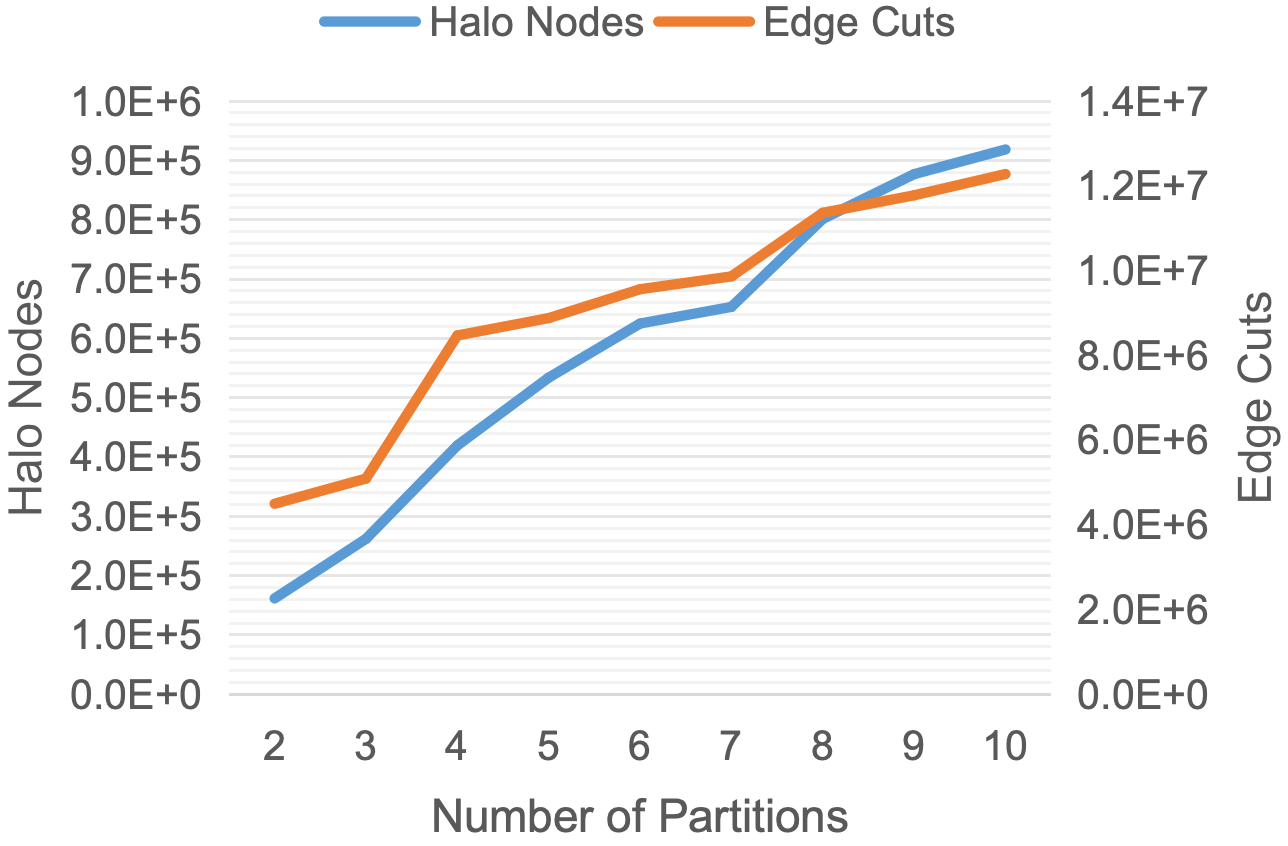}
        \caption{Reddit.}
        \label{fig:ec_hl_reddit}
    \end{subfigure}
    \hfill
    \begin{subfigure}[t]{0.19\linewidth}
        \centering
        \includegraphics[width=\linewidth]{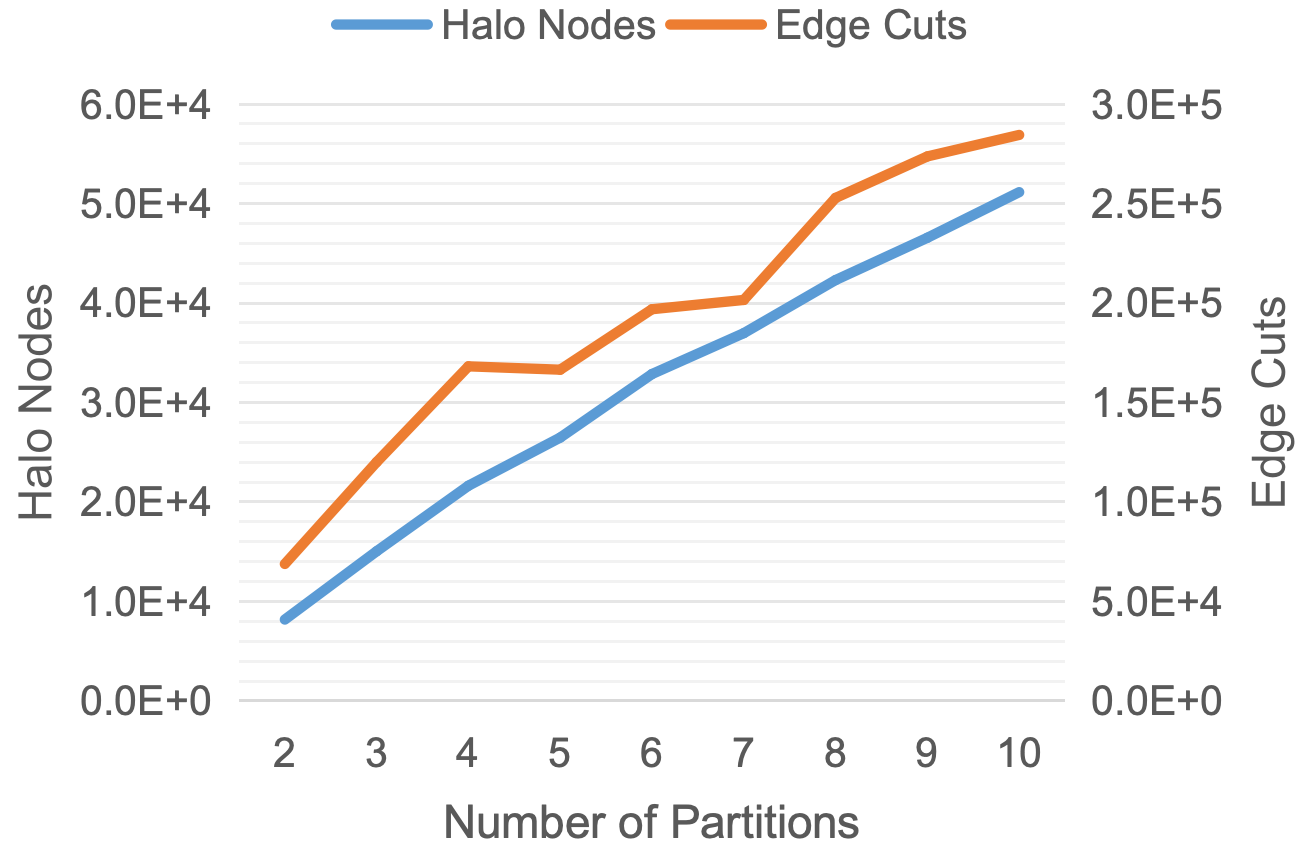}
        \caption{Tolokers.}
        \label{fig:ec_hl_tolokers}
    \end{subfigure}
    \hfill
    \begin{subfigure}[t]{0.19\linewidth}
        \centering
        \includegraphics[width=\linewidth]{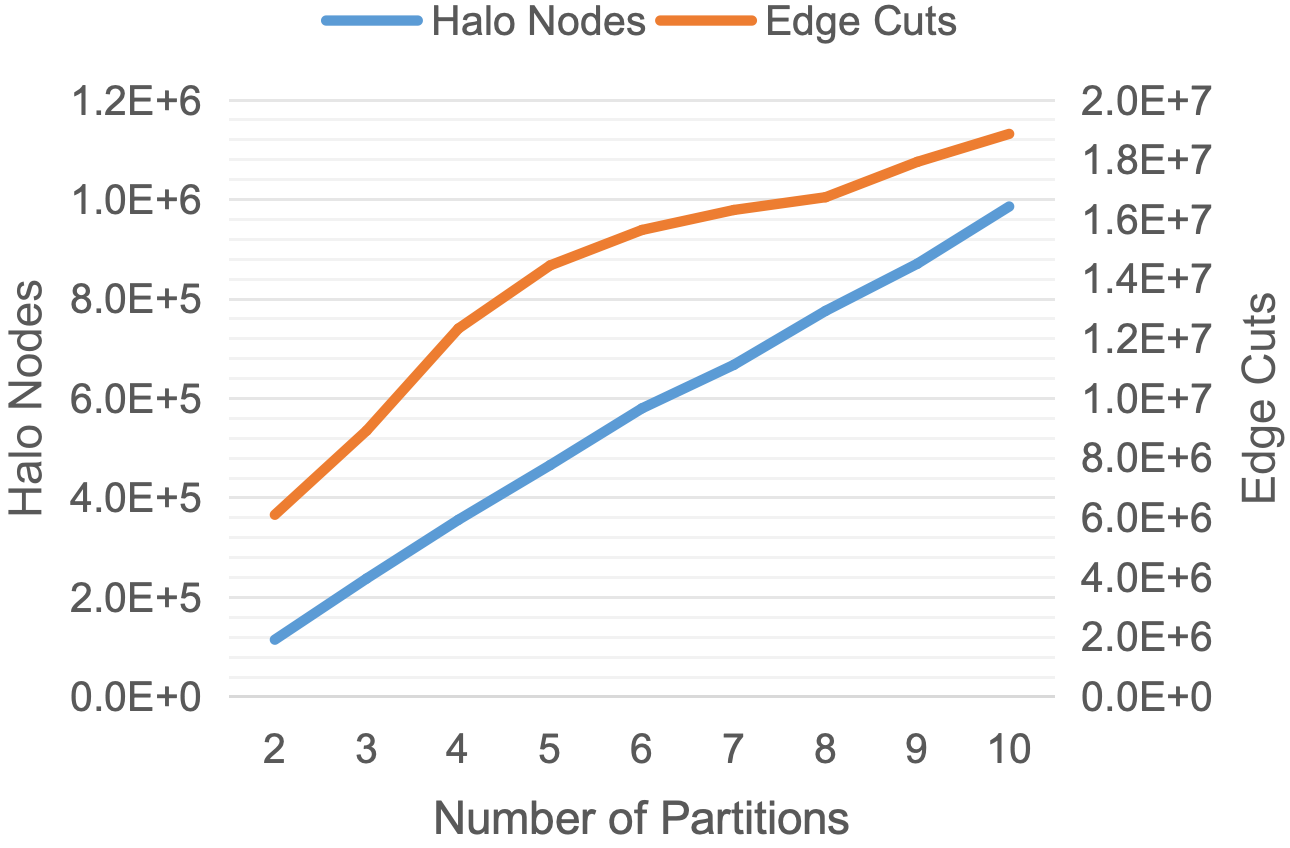}
        \caption{ogbn-proteins.}
        \label{fig:ec_hl_proteins}
    \end{subfigure}
    \caption{Correlation between edge cut and total 1 hop halo vertex count across datasets and partition numbers. Here, edge cut counts unique inter-partition edges, where each bidirectional pair is counted once.}
    \label{fig:edgecuts_to_halo}
    \vspace{-1em}
\end{figure*}

We further observe a clear positive correlation between the number of edge cuts and the total 1-hop halo vertex counts as the number of partitions increases, as shown in Fig~\ref{fig:edgecuts_to_halo}. This indicates that minimizing edge cuts tends to reduce halo size, while reinforcing halo vertices as a more communication-relevant indicator.

\begin{figure*}[htbp]
    \centering
    \setlength{\abovecaptionskip}{0pt}
    \setlength{\belowcaptionskip}{0pt}
    \begin{subfigure}[t]{0.24\linewidth}
        \centering
        \setlength{\abovecaptionskip}{0pt}
        \setlength{\belowcaptionskip}{0pt}
        \includegraphics[width=\linewidth]{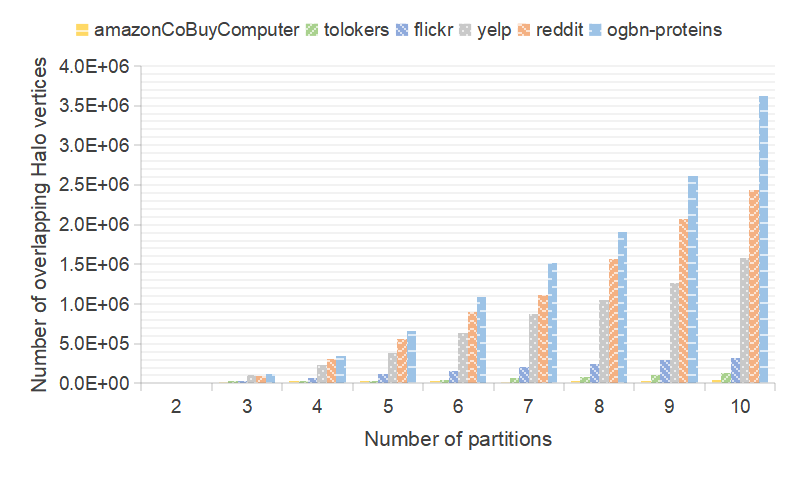}
        \caption{METIS, 1-hop neighbors.}
        \label{fig:overlap_1hop_metis}
    \end{subfigure}
    \hfill
    \begin{subfigure}[t]{0.24\linewidth}
        \centering
        \setlength{\abovecaptionskip}{0pt}
        \setlength{\belowcaptionskip}{0pt}
        \includegraphics[width=\linewidth]{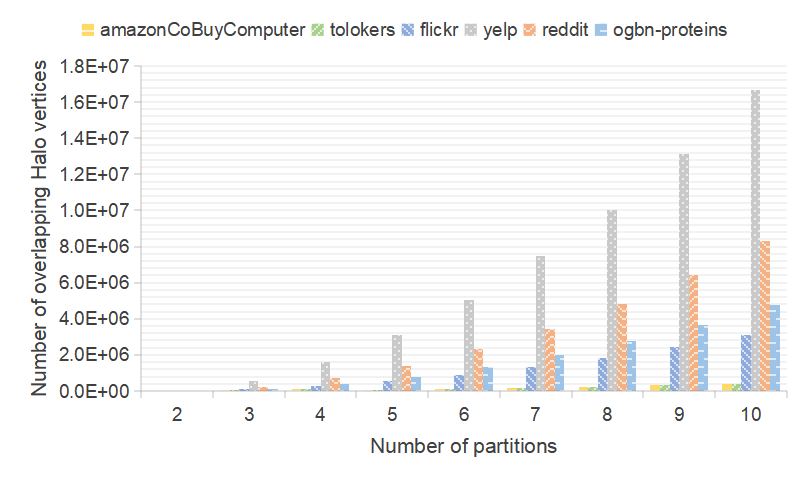}
        \caption{METIS, 2-hop neighbors.}
        \label{fig:overlap_2hop_metis}
    \end{subfigure}
    \hfill
    \begin{subfigure}[t]{0.24\linewidth}
        \centering
        \setlength{\abovecaptionskip}{0pt}
        \setlength{\belowcaptionskip}{0pt}
        \includegraphics[width=\linewidth]{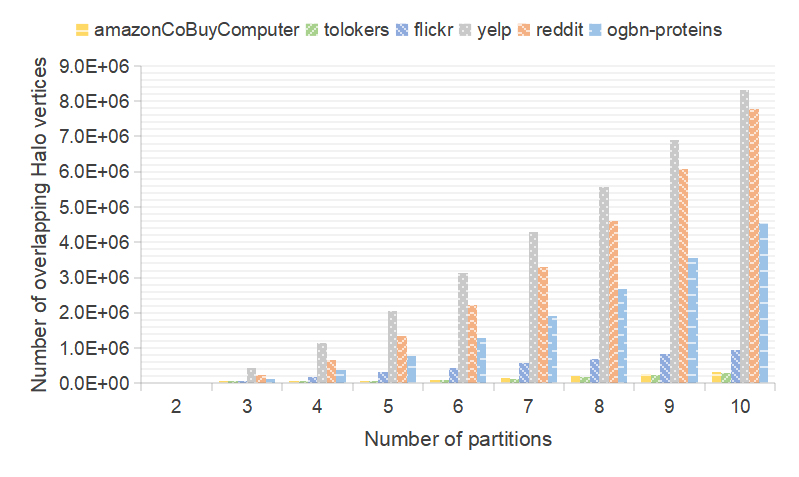}
        \caption{Random, 1-hop neighbors.}
        \label{fig:overlap_1hop_random}
    \end{subfigure}
    \hfill
    \begin{subfigure}[t]{0.24\linewidth}
        \centering
        \setlength{\abovecaptionskip}{0pt}
        \setlength{\belowcaptionskip}{0pt}
        \includegraphics[width=\linewidth]{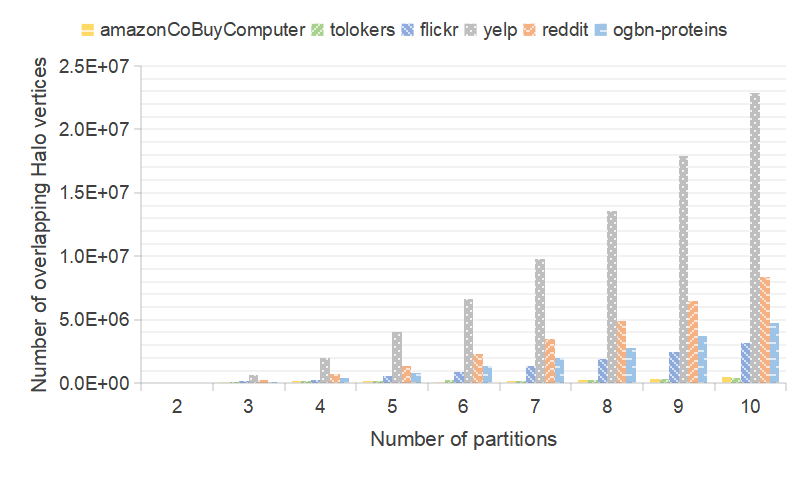}
        \caption{Random, 2-hop neighbors.}
        \label{fig:overlap_2hop_random}
    \end{subfigure}
    \caption{Number of overlapping halo vertices for different number of partitions, hops, datasets, and partition methods.}
    \label{fig:ratio_halo_copy}
    \vspace{-1em}
\end{figure*}

\textbf{\textit{Observation 2}: Duplicate halo vertices between partitions increase with the number of partitions and hops.}
In the Edge-centric partitioning method, vertices connected to the edges that span multiple partitions are duplicated across partitions, thereby becoming halo vertices for those partitions. Using the same experimental setup as in Observation 1, we evaluate the overlap rate of halo vertices under varying numbers of partitions and hops.

As shown in Fig.~\ref{fig:ratio_halo_copy}, the overlap of halo vertices between partitions becomes increasingly severe as the number of partitions and hops increases. This redundancy results in excessive data storage and significantly increases inter-partition communication. To address this issue, we propose a hierarchical caching structure that leverages both GPU and CPU memory to optimize the storage and access efficiency of halo vertices.

\textbf{\textit{Observation 3}: GPUs exhibit varying computational and communication capabilities.} 
Even for the same GPU model, subtle performance variations may arise due to hardware environment or system-level factors. To visualize these differences, we conduct experiments on various GPUs to evaluate their performance in computational tasks, such as \textit{Matrix Multiplication} (MM) and \textit{Sparse Matrix-Matrix Multiplication} (SpMM, sparsity=99.6\%), and communication tasks, including \textit{Host-to-Device} (H2D), \textit{Device-to-Host} (D2H), and \textit{Intra-Device Transfer} (IDT). Each task is performed using a $16384 \times 16384$ \textit{float32} matrix and repeated 50 times to minimize randomness. 

\begin{table}[!htbp]
    \centering
    \setlength{\abovecaptionskip}{0em}
    \setlength{\belowcaptionskip}{0em}
    \caption{Comparative performance of GPUs in computation tasks (MM and SpMM) and communication tasks (H2D, D2H, IDT), measured in seconds.}
    \label{tab:gpu_extended_performance}
    \resizebox{\linewidth}{!}{
        \begin{tabular}{lcccccc}
            \Xhline{1.5pt}
            \textbf{GPU}       & \textbf{ID} & \textbf{MM}       & \textbf{SpMM} & \textbf{H2D}       & \textbf{D2H}       & \textbf{IDT}       \\ 
            \Xhline{1.5pt}
            \multirow{6}{*}{\textbf{RTX 3090}}  & 1  & 0.1409 ± 0.0005 & 0.1067 ± 0.0057 & 0.1184 ± 0.0031 & 0.1217 ± 0.0043 & 0.0014 ± 0.0 \\
                                       & 2  & 0.1351 ± 0.0008 & 0.1054 ± 0.0078 & 0.1196 ± 0.0076 & 0.1207 ± 0.0031 & 0.0014 ± 0.0 \\
                                       & 3  & 0.1389 ± 0.0005 & 0.1069 ± 0.0059 & 0.1198 ± 0.0090 & 0.1207 ± 0.0035 & 0.0014 ± 0.0 \\
                                       & 4  & 0.1372 ± 0.0005 & 0.1060 ± 0.0057 & 0.1188 ± 0.0059 & 0.1198 ± 0.0081 & 0.0014 ± 0.0 \\
                                       & 5  & 0.1393 ± 0.0006 & 0.1072 ± 0.0056 & 0.1224 ± 0.0093 & 0.1242 ± 0.0031 & 0.0014 ± 0.0 \\
                                       & 6  & 0.1385 ± 0.0008 & 0.1058 ± 0.0055 & 0.1194 ± 0.0077 & 0.1208 ± 0.0031 & 0.0014 ± 0.0 \\ 
            \Xhline{1pt}
            \multirow{2}{*}{\textbf{Tesla A40}} & 7  & 0.1416 ± 0.0011 & 0.1195 ± 0.0035 & 0.1175 ± 0.0077 & 0.1184 ± 0.0003 & 0.0021 ± 0.0 \\
                                       & 8  & 0.1426 ± 0.0011 & 0.1201 ± 0.0050 & 0.1198 ± 0.0091 & 0.1193 ± 0.0003 & 0.0021 ± 0.0 \\ 
            \Xhline{1pt}
            \multirow{2}{*}{\textbf{RTX 3060}}  & 9  & 0.3393 ± 0.0016 & 0.1948 ± 0.0042 & 0.1223 ± 0.0054 & 0.1240 ± 0.0027 & 0.0038 ± 0.0 \\
                                       & 10 & 0.3485 ± 0.0019 & 0.1975 ± 0.0044 & 0.1217 ± 0.0064 & 0.1232 ± 0.0034 & 0.0038 ± 0.0 \\ 
            \Xhline{1pt}
            \multirow{2}{*}{\textbf{RTX 2060}}  & 11 & 0.4942 ± 0.0030 & 0.2934 ± 0.0044 & 0.1188 ± 0.0083 & 0.1186 ± 0.0034 & 0.0033 ± 0.0 \\
                                       & 12 & 0.5002 ± 0.0016 & 0.2975 ± 0.0043 & 0.1195 ± 0.0061 & 0.1204 ± 0.0033 & 0.0033 ± 0.0 \\ 
            \Xhline{1pt}
            \multirow{2}{*}{\textbf{GTX 1660Ti}}& 13 & 0.9791 ± 0.0069 & 0.3370 ± 0.0034 & 0.1220 ± 0.0081 & 0.1232 ± 0.0031 & 0.0057 ± 0.0 \\
                                       & 14 & 1.0084 ± 0.0166 & 0.3447 ± 0.0030 & 0.1255 ± 0.0112 & 0.1256 ± 0.0031 & 0.0057 ± 0.0 \\ 
            \Xhline{1pt}
            \multirow{2}{*}{\textbf{GTX 1650}}  & 15 & 1.2730 ± 0.0028 & 0.6316 ± 0.0022 & 0.1246 ± 0.0072 & 0.1249 ± 0.0032 & 0.0094 ± 0.0 \\
                                       & 16 & 1.2755 ± 0.0030 & 0.6330 ± 0.0025 & 0.1259 ± 0.0067 & 0.1257 ± 0.0033 & 0.0094 ± 0.0 \\ 
            \Xhline{1.5pt}
        \end{tabular}
    }
\end{table}

The experimental results are presented in Table~\ref{tab:gpu_extended_performance}. These performance differences underscore the key considerations for achieving balanced load distribution in parallel systems. To address this, we propose a resource-aware graph partitioning algorithm that adjusts subgraph sizes to match the performance of each GPU closely.

\section{System Design}
\label{sec:system_design}
\begin{figure}[htbp]
    \centering
    \setlength{\abovecaptionskip}{0pt}
    \setlength{\belowcaptionskip}{0pt}
    \includegraphics[width=0.9\linewidth]{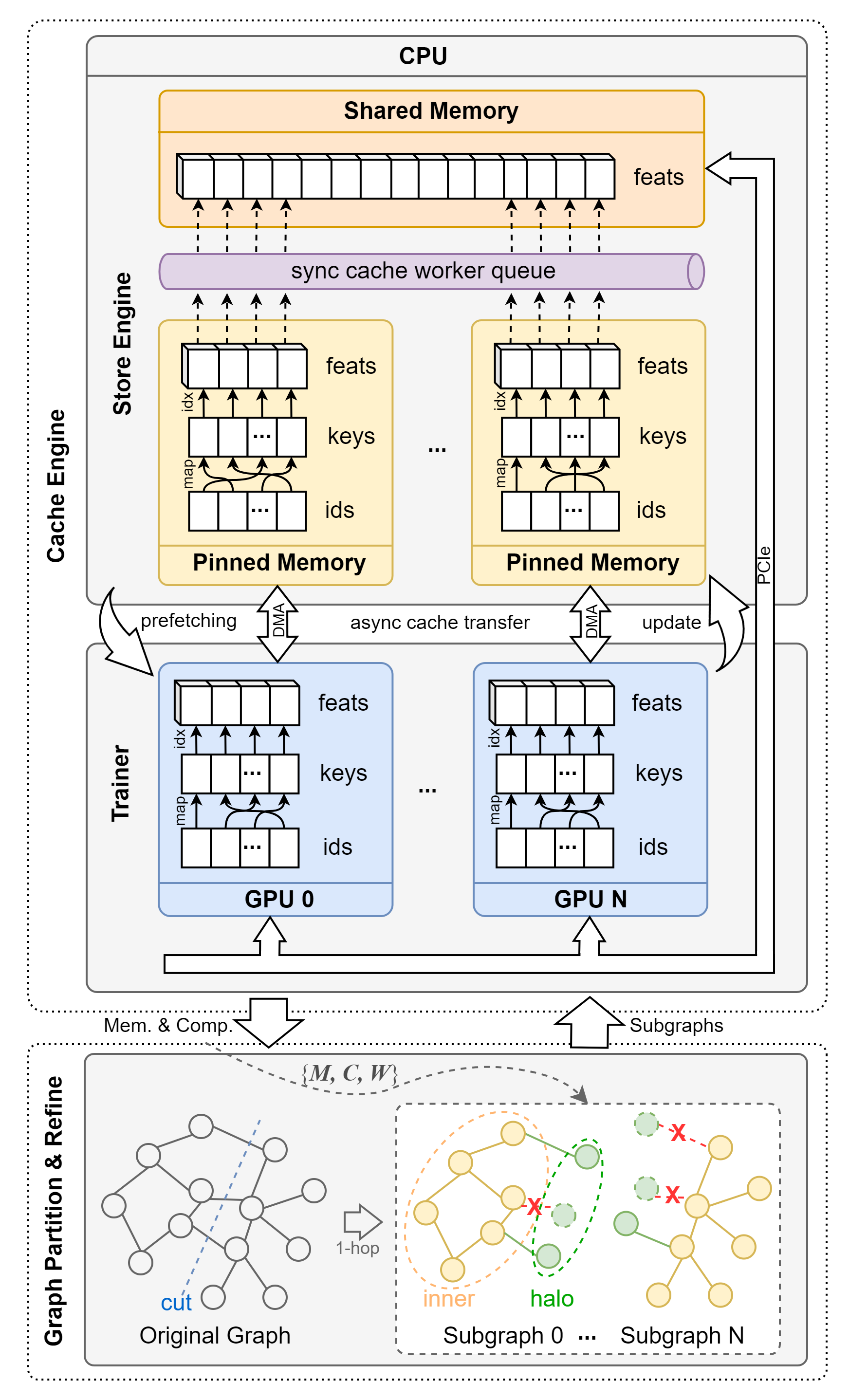}
    \caption{Overall architecture of CaPGNN.}
    \label{fig:overall_architecture}
    \vspace{-1em}
\end{figure}

We propose CaPGNN, a novel {\it \textbf{C}aching \textbf{a}nd \textbf{P}artitioning} framework for parallel full-batch GNN training. CaPGNN reduces redundant communication and ensures as much load balancing as possible. Training is conducted on a single machine with multiple GPUs (i.e., workers), with each GPU responsible for training a subgraph. Fig.~\ref{fig:overall_architecture} shows the overall architecture. Note that the design principles of CaPGNN is easily extensible to distributed systems, as we demonstrate in the experiments.

\begin{table}[!htbp]
    \centering
    \caption{Summary of the key symbols and notations.}
    \setlength{\abovecaptionskip}{0em}
    \setlength{\belowcaptionskip}{0em}
    \label{tab:notations}
    \resizebox{\columnwidth}{!}{%
        \renewcommand{\arraystretch}{1.4}
        \begin{tabular}{c||l}
            \Xhline{1.5pt}
            \textbf{Notation} & \textbf{Description} \\
            \Xhline{1pt}
            $G, G_i$ & The original graph and the $i$-th subgraph (partition) of $G$ \\
            $V, E$ & The vertex set and the edge set of graph $G$ \\ 
            $|V|, |E|$ & The order and size of graph $G$ \\
            $N(i)$ & The neighbor set of vertex $v_i$ \\
            $v_i, e_{ij}$ & The $i$-th vertex, and the edge between $v_i$ and $v_j$ \\
            $R(v_k)$ & The overlap ratio of vertex \( v_k \) across partitions \\
            $P, T$ & The total number of subgraphs (partitions) and epochs \\
            $H, H(G_i)$ & The set of halo vertices across all partitions and in subgraph $G_i$ \\
            $\mathbb{I}(\cdot)$ & The indicator function \\
            $M_{\text{GPU}, i}$ & The available GPU memory for subgraph $G_i$ \\
            $M_{\text{CPU}}$ & The available CPU memory \\
            $C_{\text{GPU}}$ & The set of GPU cache capacities \\
            $M_{\text{CPU}}^{\text{res}}, M_{\text{GPU}}^{\text{res}}$ & The reserved CPU and GPU memory \\
            $C_{\text{max}, i}$ & The maximum cache capacity for subgraph $G_i$ \\
            $\mathbf{\hat{A}}_i$ & The adjacency matrix after symmetric normalization of $G_i$ \\
            $\epsilon_H$ & The staleness bound for embeddings \\
            $\sigma(\cdot), \tau$ & The activation function and the learning rate \\
            $\mathbf{H}_i^{(\ell)}, \mathbf{\hat{H}}_i^{(\ell)}$ & The exact and cached embedding for worker $i$ at layer $\ell$ \\
            $\mathbf{Z}_i^{(\ell)}, \mathbf{\tilde{Z}}_i^{(\ell)}$ & The intermediate and cached result for worker $i$ at layer $\ell$ \\
            $\mathbf{W}_t$, $\mathbf{W}_\star$ & The model weight at iteration $t$ and the optimal weight\\
            $\mathcal{L}(\mathbf{W}), \nabla \mathcal{L}(\mathbf{W})$ & The loss function and the gradient of $\mathcal{L}(\mathbf{W})$\\
            \Xhline{1.5pt}
        \end{tabular}%
    }
    \vspace{-1em}
\end{table}

\subsection{Overview}
CaPGNN consists of two key components: the {\it Joint Adaptive Caching Algorithm} (JACA) and the {\it Resource-Aware Partitioning Algorithm} (RAPA). The original graph is first partitioned into subgraphs using RAPA, a heuristic algorithm that considers vertex importance and constraints from computational, communication, and memory to adjust the subgraph size. Subgraphs are distributed across multiple GPUs for parallel full-batch training. During training, the JACA dynamically manages data transfers between the CPU and GPUs. The JACA consists of two components: \textit{StoreEngine} handles unified memory management and \textit{CacheEngine} executes policies such as caching, prefetching, and updating.

\subsection{Joint Adaptive Caching Algorithm}
\label{sec:joint_cache}
\begin{figure}[!hbtp]
    \centering
    \setlength{\abovecaptionskip}{0pt}
    \setlength{\belowcaptionskip}{0pt}
    \includegraphics[width=0.55\linewidth]{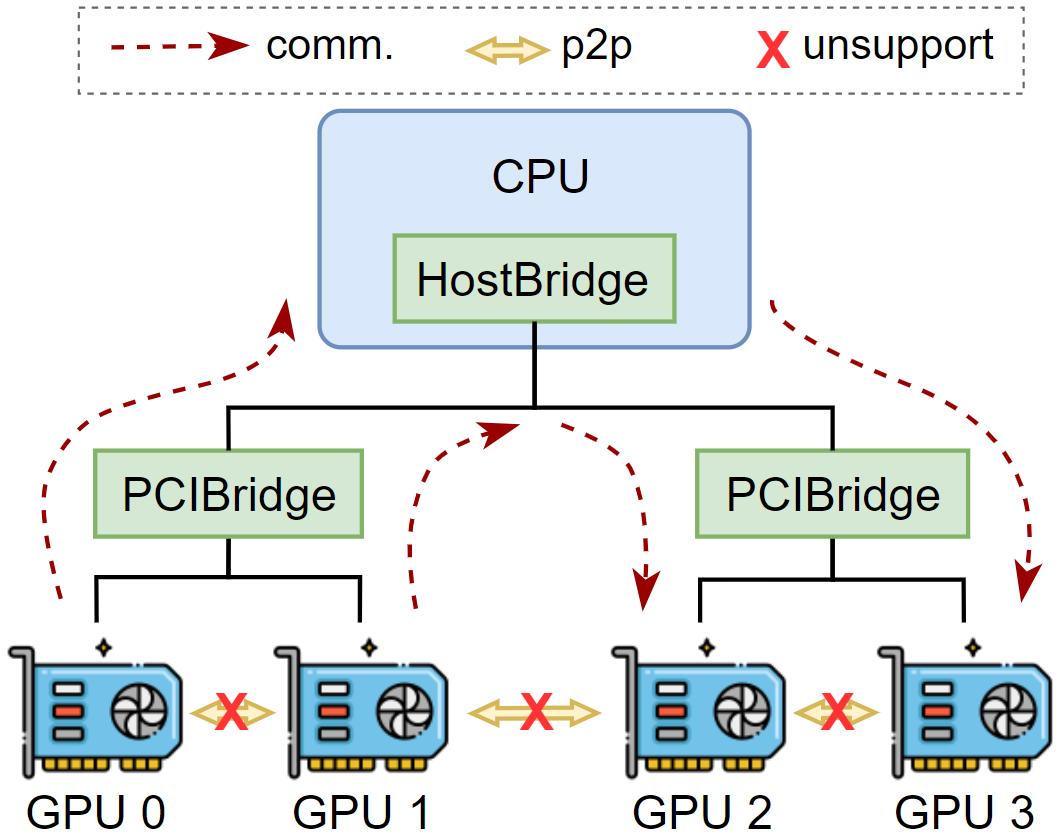}
    \caption{An example of the topology for a single machine with multiple GPUs. For GPUs that do not support P2P communication, data transfer must be routed through the CPU.}
    \label{fig:gpu_topo}
    \vspace{-1em}
\end{figure}

The GPUDirect Peer-to-Peer (P2P) technology facilitates direct data transfers between GPUs.\footnote{In CUDA, the \textit{cudaDeviceCanAccessPeer} function checks if two GPUs support P2P communication.} As shown in Fig.~\ref{fig:gpu_topo}, for GPUs lacking P2P support, this process requires two memory copies: first, copying data from one GPU to CPU memory via PCIe, and then transferring the data from CPU memory to the target GPU. In Section~\ref{subsec:motivation}, we experimentally validate the substantial presence of inter-partition halo vertex data transfers, where the communication volume is influenced by factors such as the number of partitions and feature dimensions. Consequently, for GPUs that do not support P2P communication, this data transfer mode results in a significant amount of communication and bandwidth contention, as multiple GPUs perform data transfers simultaneously~\cite{8763922}.

\begin{figure}[htbp]
    \centering
    \setlength{\abovecaptionskip}{0pt}
    \setlength{\belowcaptionskip}{0pt}
    \includegraphics[width=\linewidth]{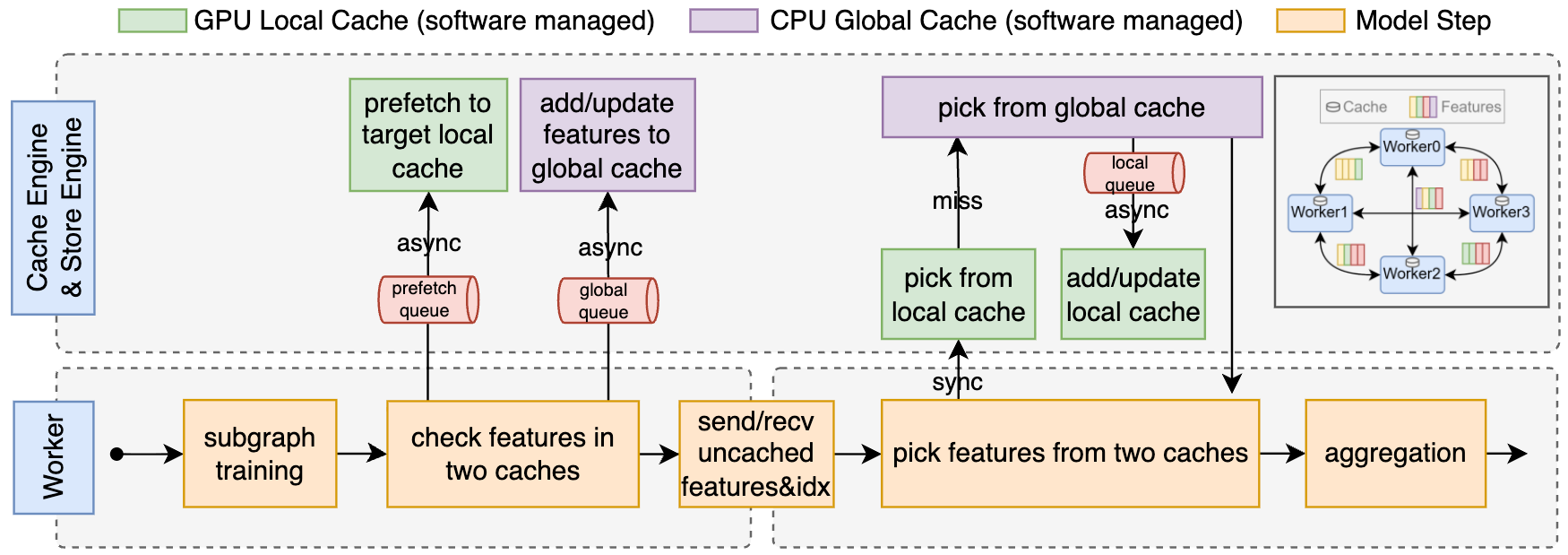}
    \caption{The workflow of the joint caching algorithm. The CPU global cache is a software-managed caching mechanism in CPU memory, while the GPU local cache resides in device memory.}
    \label{fig:cache_strategy}
\end{figure}

Caching vertices effectively mitigates the aforementioned issues. Unlike existing caching approaches that are primarily designed for sampling-based training or treat CPU memory as a passive backing store~\cite{zhao2024neutroncache, zhang2023two}, JACA is designed for full-batch GNN training and focuses on caching redundant halo vertices across partitions. For simplicity, we use \textit{vertex features} to collectively denote the \textit{read-only input vertex features} and the \textit{read-write intermediate embeddings} during training. Based on this, unless otherwise specified, ``caching'' in CaPGNN refers to caching input halo vertex features and intermediate embeddings across layers. In addition, we propose strategies such as priority ranking based on vertex importance, data prefetching, and lightweight vertex updates under bounded staleness. These design choices reduce redundant halo transfers and better overlap communication with computation. Fig.~\ref{fig:cache_strategy} shows the workflow of the caching process.

\textbf{Joint Caching and Data Prefetching Strategy.}
To leverage the high-speed access of GPU memory and the large capacity of CPU memory, we propose a joint caching algorithm that uses GPU (i.e., \textit{device}) memory as a local cache and CPU (i.e., \textit{host})\footnote{For convenience of discussion in this paper, we use {\it device} to refer to GPU and {\it host} to refer to CPU.} memory as a software-managed global cache. Throughout the paper, we use ``global cache'' to denote this software-managed CPU memory caching mechanism, rather than a hardware cache. Specifically, during training, each layer in forward propagation consists of aggregation and combination operations. In the aggregation phase, GPUs exchange halo vertex features to compute weights locally and update vertex features. We use the \textit{all-to-all} communication pattern to exchange vertex features and model parameters implemented with PyTorch’s Gloo backend.

With the caching algorithm, before sending features, a worker first checks whether the vertices are already present in the local or global cache. If vertices are cached, they will not be sent. Additionally, for vertices that exist only in the local cache, their features are copied to the global cache and prefetched into the target GPU memory. The receiving worker then looks up the cached vertex features from local and global caches, combines them with features received through communication, and proceeds with the subsequent aggregation operations. For clarity, we describe JACA mainly in the context of the forward aggregation stage, where halo exchanges dominate the communication volume. In our implementation, the cached vertex features are also used by backward computation through the standard autograd pipeline.

Different from the caching approaches~\cite{lin2020pagraph} that keeps the full features in CPU memory as a passive backing store, CaPGNN allocates a capacity-limited shared-memory as a global cache across GPUs, and on top of that, strategically manages and schedules the CPU/GPU memory, including priority, update, and eviction, for high-reuse halo vertex representations, coordinated with local GPU caches and prefetch queues. Therefore, CPU-level caching and GPU-level caching have similar behavior.

\textbf{Efficient Memory Management.} 
We design an efficient data access and storage structure. First, considering the large feature dimensions, we decouple the graph structure from the vertex features and use hash-based feature retrieval. On the other hand, conventional CPU-GPU communication is synchronous, which can block execution. To address this, we employ \textit{pinned memory} to achieve \textit{Direct Memory Access} (DMA) for asynchronous transfers, thereby improving efficiency. Since resources are isolated between processes, we allocate a dedicated pinned memory region in the CPU for each GPU. Additionally, to introduce a global cache among GPUs, we use a shared memory region in the CPU. The upper half of Fig.~\ref{fig:parallel_gnn_workflow} illustrates this structure.

\begin{figure}[htbp]
    \centering
    \setlength{\abovecaptionskip}{0pt}
    \setlength{\belowcaptionskip}{0pt}
    \includegraphics[width=\linewidth]{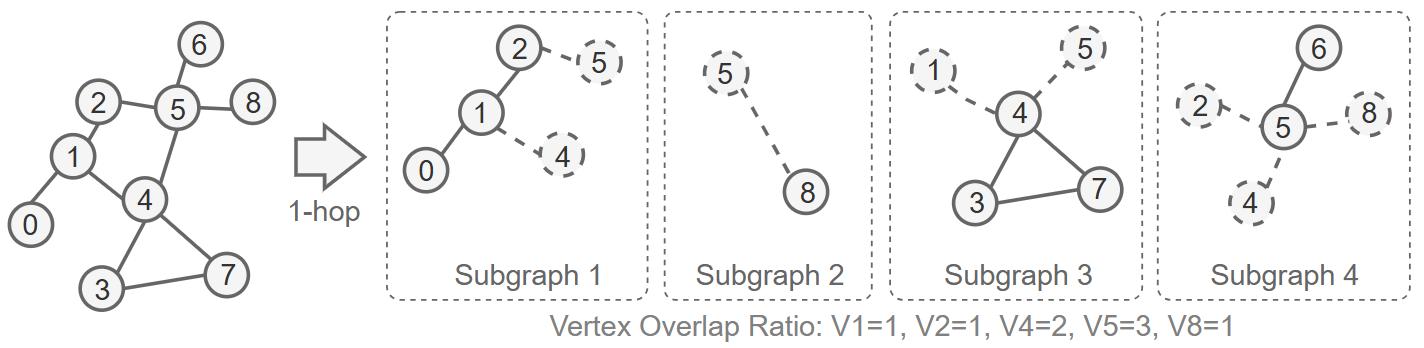}
    \caption{Example of vertex overlap ratio. The graph is partitioned using METIS. Dashed nodes indicate halo vertices.}
    \label{fig:example_of_vor}
    \vspace{-1em}
\end{figure}

\textbf{Vertex Importance and Vertex Update.} 
To improve cache hit rates, it is effective to prioritize vertices with higher reuse potential. In sampling-based training, caching decisions require modeling of complex and dynamic access patterns. By contrast, in full-batch training, where all vertices are accessed in each epoch, we define a simple and effective metric, the \textbf{vertex overlap ratio}, to quantify reuse. This ratio reflects how many partitions a vertex appears in. A higher overlap ratio implies more frequent access from multiple GPUs, leading to redundant transmission. As shown in Fig.~\ref{fig:example_of_vor}, some vertices (e.g., V5) appear in multiple subgraphs due to 1-hop expansion. The number of subgraphs a vertex belongs to defines its overlap ratio. By caching such high-overlap vertices, we can significantly reduce communication cost and improve cache efficiency. The vertex overlap ratio is defined as follows:
\begin{equation}
    \label{eq:vertex_overlap}
    R(v_k) = \sum_{i=1}^{P} \mathbb{I}(v_k \in H(G_i)),
\end{equation}
where \( R(v_k) \) represents the overlap count of the vertex \( v_k \), \( P \) is the total number of subgraph partitions, \( H = \bigcup_{i=1}^{P} H(G_i) \) denotes the set of halo vertices across all partitions, \( H(G_i) \) represents the set of halo vertices within the subgraph partition \( G_i \), and \( \mathbb{I}(x) \) is an indicator function that returns 1 if \( x \) is true and 0 otherwise.

However, for large-scale graphs, the static cache hit rate remains low~\cite{liu2023bgl}, even with the introduction of vertex importance. Therefore, it is necessary to dynamically update the vertices in the cache, especially when the cache capacity is less than the total number of vertices. Given the characteristics of full-batch training, we identify replaceable vertices based on their overlap ratio.

\textbf{Adaptive Cache Capacity.} 
The number of vertices and feature dimensions vary significantly across different graphs, and an inappropriate cache capacity can negatively affect cache performance or even lead to memory overflow. To address this, we design an adaptive cache capacity mechanism that takes into account both subgraph information and the memory of each GPU. Algorithm~\ref{alg:cache_capacity} outlines the process.

\textbf{Pipeline Design.}
The mismatch between data transmission rates and computational speed can lead to wasted computational resources and decreased overall efficiency. In a multi-GPU parallel computing environment, especially when GPUs have differing performance capabilities, this issue becomes even more pronounced. To address this challenge, we introduce three queues to implement pipelining, which overlap communication with computation: the \textit{local queue}, \textit{global queue}, and \textit{prefetch queue}.

\begin{algorithm}[!htbp]
    \caption{\texttt{cal\_capacity}: Cache Capacity.}
    \label{alg:cache_capacity}
    \KwIn{total partitions $P$, subgraphs $\{G_i \mid i \in \{1, \dots, P\}\}$, top $k$ vertices to select ($k = -1$ for all vertices), reserved GPU memory $M_{\text{GPU}}^{\text{res}}$, reserved CPU memory $M_{\text{CPU}}^{\text{res}}$, available CPU memory $M_{\text{CPU}}$, available GPU memory $\{M_{\text{GPU}, i} \mid i \in \{1, \dots, P\}\}$, vertex feature dimension $\mathbf{f_{\text{dim}}}$, number of model's layers $L$}
    \KwOut{CPU capacity $C_{\text{CPU}}$, GPU capacities $C_{\text{GPU}}$}
    \SetKwInOut{Initialize}{Initialize}
     \Initialize{
        $C_{\text{GPU}} \gets \emptyset$ \tcp*[r]{GPU cache capacities}\\
        $H_{\text{CPU}} \gets \emptyset$ \tcp*[r]{Halo vertices for CPU}
    }
    \textbf{GPU Cache Capacities Calculation Phase} \\
    \For{$i \in \{1, \dots, P\}$}{
        $H_i \gets \emptyset$ \tcp*[r]{Halo vertices for $G_i$}
        \For{$j \in \{1, \dots, P\} \setminus \{i\}$}{
            identify halo vertices $H(G_i)$ belonging to $G_j$\;
            compute overlap ratios $R(v_k)$ for $v_k \in H(G_i)$\;
            sort $R(v_k)$ and select top $k$ vertices\;
            update $H_i \gets H_i \cup \{v_k \mid v_k \in \text{top } k \text{ vertices}\}$\;
            update $|V_i|$ and $|E_i|$;
        }
        $C_{\text{GPU}} \gets \min (\left\lfloor \frac{(M_{\text{GPU}, i} \times 1024 - M_{\text{GPU}}^{\text{res}}) \times 1024^2}{\sum_{k=1}^{L} (\mathbf{f_{\text{dim}}}[k] \times 4)} \right\rfloor, |H_i|) $\;
        update $H_{\text{CPU}} \gets H_{\text{CPU}} \cup H_i$\;
    }
    \textbf{CPU Cache Capacity Calculation Phase} \\
    $C_{\text{CPU}}[i] \gets \min (\left\lfloor \frac{(M_\text{CPU} \times 1024 - M_{\text{CPU}}^{\text{res}}) \times 1024^2}{\sum_{k=1}^{L} (\mathbf{f_{\text{dim}}}[k] \times 4)} \right\rfloor, |H_{\text{CPU}}|) $\;
    \Return{$C_{\text{CPU}}$, $C_{\text{GPU}}$}\;
\end{algorithm}

Specifically, each worker has a local queue for fetching vertex features from the global cache and storing them in the local cache, ensuring that the data required during computation can be accessed quickly. The global queue is a single queue that receives vertex features from each worker and aggregates these features into the global cache for use by other workers. Additionally, each worker has a prefetch queue to preemptively transfer vertex features to a designated worker, reducing waiting time during computation and enabling data preloading.
Note that, to further reduce waiting overhead, we employ a lightweight vertex update and prefetch strategy using optimistic locks instead of traditional mutex locks.

\textbf{Staleness in CaPGNN.}
In single-GPU full-batch training, the representations of all vertices used at each iteration are computed from the latest neighbor representations. In partition-parallel multi-GPU training, however, each GPU only owns a subgraph, and the representations of remote neighbors must be obtained from other partitions as halo vertices. To avoid fetching halo vertex representations at every step, CaPGNN caches halo vertex representations and overlaps communication with computation via the local/global/prefetch queues described above. As a result, in some iterations, a worker may reuse halo representations that were written into the cache in previous iterations, instead of waiting for the latest ones. This leads to stale embeddings being used in aggregation, and thus the computed gradient can deviate from the ideal gradient under fully up-to-date neighbor information, potentially harming convergence. Therefore, our analysis introduces a bounded-staleness. In CaPGNN, we enforce this boundedness by periodically synchronizing cached halo representations after a configurable number of iterations, preventing any halo entry from remaining stale for an unbounded duration.

\textbf{Proof of Convergence.}
From a system perspective, caching turns synchronous training into an approximate training process where some halo embeddings can be slightly stale. The analysis below shows that this approximation can be viewed as introducing a bounded perturbation into gradient descent, as long as the embedding error is bounded by $\epsilon_H$, the training process remains convergent. Next, we formalize the staleness bound and incorporate the induced errors into the gradient-descent framework. Specifically, we bound the embedding error when stale embeddings are used, show that the per-layer gradient error is also bounded, and establish the global convergence of the overall training process.

\begin{lemma}
    \label{mth:lemma_1}
    Let the infinity norms of matrices \(\mathbf{A}\) and \(\mathbf{B}\) be defined as \(\|\mathbf{A}\|_\infty = \max_{i,j} |\mathbf{A}_{i,j}|\) and \(\|\mathbf{B}\|_\infty = \max_{i,j} |\mathbf{B}_{i,j}|\). The following inequalities are satisfied~\cite{xue2023sugar}: (a) \(\|\mathbf{A}\mathbf{B}\|_\infty \leq \text{col}(\mathbf{A}) \|\mathbf{A}\|_\infty \|\mathbf{B}\|_\infty\), (b) \(\|\mathbf{A} \odot \mathbf{B}\|_\infty \leq \|\mathbf{A}\|_\infty \|\mathbf{B}\|_\infty\), and (c) \(\|\mathbf{A} + \mathbf{B}\|_\infty \leq \|\mathbf{A}\|_\infty + \|\mathbf{B}\|_\infty\). The \(\text{col}(\mathbf{A})\) represents the number of columns in matrix \(A\), and \(\odot\) is the element-wise product.
\end{lemma}

\begin{lemma}
    \label{mth:lemma_2}
    For \(p\) workers, if: (a) the activation function $\sigma(\cdot)$ is $\rho$-Lipschitz continuous, (b) the related matrices $\mathbf{\hat{A}}_i$, $\mathbf{\hat{H}}_i$, $H_i$, and $W_i$ are bounded, (c) the difference between the historical embedding $\mathbf{\hat{H}}_i^{(\ell)}$ and the exact embedding $\mathbf{H}_i^{(\ell)}$ satisfies a staleness bound, i.e. $\|\mathbf{H}_i^{(\ell)} - \mathbf{\hat{H}}_i^{(\ell)}\|_\infty \leq \epsilon_H$, then the approximation error of the intermediate result $\mathbf{\tilde{Z}}_i^{(\ell)}$ is also bounded and satisfies $\|\mathbf{\tilde{Z}}_i^{(\ell)} - \mathbf{Z}_i^{(\ell)}\|_\infty \leq \eta^2 \beta^2 \epsilon_H$. Here, $\epsilon_H$ represents the staleness bound, and $\ell \in [1, L]$ denotes the $\ell$-th layer, \(\eta\) is the maximum number of columns that exists in the proof. In addition, \(\beta\) is the constant that: $\|\mathbf{\hat{A}}_i\|_\infty \leq \beta, \|\mathbf{\hat{H}}_i\|_\infty \leq \beta, \|\mathbf{H}_i\|_\infty \leq \beta, \|\mathbf{W}_i\|_\infty \leq \beta$. 
\end{lemma}

\begin{proof}
    In the $i$-th worker, for the GCN model, the forward propagation rule of the \(\ell\)-th layer is given by: 
    \begin{equation}
        \mathbf{Z}_i^{(\ell)} = \mathbf{\hat{A}}_i \mathbf{H}_i^{(\ell-1)} \mathbf{W}_i^{(\ell)},
        \quad \mathbf{H}_i^{(\ell)} = \sigma(\mathbf{Z}_i^{(\ell)}),
    \end{equation}
    where, \(\mathbf{\hat{A}}_i\) represents the adjacency matrix after symmetric normalization. If a caching mechanism is adopted, the cached intermediate result is $\mathbf{\tilde{Z}}_i^{(\ell)}$, and the forward propagation is given by:
    \begin{equation}
        \mathbf{\tilde{Z}}_i^{(\ell)} = \mathbf{\hat{A}}_i \mathbf{\tilde{H}}_i^{(\ell-1)} \mathbf{W}_i^{(\ell)},
        \quad \mathbf{\tilde{H}}_i^{(\ell)} = \sigma(\mathbf{\tilde{Z}}_i^{(\ell)}).
    \end{equation}

    According to Lemma~\ref{mth:lemma_1}, we have:
    \begin{equation}
        \resizebox{0.9\linewidth}{!}{$
            \begin{aligned}
                \|\mathbf{\tilde{Z}}_i^{(\ell)} - \mathbf{Z}_i^{(\ell)}\|_\infty 
                &= \|\mathbf{\hat{A}}_i \mathbf{\tilde{H}}_i^{(\ell-1)} \mathbf{W}_i^{(\ell-1)} - \mathbf{\hat{A}}_i \mathbf{H}_i^{(\ell-1)} \mathbf{W}_i^{(\ell-1)}\|_\infty \\
                &\leq \eta \|\mathbf{\hat{A}}_i \mathbf{\tilde{H}}_i^{(\ell-1)} - \mathbf{\hat{A}}_i \mathbf{H}_i^{(\ell-1)}\|_\infty \|\mathbf{W}_i^{(\ell-1)}\|_\infty \\
                &\leq \eta^2 \|\mathbf{\hat{A}}_i\|_\infty \|\mathbf{\tilde{H}}_i^{(\ell-1)} - \mathbf{H}_i^{(\ell-1)}\|_\infty \|\mathbf{W}_i^{(\ell-1)}\|_\infty \\
                &\leq \eta^2 \beta^2 \epsilon_H.
            \end{aligned}
        $}
    \end{equation}
\end{proof}

Next, we will prove that if cached values are used during the forward propagation phase, the error of the model weight gradients obtained in the backward propagation phase is also bounded.

\begin{lemma}
    \label{mth:lemma_3}
    We first make the following assumptions. In the $i$-th worker, (a) the activation function $\sigma(\cdot)$ and the gradient $\nabla \mathcal{L}$ are $\rho$-Lipschitz continuous, (b) the matrices $\hat{A}_i$, $W_i$, $\delta_i^{(\ell)}$ and $\sigma'(Z_i)$ are bounded. Thus, we have $\|\nabla_{\mathbf{\tilde{Z}}_i^{(\ell)}} \tilde{\mathcal{L}} - \nabla_{\mathbf{Z}_i^{(\ell)}} \mathcal{L}\|_\infty \leq {\rho \eta^2 \beta^2 \epsilon_H}$. Here, $\delta_i^{(\ell)}$ denotes the exact gradients, $\sigma'(\mathbf{Z}_i)$ represents the derivative of $\sigma(\mathbf{Z}_i)$ with respect to the input $\mathbf{Z}_i$, $\ell \in [1, L]$ represents the $\ell$-th layer, and $\nabla_{\mathbf{Z}_i^{(\ell)}} \mathcal{L}$ is the gradient matrix of loss $\mathcal{L}$ with respect to $\mathbf{Z}_i^{(\ell)}$ at the $\ell$-th layer.
\end{lemma}

\begin{proof}
    The last layer of a GNN is typically directly used for the computation of the loss function, and its gradient depends only on the output of that layer. As a result, the error can be directly controlled using the $\rho$-Lipschitz continuity property. Given that the activation function $\sigma(\cdot)$ and the gradient $\nabla \mathcal{L}$ are $\rho$-Lipschitz continuous, and based on Lemma 1 and Lemma 2, for the $L$-th layer, we have:
    \begin{equation}
        \|\nabla_{\mathbf{\tilde{Z}}_i^{(L)}} \tilde{\mathcal{L}} - \nabla_{\mathbf{Z}_i^{(L)}} \mathcal{L}\|_\infty 
    \leq \rho \|\mathbf{\tilde{Z}}_i^{(L)} - \mathbf{Z}_i^{(L)}\|_\infty 
    \leq \rho \eta^2 \beta^2 \epsilon_H.
    \end{equation}

    According to the characteristics of GNNs, the gradient computation at $\ell$-th layer depends on the results of $(\ell+1)$-th layer. Additionally, assume that $\|\nabla_{\mathbf{\tilde{Z}}_i^{(\ell')}} \tilde{\mathcal{L}} - \nabla_{\mathbf{Z}_i^{(\ell')}} \mathcal{L}\|_\infty \leq K^{(\ell')}, \forall \ell' > \ell$. Therefore, using the induction hypothesis, the gradient error at $\ell$-th layer can be expressed as:
    \begin{equation}
        \resizebox{0.9\linewidth}{!}{$
            \delta_i^{(\ell)} = \nabla_{\mathbf{Z}_i^{(\ell)}} \mathcal{L} = \frac{\partial \mathcal{L}}{\partial \mathbf{Z}_i^{(l)}} = \delta_i^{(\ell+1)} \hat{\mathbf{A}}_i \left(\mathbf{W}_i^{(\ell)}\right)^\top \odot \sigma'\left(\mathbf{Z}_i^{(\ell)}\right),
        $}
    \end{equation}
    \begin{equation}
        \resizebox{0.9\linewidth}{!}{$
            \begin{aligned}
                &\|\nabla_{\mathbf{\tilde{Z}}_i^{(\ell)}} \tilde{\mathcal{L}} - \nabla_{\mathbf{Z}_i^{(\ell)}} \mathcal{L}\|_\infty = \|\tilde{\delta}_i^{(\ell)} - \delta_i^{(\ell)}\|_\infty \\
                &\qquad\qquad= \|\tilde{\delta}_i^{(\ell+1)} \mathbf{\hat{A}}_i \big(\mathbf{W}^{(\ell)}_i\big)^\top \odot \sigma'\big(\mathbf{\tilde{Z}}_i^{(\ell)}\big) \\
                &\qquad\qquad\quad - \delta_i^{(\ell+1)} \mathbf{\hat{A}}_i \big(\mathbf{W}^{(\ell)}_i\big)^\top \odot \sigma'\big(\mathbf{Z}_i^{(\ell)}\big)\|_\infty \\
                &\qquad\qquad\leq \eta^2 \Big\{\|\tilde{\delta}_i^{(\ell+1)}\|_\infty \|\mathbf{\hat{A}}_i\|_\infty \|\big(\mathbf{W}_i^{(\ell)}\big)^\top\|_\infty 
                   \|\sigma'\big(\mathbf{\tilde{Z}}_i^{(\ell)}\big) - \sigma'\big(\mathbf{Z}_i^{(\ell)}\big)\|_\infty \\
                &\qquad\qquad\quad + \|\tilde{\delta}_i^{(\ell+1)} - \delta_i^{(\ell+1)}\|_\infty \|\mathbf{\hat{A}}_i\|_\infty \|\big(\mathbf{W}_i^{(\ell)}\big)^\top\|_\infty 
                   \|\sigma'\big(\mathbf{Z}_i^{(\ell)}\big)\|_\infty\Big\} \\
                &\qquad\qquad\leq \eta^2 \Big(\beta^3 \rho(\eta^2 \beta^2 \epsilon_H) + K_{(\ell+1)} \beta^3\Big) \\
                &\qquad\qquad= \eta^2 \beta^3 \big(K_{(L)} + K_{(\ell+1)}\big).
            \end{aligned}
        $}
    \end{equation}

    Define the error bound for $\ell$-th layer as $K_{(\ell)}=\eta^2 \beta^3 \big(K_{(L)} + K_{(\ell+1)}\big)$. By induction, starting from the final layer $L$ and iterating backward, it can be shown that the gradient error for all layers is bounded, and the error bounds accumulate progressively across layers. This establishes that the training process retains theoretical convergence even in the presence of stale embeddings.
\end{proof}

Next, we will prove the convergence guarantee of GNN training with cached embeddings under bounded staleness conditions.

\begin{theorem}
    For a model with $L$ layers, let $\mathbf{W}_t$ denote the model parameters at the $t$-th training iteration, and $\mathbf{W}_\star$ represent the optimal model weights. Assume the following conditions are satisfied: (a) the activation function $\sigma(\cdot)$ and the gradient of loss $\nabla \mathcal{L}$ are $\rho$-Lipschitz, (b) the matrices $\mathbf{\hat{A}}$, $\mathbf{W}$, $\mathbf{H}$, and the gradients of the loss are bounded by the constant $\gamma$: $\|\mathbf{\hat{A}}\|_\infty \leq \gamma, \|\mathbf{W}\|_\infty \leq \gamma, \|\mathbf{H}\|_\infty \leq \gamma, \|\nabla_{\mathbf{W}} \tilde{\mathcal{L}}\|_\infty \leq \gamma, \|\nabla_{\mathbf{W}} \mathcal{L}\|_\infty \leq \gamma, \|\mathcal{L}(\mathbf{W})\|_\infty \leq \gamma$ , (c) the loss $\mathcal{L}(\mathbf{W})$ is $\rho$-smooth. Then, there exists $\alpha>0$, s.t., $\forall T > L \epsilon_H$, if the GNN is trained in parallel with bounded staleness for at most $R \leq T$ iterations, where $R$ is chosen uniformly from $T$ and the learning rate is defined as $\tau = \min\left\{\frac{1}{\rho}, \frac{1}{\sqrt{T}}\right\}$. Thus, the the following bound holds:
    \begin{equation}
        \mathbb{E}_R \left[\|\nabla \mathcal{L}(\mathbf{W}_{R})\|_F^2\right] \leq \frac{2\left(\mathcal{L}(\mathbf{W}_{1}) - \mathcal{L}(\mathbf{W}_\star)\right)}{\sqrt{T}} + \frac{\rho \alpha}{2\sqrt{T}},
    \end{equation}
    where, $\|\cdot\|_F$ is the Frobenius norm, $T$ is the total epochs.
\end{theorem}

\begin{proof}
    Let $\delta_t = \nabla_{\mathbf{W}_t}\tilde{\mathcal{L}} - \nabla \mathcal{L}(\mathbf{W}_t)$ denote the differences between gradients at epoch $t$. By the $\rho$-smoothness of $\mathcal{L}(\mathbf{W})$, the model update rule $\mathbf{W}_{t+1} = \mathbf{W}_t - \tau \nabla \tilde{\mathcal{L}}(\mathbf{W}_t)$, and Lemma~\ref{mth:lemma_1}, we have:

    \begin{equation}
        \resizebox{0.85\linewidth}{!}{$
            \begin{aligned}
                &\mathcal{L}(\mathbf{W}_{t+1}) \leq \mathcal{L}(\mathbf{W}_{t}) + \langle \nabla \mathcal{L}(\mathbf{W}_{t}), \mathbf{W}_{t+1} - \mathbf{W}_{t} \rangle + \frac{\rho}{2} \|\mathbf{W}_{t+1} - \mathbf{W}_{t}\|_F^2, \\
                &\qquad= \mathcal{L}(\mathbf{W}_{t}) + \langle \nabla \mathcal{L}(\mathbf{W}_{t}), \mathbf{W}_{t+1} - \mathbf{W}_{t} \rangle + \frac{\rho}{2} \tau^2 \|\nabla_{\mathbf{W}_{t}} \tilde{\mathcal{L}}\|_F^2 \\
                &\qquad= \mathcal{L}(\mathbf{W}_{t}) - \tau \langle \nabla \mathcal{L}(\mathbf{W}_{t}), \nabla_{\mathbf{W}_{t}} \tilde{\mathcal{L}} \rangle + \frac{\rho}{2} \tau^2 \|\nabla_{\mathbf{W}_{t}} \tilde{\mathcal{L}}\|_F^2 \\
                &\qquad= \mathcal{L}(\mathbf{W}_{t}) - \tau \langle \nabla \mathcal{L}(\mathbf{W}_{t}), \delta_{t} \rangle - \tau \|\nabla \mathcal{L}(\mathbf{W}_{t})\|_F^2 \\
                &\qquad\quad + \frac{\rho}{2} \tau^2 \left( \|\delta_{t}\|_F^2 + \|\nabla \mathcal{L}(\mathbf{W}_{t})\|_F^2 + 2 \langle \delta_{t}, \nabla \mathcal{L}(\mathbf{W}_{t}) \rangle \right) \\
                &\qquad\leq \mathcal{L}(\mathbf{W}_{t}) - \left(\tau - \frac{\rho}{2} \tau^2 \right) \|\nabla \mathcal{L}(\mathbf{W}_{t})\|_F^2 + \frac{\rho}{2} \tau^2 \|\delta_{t}\|_F^2 \\
                &\qquad\leq \mathcal{L}(\mathbf{W}_{t}) - \left(\tau - \frac{\rho}{2} \tau^2 \right) \|\nabla \mathcal{L}(\mathbf{W}_{t})\|_F^2 \\
                &\qquad\quad + \frac{\rho}{2} \tau^2 (\|\nabla_{\mathbf{W}_{t}} \tilde{\mathcal{L}}\|_\infty + \|\nabla \mathcal{L}(\mathbf{W}_{t})\|_\infty) \\
                &\qquad\leq \mathcal{L}(\mathbf{W}_{t}) - \left(\tau - \frac{\rho}{2} \tau^2 \right) \|\nabla \mathcal{L}(\mathbf{W}_{t})\|_F^2 + \frac{\rho}{2} \tau^2 (2\gamma^2) \\
                &\qquad\leq \mathcal{L}(\mathbf{W}_{t}) - \left(\tau - \frac{\rho}{2} \tau^2 \right) \|\nabla \mathcal{L}(\mathbf{W}_{t})\|_F^2 + \frac{\rho}{2} \tau^2 \alpha,
            \end{aligned}
        $}
    \end{equation}
    where, $\langle \cdot, \cdot \rangle$ denotes the inner product. Furthermore, we have:
    \begin{equation}
        \label{eq:lm_sum}
        \resizebox{0.85\linewidth}{!}{$
            \begin{aligned}
                \forall t,\quad\left(\tau - \frac{\rho}{2} \tau^2 \right) \sum_{i=1}^T \|\nabla \mathcal{L}(\mathbf{W}_{t})\|_F^2 \leq \mathcal{L}(\mathbf{W}_{1}) - \mathcal{L}(\mathbf{W}_\star) + \frac{\rho}{2} \tau^2 \alpha T.
            \end{aligned}
        $}
    \end{equation}

    According to $\tau = \min\left\{\frac{1}{\rho}, \frac{1}{\sqrt{T}} \right\}$, we can convert Equation~\ref{eq:lm_sum} to:
    \begin{equation}
        \label{eq:lm_e}
        \resizebox{0.85\linewidth}{!}{$
            \begin{aligned}
                \mathbb{E}_R \|\nabla \mathcal{L}(\mathbf{W}_{R})\|_F^2 &= \frac{1}{T} \sum_{i=1}^T \|\nabla \mathcal{L}(\mathbf{W}_{t})\|_F^2 \leq \frac{\mathcal{L}(\mathbf{W}_{1}) - \mathcal{L}(\mathbf{W}_\star) + \frac{\rho}{2} \tau^2 \alpha T}{T \tau \left(2 - \rho \tau\right)} \\
                &\leq \frac{2 \left(\mathcal{L}(\mathbf{W}_{1}) - \mathcal{L}(\mathbf{W}^\star)\right)}{T \tau} + \rho \tau \alpha \\
                &\leq \frac{2 \left(\mathcal{L}(\mathbf{W}_{1}) - \mathcal{L}(\mathbf{W}^\star)\right)}{\sqrt{T}} + \frac{\rho \alpha}{2 \sqrt{T}}.
            \end{aligned}
        $}
    \end{equation}

    As $T \to \infty$, we can find that $\mathbb{E}_R \|\nabla \mathcal{L}(\mathbf{W}_{R})\|_F^2 \to 0$. The result demonstrates that even with stale embeddings in the parallel GNN training, the optimization process maintains theoretical convergence. Additionally, the error introduced by stale embeddings and randomness, represented by the term $\frac{\rho \alpha}{2 \sqrt{T}}$, decreases as $T$ increases, showing that the system's approximation error is bounded and controlled.  
\end{proof}

Note that our convergence analysis is stated using a generic message-passing form and standard boundedness/Lipschitz smoothness assumptions; therefore it covers common architectures such as GCN, GraphSAGE, and GIN whose layer update can be written as a composition of neighbor aggregation, a linear transform, and a Lipschitz nonlinearity. For attention-based variants such as GAT, staleness may additionally affect the attention coefficients because they are feature-dependent. Our argument can still apply when the attention computation is bounded and Lipschitz over the domain encountered during training; otherwise, the sensitivity to historical embeddings may increase and a tighter staleness bound, i.e., more frequent refresh, is required. Moreover, the perturbation bound can accumulate across layers, implying that deeper GNNs may be more sensitive to the same staleness level. This suggests using a smaller staleness bound or a shorter refresh interval for deeper models to preserve accuracy.

\subsection{Resource-Aware Partitioning Algorithm}
\label{sec:graph_partition}
\begin{figure}[htbp]
    \centering
    \setlength{\abovecaptionskip}{0pt}
    \setlength{\belowcaptionskip}{0pt}
    \includegraphics[width=\linewidth]{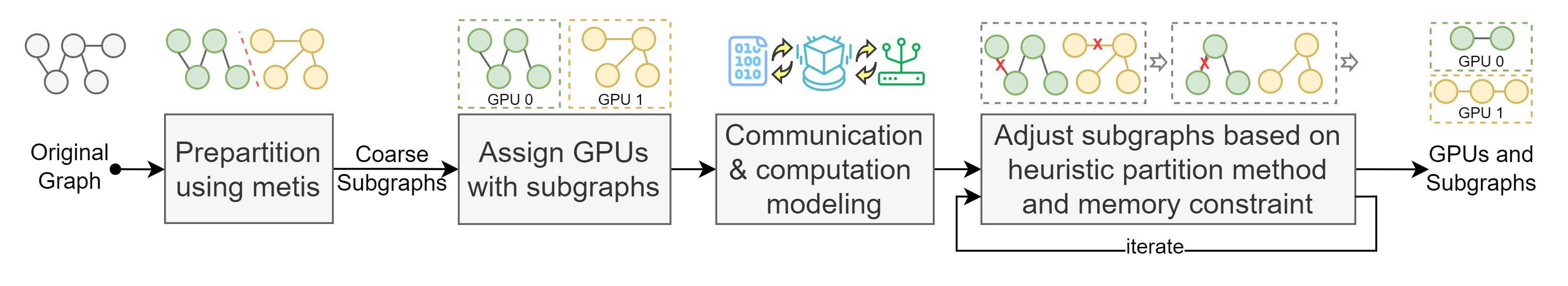}
    \caption{The workflow of the proposed RAPA.}
    \label{fig:graph_partition_workflow}
    \vspace{-1em}
\end{figure}

To further reduce communication overhead, we propose a resource-aware heuristic graph partitioning method. The goal is to adjust the size of the subgraphs to match the computational and communication capabilities of the corresponding GPUs, thereby minimizing the total computation cost under memory constraints while ensuring that the computational load across subgraphs remains balanced. A sample workflow is shown in Fig.~\ref{fig:graph_partition_workflow}. Note that RAPA only adjusts halo vertices without discarding original graph data, so the framework still belongs to full-batch training.

\textbf{Prepartitioning and Subgraph Assignment.}
We use a partition method such as METIS~\cite{karypis1998metis} to perform an initial partition of the original graph, generating coarse subgraphs. These subgraphs are of similar size and include halo vertices. To enable finer adjustments based on computational and communication resources in subsequent stages, the coarse subgraphs are randomly assigned to different GPUs. Note that since each GPU corresponds to one subgraph, thus the \textit{i}-th GPU corresponds to the \textit{i}-th subgraph.

\begin{figure}[htbp]
    \centering
    \setlength{\abovecaptionskip}{0pt}
    \setlength{\belowcaptionskip}{0pt}
    \includegraphics[width=0.55\linewidth]{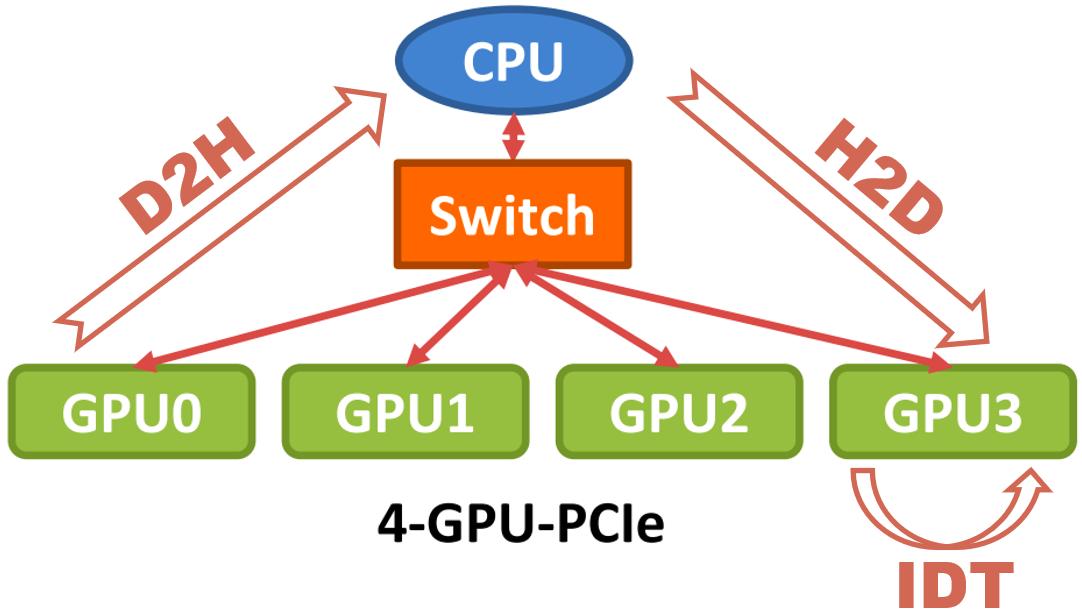}
    \caption{Three communication patterns. H2D refers to host-to-device, D2H refers to device-to-host, and IDT refers to inter-device-transfer.}
    \label{fig:comm_patterns}
\end{figure}

\textbf{Communication and Computation Modeling.}
In GNNs, communication mainly involves {\it host-to-device} (H2D), {\it device-to-host} (D2H) and {\it inter-device-transfer} (IDT), as shown in Fig.~\ref{fig:comm_patterns}. Computation consists mainly of SpMM, MM and their variants~\cite{Hamilton_Ying_Leskovec_2017, Kipf_Welling_2016, velickovic2018graph, gilmer2017neural}. Contrary to prior studies~\cite{windgp} that directly use floating-point computation time and unidirectional data transfer time to estimate computation and communication costs, we model computation and communication costs based on the test results from the above metrics, leading to a more intuitive and accurate performance evaluation.

Considering that the impact of H2D and D2H increases as the number of GPUs grows, and only cross-partition dependencies incur communication, we estimate the communication-related cost using a lightweight proxy, as shown in Eq.~\ref{eq:communication_cost}. Note that halo exchanges are typically packed by unique halo vertices in practice. However, RAPA requires an incrementally maintainable objective to evaluate candidate halo vertex pruning operations. We therefore use the number of outer edges as a cross-partition interaction proxy rather than an exact byte-level transfer model. When a halo vertex is removed from a subgraph during the adjustment stage, all cross-partition edges incident to this halo vertex are eliminated from that subgraph, so outer edges decrease accordingly. This proxy is used only inside RAPA to compare adjustment candidates; the end-to-end communication reduction is still validated by runtime measurements in our experiments.

\begin{equation}
    \label{eq:communication_cost}
    \begin{aligned}
        \resizebox{0.87\linewidth}{!}{$
            T_i^{\text{comm}} = |E_i^{\text{outer}}| \cdot \left( \left( \frac{F_i^{\text{H2D}}}{{\textbf{F}_{max}^{\text{H2D}}}} + \frac{F_i^{\text{D2H}}}{{\textbf{F}_{max}^{\text{D2H}}}} \right) \cdot \left( 1 - \frac{1}{P} \right) + \frac{F_i^{\text{IDT}}}{{\textbf{F}_{max}^{\text{IDT}}}} \cdot \frac{1}{P} \right), P \geq 1 ,
        $}
    \end{aligned}
\end{equation}
where, $|E_i^{\text{outer}}|$ represents the total number of outer edges in the $i$-th subgraph. $F_i^{\text{H2D}}$, $F_i^{\text{D2H}}$, and $F_i^{\text{IDT}}$ denote the H2D, D2H, and IDT communication capabilities of the $i$-th GPU, respectively. These communication capabilities are measured by transmitting a $11585 \times 11585$ matrix in the corresponding direction, with the results averaged over 50 runs to reduce error. $\textbf{F}_{max}^{\text{H2D}}$, $\textbf{F}_{max}^{\text{D2H}}$, and $\textbf{F}_{max}^{\text{IDT}}$ represent the maximum H2D, D2H, and IDT communication capabilities among all GPUs, respectively. $P$ is the number of subgraphs.

On the other hand, SpMM is related to the number of edges, while MM is related to the number of vertices. Therefore, the formula for measuring the computation cost is as follows:
\begin{equation}
    \label{eq:computation_cost}
    \begin{aligned}
        \resizebox{0.85\linewidth}{!}{$
            T_i^{\text{comp}} = \alpha \cdot |E_i^{\text{all}}| \cdot \frac{t_i^{\text{SpMM}}}{t_{max}^{\text{SpMM}}} 
+ \left( 1 - \alpha \right) \cdot \left| V_i^{\text{inner}} \right| \cdot \frac{t_i^{\text{MM}}}{t_{max}^{\text{MM}}},
        $}
    \end{aligned}
\end{equation}
where, $|E_i^{\text{all}}|$ represents the total number of edges and $\left| V_i^{\text{inner}} \right|$ denotes the number of inner vertices. $t_i^{\text{SpMM}}$ and $t_i^{\text{MM}}$ represent the computational capability of the $i$-th GPU, respectively. These capabilities are measured by performing SpMM and MM operations on a $11585 \times 11585$ matrix, with the results averaged over 50 runs to reduce error. $t_{\text{max}}^{\text{SpMM}}$ and $t_{\text{max}}^{\text{MM}}$ represent the maximum SpMM and MM capabilities among all GPUs, respectively. $\alpha$ is a factor that controls which type of computation is prioritized based on the computational model.

\textbf{Heuristic Partition Method.}
Based on the previous discussion, we propose a heuristic partitioning method aimed at minimizing the total cost while ensuring that the costs of different subgraphs are as balanced as possible. Additionally, to prevent memory overflow, the memory usage of each subgraph should be less than or equal to the available memory of the corresponding GPU. The formula is given as follows:
\begin{equation}
    \resizebox{0.85\linewidth}{!}{$
    \begin{aligned}
        &\text{minimize} \quad \lambda + \text{Std}(\lambda_i), \\
        &\text{subject to} \\ 
        &\qquad \qquad \lambda = \max \{\lambda_i\}, \\
        &\qquad \qquad \text{Std}(\{\lambda_i\}) \leq \epsilon, \\
        &\qquad \qquad \lambda_i = T_i^{\text{comp}} + T_i^{\text{comm}}, \\
        &\qquad \qquad M_i \geq |V_i| \cdot M^{\text{vertex}} + |E_i| \cdot M^{\text{edge}} + f_{dim} \cdot M^{\text{feat}} + \beta.
    \end{aligned}
    $}
    \label{eq:heuristic_partition}
\end{equation}
where $\lambda_i$ represents the computational cost of the $i$-th GPU and the subgraph, while $\lambda$ denotes the total cost. $\text{Std}(\cdot)$ represents the standard deviation operation and $\epsilon$ is the threshold for the stopping condition. $M_i$ refers to the available memory of the $i$-th GPU, with $|V_i|$ and $|E_i|$ representing the number of vertices and edges in the $i$-th subgraph, respectively. $M^{\text{vertex}}$ and $M^{\text{edge}}$ denote the storage size of a single vertex and a single edge, respectively. $\beta$ represents the reserved memory for storing data such as gradients.

\textbf{Subgraph adjustment strategy.}
Graph sparsification simplifies the graph structure by removing a subset of edges. However, it focuses primarily on edge reduction rather than node reduction, meaning that halo vertices still need to store features and participate in training. Inspired by this and to ensure that each subgraph meets the aforementioned constraints, we adjust the number of halo vertices within each partition. The algorithm~\ref{alg:do_partition} and algorithm~\ref{alg:adjust_subgraph} show the adjustment process.

\begin{algorithm}[htbp]
    \caption{\texttt{do\_partition}: Partition Adjustment.}
    \label{alg:do_partition}
    \KwIn{$\{G_i\}_{i=1}^P$: Initial graph partitions across $P$ GPUs; threshold $\epsilon$ for convergence}
    \KwOut{Balanced partitions $\{G_i\}_{i=1}^P$}
    \While{True}{
        $(\{G_i\}, \mathbf{r}) \leftarrow \texttt{adjust\_subgraph}(\{G_i\})$\;
        $\quad \sigma_\lambda = \sqrt{\frac{1}{P} \sum_{i=1}^P (\lambda_i - \bar{\lambda})^2}$\;
        \If{$\sigma_\lambda < \epsilon$}{
            \textbf{break}\tcp*[f]{cost is balanced enough}
        }
        \If{$\mathbf{r} = \mathbf{1}$}{
            \textbf{break}\tcp*[f]{No further improvements possible}
        }
    }
    \Return{$\{G_i\}_{i=1}^P$}\;
\end{algorithm}

\begin{algorithm}[htbp]
    \caption{\texttt{adjust\_subgraph}: Adjust Graph Assignments Across GPUs.}
    \label{alg:adjust_subgraph}
    \KwIn{$\{G_i\}_{i=1}^P$: A set of graph partitions assigned to $P$ GPUs}
    \KwOut{Updated $\{G_i\}_{i=1}^P$ and adjustment status vector $\mathbf{r} \in \{0,1\}^P$}
    Initialize $\mathbf{r} \leftarrow \mathbf{0} \in \mathbb{R}^P$\;
    \For{$i \leftarrow 1$ \KwTo $P$ (from weakest GPU)}{
        Let $G_i = (V_i, E_i)$ be the local subgraph on GPU $i$\;
        Compute current cost $\lambda_i$ and average cost $\bar{\lambda} \leftarrow \frac{1}{P} \sum_{j=1}^P \lambda_j$\;
        \If{$\lambda_i \leq \bar{\lambda}$ \textbf{and} $M_{\text{GPU}, i}$ is sufficient}{
            Set $r_i \leftarrow 1$ and \textbf{continue}\;
        }
        Identify $H(G_i) \subseteq V_i$: the set of halo nodes\;
        For each $v_k \in H(G_i)$, compute score $S_k$\;
        Sort halo nodes $\{v_k\}$ in ascending order by $S_k$\;
        Initialize: $V_i^{\text{rem}} \leftarrow \emptyset$, $E_i^{\text{rem}} \leftarrow \emptyset$\;
        \While{Estimated cost $\hat{\lambda}_i > \bar{\lambda}$}{
            Select $v_k \in H(G_i)$ with lowest $S_k$ not yet considered\;
            Add $v_k$ to $V_i^{\text{rem}}$, and all $e_{kj} \in E_i$ to $E_i^{\text{rem}}$\;
            Recompute $\hat{\lambda}_i$ using updated $|V_i|$ and $|E_i|$\;
    
            \If{$\hat{\lambda}_i \leq \frac{1}{2} (\lambda_i + \bar{\lambda})$ \textbf{and} $M_{\text{GPU}, i}$ is sufficient}{
                \textbf{break}\;
            }
        }
        \If{$V_i^{\text{rem}} \neq \emptyset$}{
            Remove $V_i^{\text{rem}}$ and $E_i^{\text{rem}}$ from $G_i$\;
            Update node features and graph statistics\;
        }
        \Else{
            Set $r_i \leftarrow 1$\;
        }
    }
    \Return{$\{G_i\}_{i=1}^P$, $\mathbf{r}$}\;
\end{algorithm}

To mitigate potential accuracy degradation caused by halo-replica reduction, we introduce a lightweight vertex influence score to rank halo replicas during adjustment, as defined in Eq.~\ref{eq:score}. This score integrates structural and system-relevant signals, including relative in-degree, relative out-degree, and repetition count, to estimate the structural impact of removing a halo replica from a subgraph. Halo vertices with lower impact scores are prioritized for removal.
\begin{equation}
    \resizebox{0.85\linewidth}{!}{$
    \begin{aligned}  
        S_i &= \left( \sum_{j \in N^{out}(i)} \frac{1}{\sqrt{D^{in}_j}} \cdot \frac{1}{\sqrt{D^{out}_i}} + \sum_{j \in N^{in}(i)} \frac{1}{\sqrt{D^{out}_j}} \cdot \frac{1}{\sqrt{D^{in}_i}} \right) \cdot C_i,
    \end{aligned}
    $}
    \label{eq:score}
\end{equation}
where \( N^{out}(i) \) denotes the set of outgoing neighbors of vertex \(i\), and \( N^{in}(i) \) denotes the set of incoming neighbors of vertex \(i\). \( D^{in}_j \) represents the in-degree of vertex \(j\) in the original graph, and \( D^{out}_i \) represents the out-degree of vertex \(i\) in the subgraph. \( C_i \) denotes the frequency with which vertex \(i\) appears across subgraphs.

\begin{figure}[htbp]
    \centering
    \includegraphics[width=0.5\linewidth]{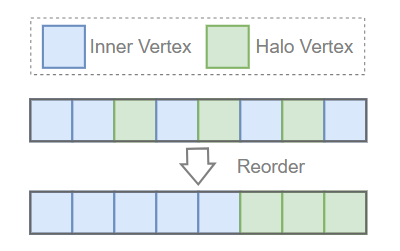}
    \caption{Effect of graph reorder. It optimizes the storage order of vertices and improves memory access efficiency.}
    \label{fig:reorder}
\end{figure}

Furthermore, to optimize the storage order of vertices for better memory access efficiency, we reorder the subgraphs~\cite{merkel2024can} as shown in Fig.~\ref{fig:reorder}. Finally, the adjusted subgraphs begin full-batch training on their corresponding GPUs.

\section{Experiments}
\label{sec:experiments}
To systematically evaluate the effectiveness of the proposed CaPGNN, we conduct extensive experiments on several widely used public datasets and compare it against representative existing methods. Our experimental design consists of the following parts: we first introduce the experimental setup and datasets; Sections~\ref{ssec:vor_cache} to~\ref{ssec:cc_cache_system} evaluate the performance of the caching strategy JACA; Sections~\ref{ssec:pr_rapa} to~\ref{ssec:eff_rapa} analyze the effectiveness of the partitioning algorithm RAPA; Sections~\ref{ssec:convergence_rate} to~\ref{ssec:overall_performance} validate the convergence behavior and the overall performance of CaPGNN; finally, Section~\ref{ssec:ablation_study} presents an ablation study to examine the individual contributions of each component to the overall system.

\subsection{Lab Setup and Datasets}
\label{ssec:lab_setup}
We implement CaPGNN based on DGL 2.3.0 and PyTorch 2.3.0, using the GLOO backend and PCIe for CPU-GPU communication. We conduct experiments on a server with a dual-core Intel® Xeon® Gold 6230 CPU, 768GB RAM, two NVIDIA Tesla A40 GPUs, four NVIDIA RTX 3090 GPUs, two NVIDIA RTX 3060 GPUs, and two NVIDIA GTX 1660Ti GPUs. The GPUs are connected to the CPU via PCIe 3.0 x8. The GPU parameters are detailed in Table~\ref{tab:gpu_info} and the GPU group configurations are detailed in Table~\ref{tab:gpu_grouping}.

\begin{table}[htb]
    \centering
    \setlength{\abovecaptionskip}{0pt}
    \setlength{\belowcaptionskip}{0pt}
    \caption{Detailed information about NVIDIA GPUs.}
    \resizebox{\columnwidth}{!}{%
        \begin{tabular}{lccccccc}
            \Xhline{1.5pt}
            \textbf{Name} &
              \textbf{Label} &
              \textbf{Arch.} &
              \textbf{\makecell[c]{CUDA\\ Cores}} &
              \textbf{\makecell[c]{Tensor\\ Cores}} &
              \textbf{\makecell[c]{FP32\\ (TFLOPS)}} &
              \textbf{\makecell[c]{Memory\\ (GB)}} &
              \textbf{\makecell[c]{Bandwidth\\ (GB/s)}} \\
            \Xhline{1pt}
            \textbf{RTX 3090} & R9 & Ampere & 10496 & 328 & 35.58 & 24 & 936.2  \\
            \textbf{Tesla A40} & T4 & Ampere & 10752 & 336 & 37.42 & 48 & 695.8  \\
            \textbf{RTX 3060} & R6 & Ampere & 3584  & 112 & 12.74 & 12 & 360.0  \\
            \textbf{GTX 1660Ti} & G6 & Turing & 1536  & N/A  & 5.43  & 6  & 288.0  \\
            \Xhline{1.5pt}
        \end{tabular}%
    }
    \label{tab:gpu_info}
\end{table}

\begin{table}[!htbp]
    \vspace{-1em}
    \centering
    \setlength{\abovecaptionskip}{0em}
    \setlength{\belowcaptionskip}{0em}
    \caption{Grouping of GPUs.}
    \label{tab:gpu_grouping}
    \setlength\tabcolsep{6pt}
    \resizebox{\columnwidth}{!}{%
        \begin{tabular}{ccccccccc}
            \Xhline{1.5pt}
            \noalign{\vskip 1pt}
            \textbf{Group} & \textbf{R9(1)} & \textbf{R9(2)} & \textbf{T4(1)} & \textbf{T4(2)} & \textbf{R6(1)} & \textbf{R6(2)} & \textbf{G6(1)} & \textbf{G6(2)} \\
            \noalign{\vskip 1pt}
            \Xhline{1pt}
            \noalign{\vskip 1pt}
            \textbf{x2} & \checkmark & \checkmark & $\times$ & $\times$ & $\times$ & $\times$ & $\times$ & $\times$ \\
            \textbf{x3} & \checkmark & \checkmark & \checkmark & $\times$ & $\times$ & $\times$ & $\times$ & $\times$ \\
            \textbf{x4} & \checkmark & \checkmark & \checkmark & \checkmark & $\times$ & $\times$ & $\times$ & $\times$ \\
            \textbf{x5} & \checkmark & \checkmark & \checkmark & \checkmark & \checkmark & $\times$ & $\times$ & $\times$ \\
            \textbf{x6} & \checkmark & \checkmark & \checkmark & \checkmark & \checkmark & \checkmark & $\times$ & $\times$ \\
            \textbf{x7} & \checkmark & \checkmark & \checkmark & \checkmark & \checkmark & \checkmark & \checkmark & $\times$ \\
            \textbf{x8} & \checkmark & \checkmark & \checkmark & \checkmark & \checkmark & \checkmark & \checkmark & \checkmark \\
            \Xhline{1.5pt}
        \end{tabular}%
    }
\end{table}

We evaluate CaPGNN on seven representative benchmark datasets of varying scales. Table~\ref{tab:dataset_info} lists these datasets and their corresponding characteristics. These datasets are managed by DGL and OGBN and are used for node prediction tasks. For single-label classification tasks, we use the Cross-Entropy loss and accuracy for evaluation, while for multi-label tasks, we use the Binary Cross-Entropy loss and F1 score. We perform the evaluation on classic GCN~\cite{Kipf_Welling_2016} and GraphSAGE~\cite{Hamilton_Ying_Leskovec_2017} models. To ensure fairness, we set the same parameters. Specifically, we use a three-layer GNN model with hidden layer dimensions of 256, a learning rate of 0.01, and training for 200 epochs, while other parameters are kept as recommended by the respective comparison methods. Unless otherwise stated, the time for all experiments is the total time for 200 epochs. The parameter $\epsilon$ is set to 1\% of the average $\lambda$, while $\beta$ is configured to 100MB.

\begin{table}[!hbt]
    \setlength{\abovecaptionskip}{1em}
    \setlength{\belowcaptionskip}{0pt}
    \caption{Graph datasets used in experiments.}
    \label{tab:dataset_info}
    \resizebox{\columnwidth}{!}{%
        \begin{tabular}{llllll}
            \Xhline{1.5pt}
            \textbf{Dataset} & \textbf{Label} & \textbf{\#Nodes} & \textbf{\#Edges} & \textbf{\#Features} & \textbf{\#Classes} \\
            \Xhline{1pt}
            \textbf{CoraFull}~\cite{bojchevski2018deep}            & Cl & 19,793   & 126,842    & 8,710 & 70    \\
            \textbf{Flickr}~\cite{graphsainticlr20}              & Fr & 89,250   & 899,756    & 500  & 7    \\
            \textbf{CoauthorPhysics}~\cite{shchur2018pitfalls}     & Cs & 34,493   & 495,924    & 8,415 & 5    \\
            \Xhline{0.5pt}
            \textbf{Reddit}~\cite{hamilton2017inductive}              & Rt & 232,965  & 114,615,892 & 602  & 41   \\
            \textbf{Yelp}~\cite{graphsainticlr20}                & Yp & 716,847  & 13,954,819  & 300  & 100  \\
            \textbf{AmazonProducts}~\cite{graphsainticlr20}      & As & 1,569,960 & 264,339,468 & 200  & 107  \\
            \textbf{ogbn-products}~\cite{hu2020open}       & Os & 2,449,029 & 61,859,140  & 100  & 47   \\
            \Xhline{1.5pt}
        \end{tabular}%
    }
\end{table}

We conduct a systematic comparison of CaPGNN with the baseline method Vanilla~\cite{wan2023adaptive} and state-of-the-art methods AdaQP~\cite{wan2023adaqp} and SANCUS~\cite{sancus}. All compared methods are full-batch training. Therefore, the comparison is fair in terms of training paradigm. Since DistGCN and CachedGCN from SANCUS does not provide implementations for GraphSAGE, we perform comparisons on their respective available versions. The differences in the comparison methods are detailed in Table~\ref{tab:alg_detail}. 

\begin{table}[ht]
    \centering
    \belowrulesep=0pt
    \aboverulesep=0pt
    \setlength{\abovecaptionskip}{0pt}
    \setlength{\belowcaptionskip}{0pt}
    \caption{Differences in comparison methods. \textit{Part} denotes partition, \textit{Pipe} denotes pipeline, \textit{Quant} denotes 8-bit quantification, and \textit{Comm} denotes communication. All compared methods are full-batch training.}
    \label{tab:alg_detail}
    \resizebox{\linewidth}{!}{%
        \begin{threeparttable}
            \begin{tabular}{lccccccc}
                \Xhline{1.5pt}
                \textbf{Alg} & \textbf{Part.} & \textbf{Cache} & \textbf{Pipe.} & \textbf{Quant.} & \textbf{GCN} & \textbf{GraphSAGE} & \textbf{Comm.} \\
                \Xhline{1pt}
                \textbf{DistGCN} & 2D Split & $\times$ & $\times$ & $\times$ & $\checkmark$ & $\times$ & NCCL \\
                \textbf{CachedGCN} & 2D Split & Block & $\times$ & $\times$ & $\checkmark$ & $\times$ & NCCL \\
                \textbf{Vanilla} & METIS & $\times$ & $\times$ & $\times$ & $\checkmark$ & $\checkmark$ & GLOO \\
                \textbf{AdaQP} & METIS & $\times$ & $\checkmark$ & Adaptive & $\checkmark$ & $\checkmark$ & GLOO \\
                \textbf{CaPGNN$^*$} & RAPA & JACA & $\checkmark$ & $\times$ & $\checkmark$ & $\checkmark$ & GLOO \\
                \Xhline{1.5pt}
            \end{tabular}%
            \begin{tablenotes}
        		\item $^*$This Work.
            \end{tablenotes}
        \end{threeparttable}
    }
\end{table}

\subsection{Impact of Vertex Overlap Ratio on Cache Hit Rate}
\label{ssec:vor_cache}
\begin{figure}[!htbp]
    \centering
    \setlength{\abovecaptionskip}{0pt}
    \setlength{\belowcaptionskip}{0pt}
    \begin{subfigure}[t]{0.48\linewidth}
        \centering
        \setlength{\abovecaptionskip}{0pt}
        \setlength{\belowcaptionskip}{0pt}
        \includegraphics[width=\linewidth]{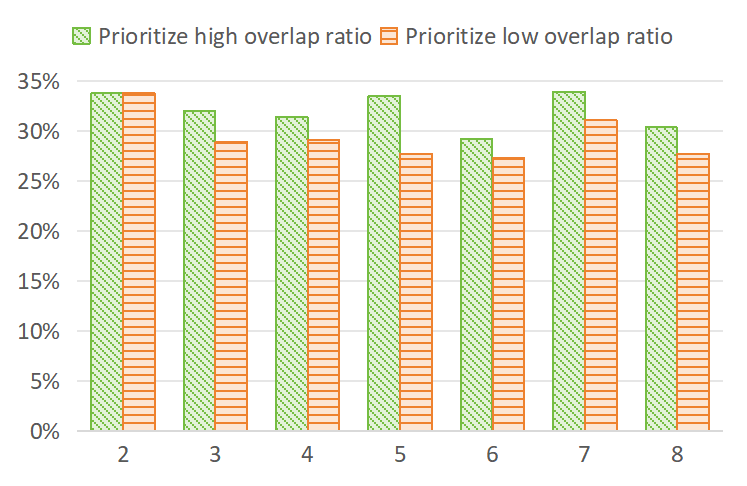}
        \caption{GCN.}
        \label{fig:cache_hit_ratio_gcn}
    \end{subfigure}
    \begin{subfigure}[t]{0.48\linewidth}
        \centering
        \setlength{\abovecaptionskip}{0pt}
        \setlength{\belowcaptionskip}{0pt}
        \includegraphics[width=\textwidth]{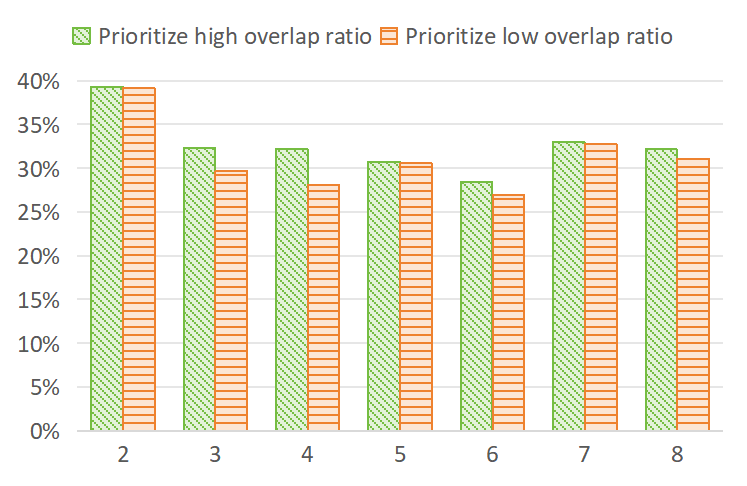}
        \caption{GraphSAGE.}
        \label{fig:cache_hit_ratio_sage}
    \end{subfigure}
    \caption{Cache hit rates on the Reddit dataset when prioritizing vertices with high vs. low overlap ratios, using GCN and GraphSAGE models. The number of partitions ranging from 2 to 8. Both global and local caches are set to 20\% of the maximum capacity.}
    \label{fig:vor_on_chr}
    \vspace{-1em}
\end{figure}

To verify the effectiveness of prioritizing caching of vertices with high overlap ratio in improving the cache hit rate as proposed in Section~\ref{sec:joint_cache}, we present in Fig.~\ref{fig:vor_on_chr} the cache hit rates on the Reddit dataset when prioritizing vertices with high versus low overlap ratios. The results are shown for both GCN and GraphSAGE models. In the experiments, the models are configured with three layers, the number of partitions increases from 2 to 8, and both global and local cache sizes are set to 20\% of their maximum capacities. We observe that, under both GCN and GraphSAGE models, prioritizing the caching of high-overlap-ratio vertices consistently leads to higher cache hit rates, validating the practical effectiveness of the overlap-ratio-based caching strategy in improving cache efficiency.

\subsection{Impact of Cache Capacity and Partition on Cache Hit Rate}
\label{ssec:cc_cache_hit}
\begin{figure*}[!htbp]
    \vspace{-1em}
    \centering
    \setlength{\abovecaptionskip}{0pt}
    \setlength{\belowcaptionskip}{0pt}
    \begin{subfigure}[b]{0.24\textwidth}
        \setlength{\abovecaptionskip}{0pt}
        \setlength{\belowcaptionskip}{0pt}
        \includegraphics[width=\textwidth]{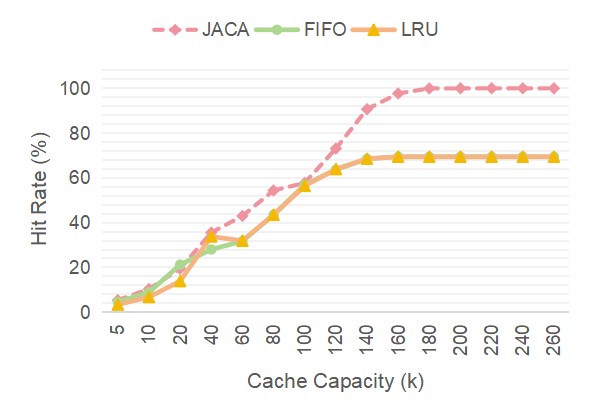}
        \caption{GCN, 2p.}
    \end{subfigure}
    \hfill
    \begin{subfigure}[b]{0.24\textwidth}
        \setlength{\abovecaptionskip}{0pt}
        \setlength{\belowcaptionskip}{0pt}
        \includegraphics[width=\textwidth]{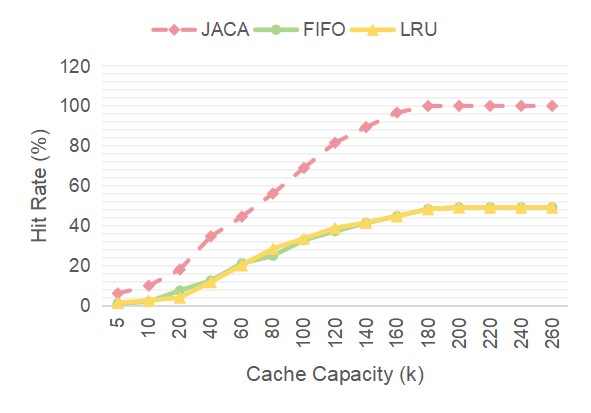}
        \caption{GCN, 4p.}
    \end{subfigure}
    \hfill
    \begin{subfigure}[b]{0.24\textwidth}
        \setlength{\abovecaptionskip}{0pt}
        \setlength{\belowcaptionskip}{0pt}
        \includegraphics[width=\textwidth]{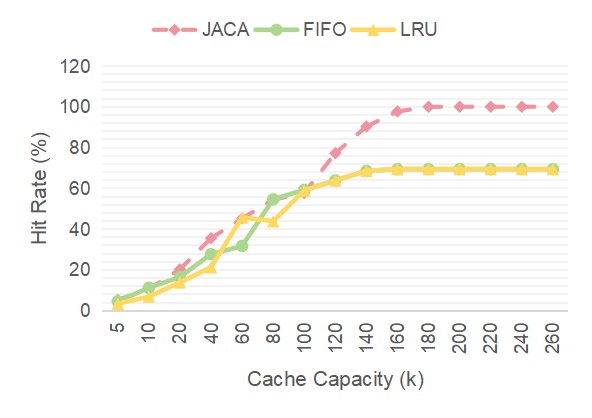}
        \caption{GraphSAGE, 2p.}
    \end{subfigure}
    \hfill
    \begin{subfigure}[b]{0.24\textwidth}
        \setlength{\abovecaptionskip}{0pt}
        \setlength{\belowcaptionskip}{0pt}
        \includegraphics[width=\textwidth]{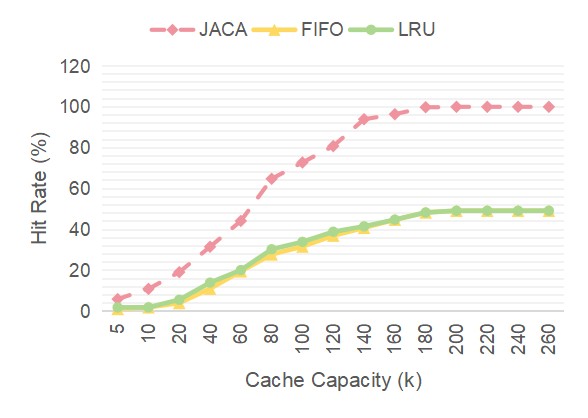}
        \caption{GraphSAGE, 4p.}
    \end{subfigure}
    \caption{Cache hit rate under different cache capacities and partitions. The number of partitions is 2 and 4, and the cache capacity ranges from 5k to 260k. The global and local cache capacities change simultaneously.
    }
    \label{fig:hit_ratio_cache_capacity}
\end{figure*}

To evaluate the impact of cache capacity and the number of partitions on cache hit rate, we compare the performance of GCN and GraphSAGE models under various configurations. To isolate the effects of caching alone, both RAPA and Pipeline techniques are excluded. The models are configured with three layers, the number of partitions is 2 and 4, and both global and local cache capacities are set equal, varying from 5K to 260K. To ensure a fair comparison, the classic cache replacement strategies FIFO and LRU are both implemented using a two-level cache structure, consisting of CPU and GPU caches, consistent with the architecture used by JACA. As shown in Fig.~\ref{fig:hit_ratio_cache_capacity}, the cache hit rate gradually increases with cache capacity and stabilizes after reaching a certain threshold value. Moreover, while there are some differences in cache hit rates across different models and partition scenarios, the overall trend remains consistent, indicating that the caching strategy adapts well to the workload characteristics of each model. Additionally, larger cache capacities generally improve hit rates, but excessive caching may lead to resource waste, highlighting the need to find an optimal cache capacity that balances performance and resource utilization. 

Upon closer examination, the JACA strategy consistently outperforms the traditional FIFO and LRU approaches. Even with relatively small cache capacities ($<$80K), JACA achieves significantly higher hit ratios compared to the other methods; as the cache size increases, its hit ratio approaches or even reaches 100\%. This highlights the effectiveness of JACA in identifying hot and cold data and managing replacements accordingly. It is also worth noting that the performance gap between FIFO and LRU is relatively small, and both fall short of JACA in high-capacity scenarios, further underscoring the importance of designing customized caching strategies tailored to the characteristics of graph computation workloads.

\begin{figure*}[!htbp]
    \centering
    \setlength{\abovecaptionskip}{0pt}
    \setlength{\belowcaptionskip}{0pt}
    \begin{subfigure}[b]{0.32\textwidth}
        \setlength{\abovecaptionskip}{0pt}
        \setlength{\belowcaptionskip}{0pt}
        \includegraphics[width=\textwidth]{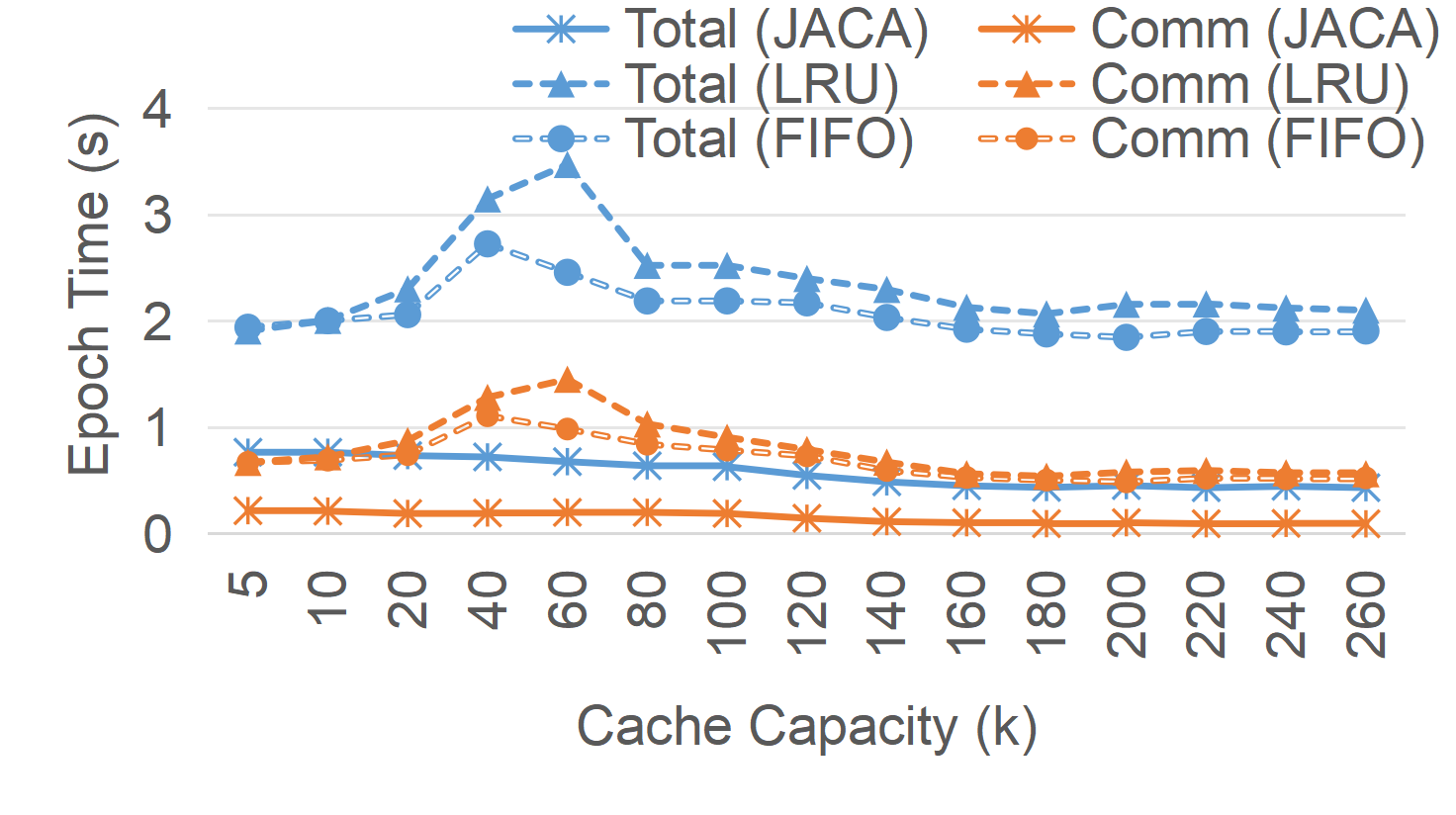}
        \caption{GCN, 2p.}
    \end{subfigure}
    \hfill
    \begin{subfigure}[b]{0.32\textwidth}
        \setlength{\abovecaptionskip}{0pt}
        \setlength{\belowcaptionskip}{0pt}
        \includegraphics[width=\textwidth]{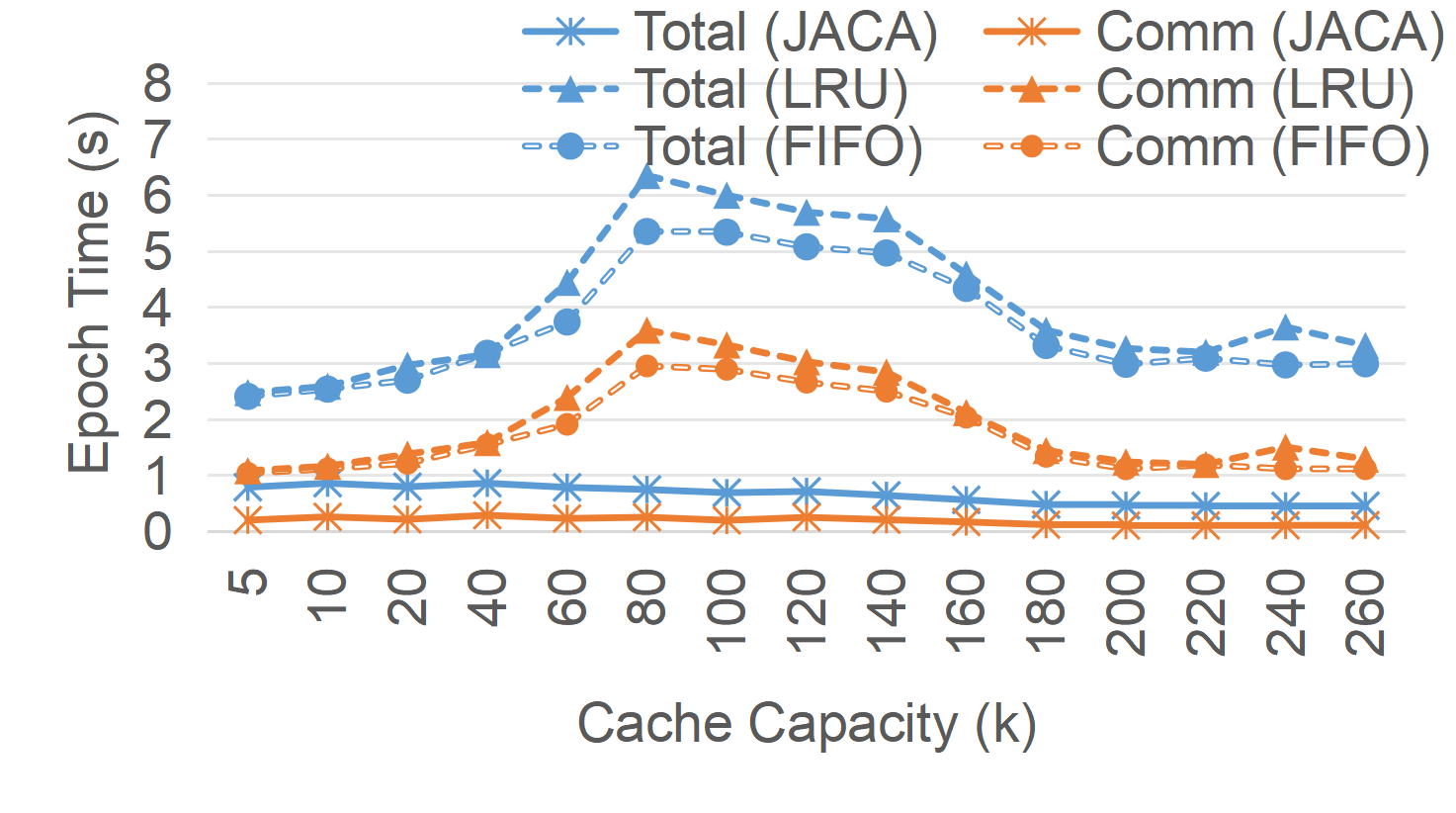}
        \caption{GCN, 3p.}
    \end{subfigure}
    \hfill
    \begin{subfigure}[b]{0.32\textwidth}
        \setlength{\abovecaptionskip}{0pt}
        \setlength{\belowcaptionskip}{0pt}
        \includegraphics[width=\textwidth]{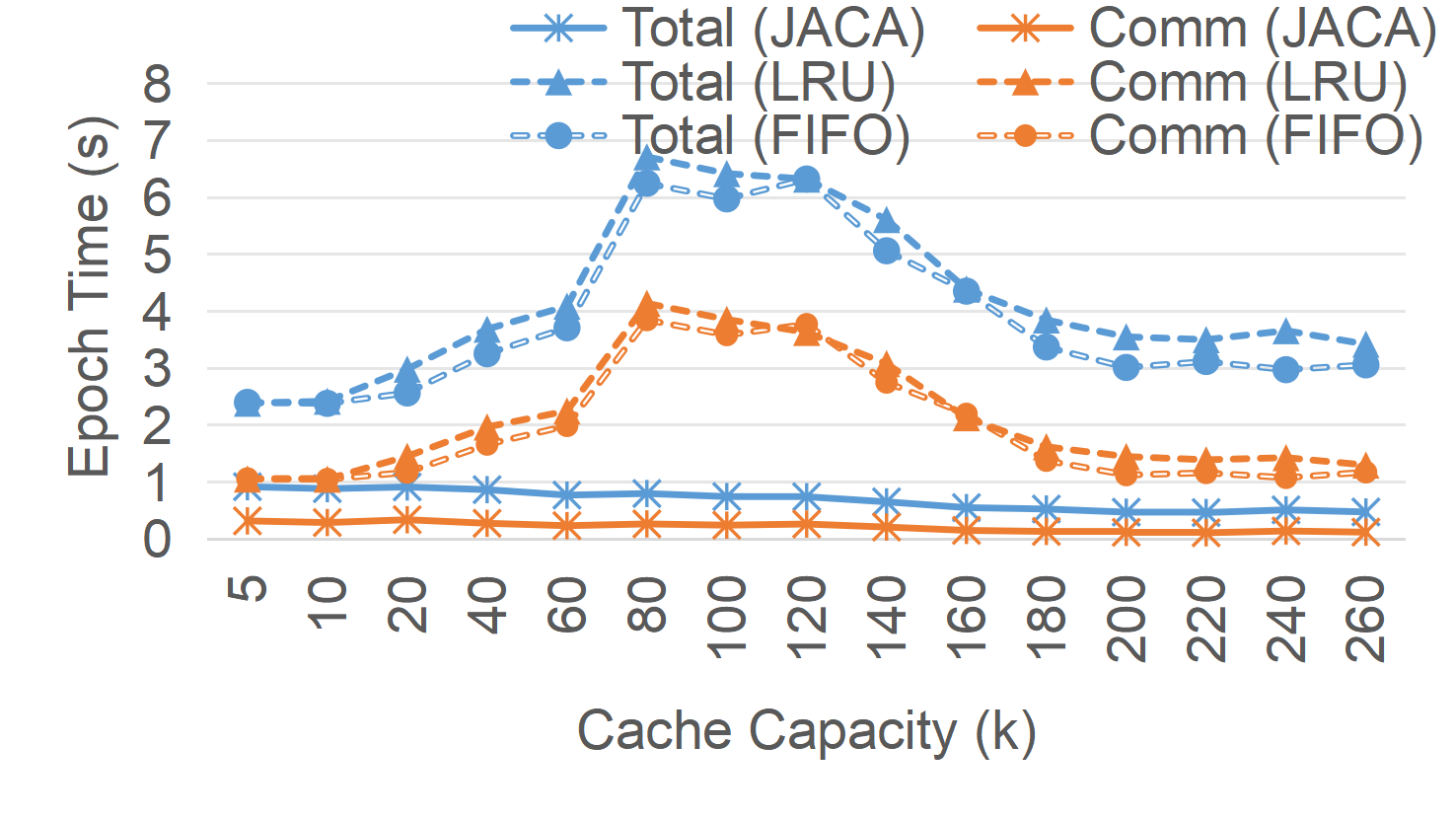}
        \caption{GCN, 4p.}
    \end{subfigure}
    \hfill
    \begin{subfigure}[b]{0.32\textwidth}
        \setlength{\abovecaptionskip}{0pt}
        \setlength{\belowcaptionskip}{0pt}
        \includegraphics[width=\textwidth]{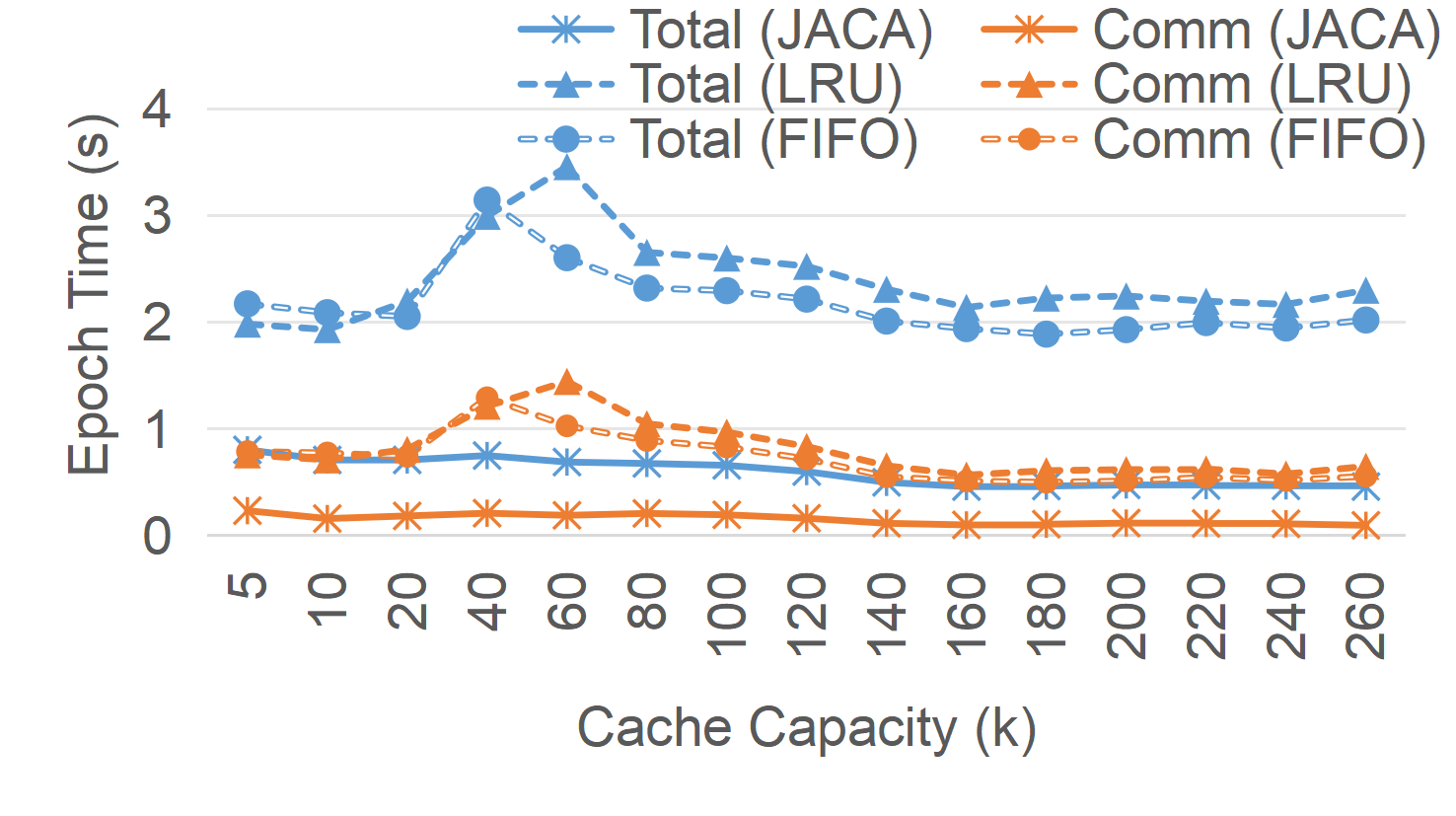}
        \caption{GraphSAGE, 2p.}
    \end{subfigure}
    \hfill
    \begin{subfigure}[b]{0.32\textwidth}
        \setlength{\abovecaptionskip}{0pt}
        \setlength{\belowcaptionskip}{0pt}
        \includegraphics[width=\textwidth]{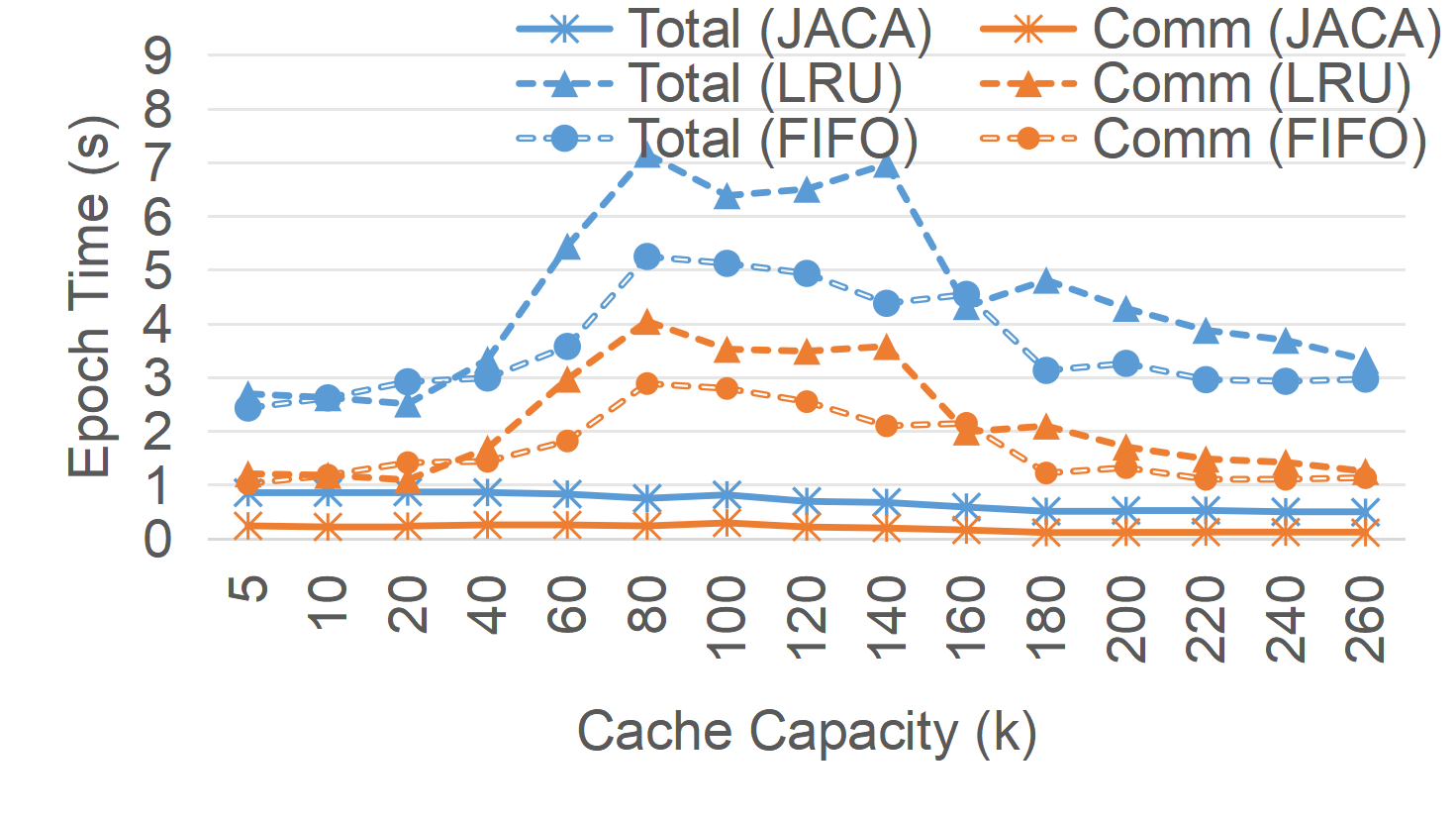}
        \caption{GraphSAGE, 3p.}
    \end{subfigure}
    \hfill
    \begin{subfigure}[b]{0.32\textwidth}
        \setlength{\abovecaptionskip}{0pt}
        \setlength{\belowcaptionskip}{0pt}
        \includegraphics[width=\textwidth]{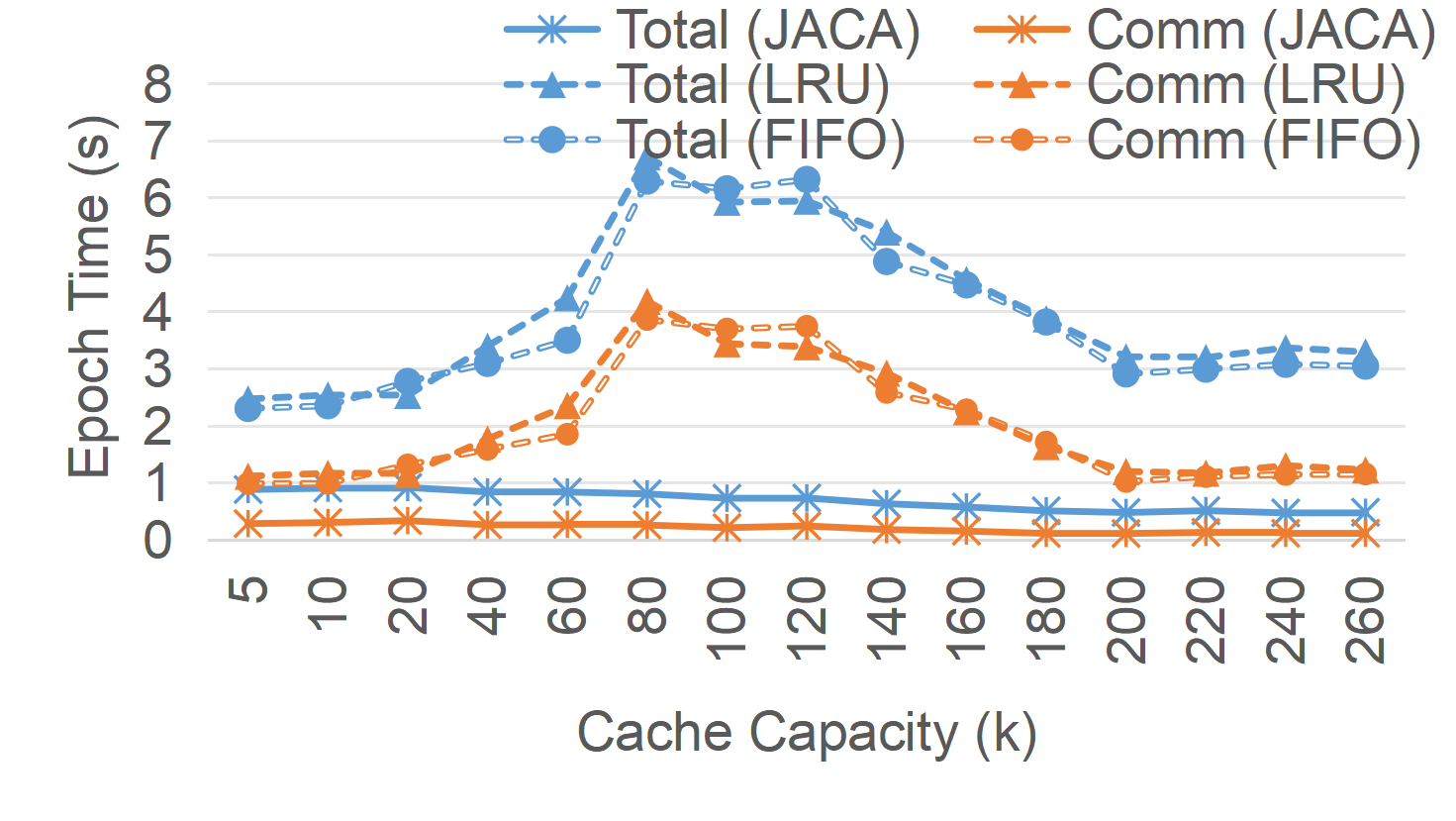}
        \caption{GraphSAGE, 4p.}
    \end{subfigure}
    \caption{Epoch time under different cache capacities and partitions. The number of partitions ranges from 2 to 4, and the cache capacity ranges from 5k to 260k. The global and local cache capacities change simultaneously.
    }
    \label{fig:etime_cache_partition}
    \vspace{-1em}
\end{figure*}

\begin{figure*}[!htbp]
    \centering
    \setlength{\abovecaptionskip}{0pt}
    \setlength{\belowcaptionskip}{0pt}
    \begin{subfigure}[b]{0.32\textwidth}
        \setlength{\abovecaptionskip}{0pt}
        \setlength{\belowcaptionskip}{0pt}
        \includegraphics[width=\textwidth]{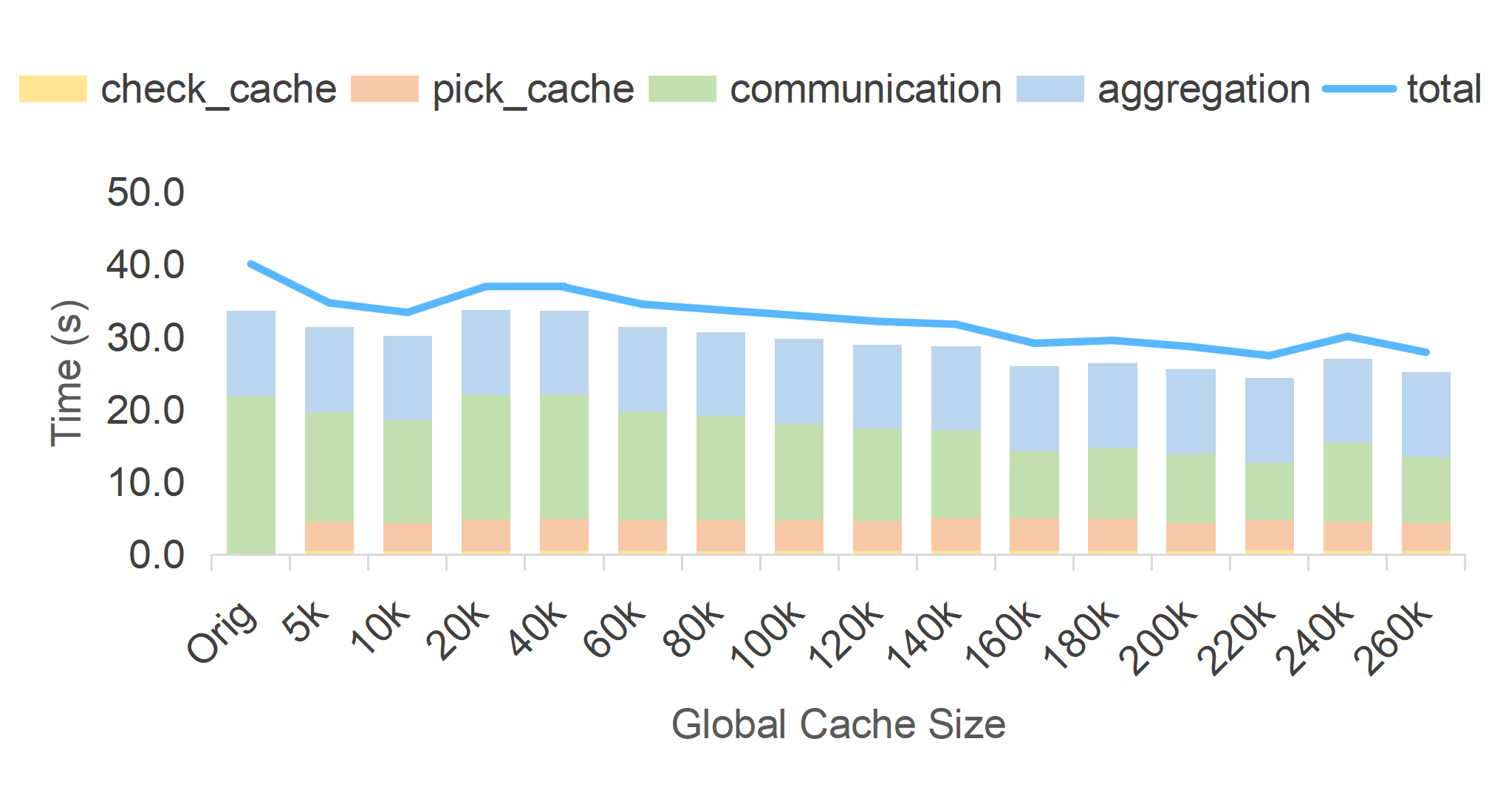}
        \caption{Fix local, 2p.}
    \end{subfigure}
    \hfill
    \begin{subfigure}[b]{0.32\textwidth}
        \setlength{\abovecaptionskip}{0pt}
        \setlength{\belowcaptionskip}{0pt}
        \includegraphics[width=\textwidth]{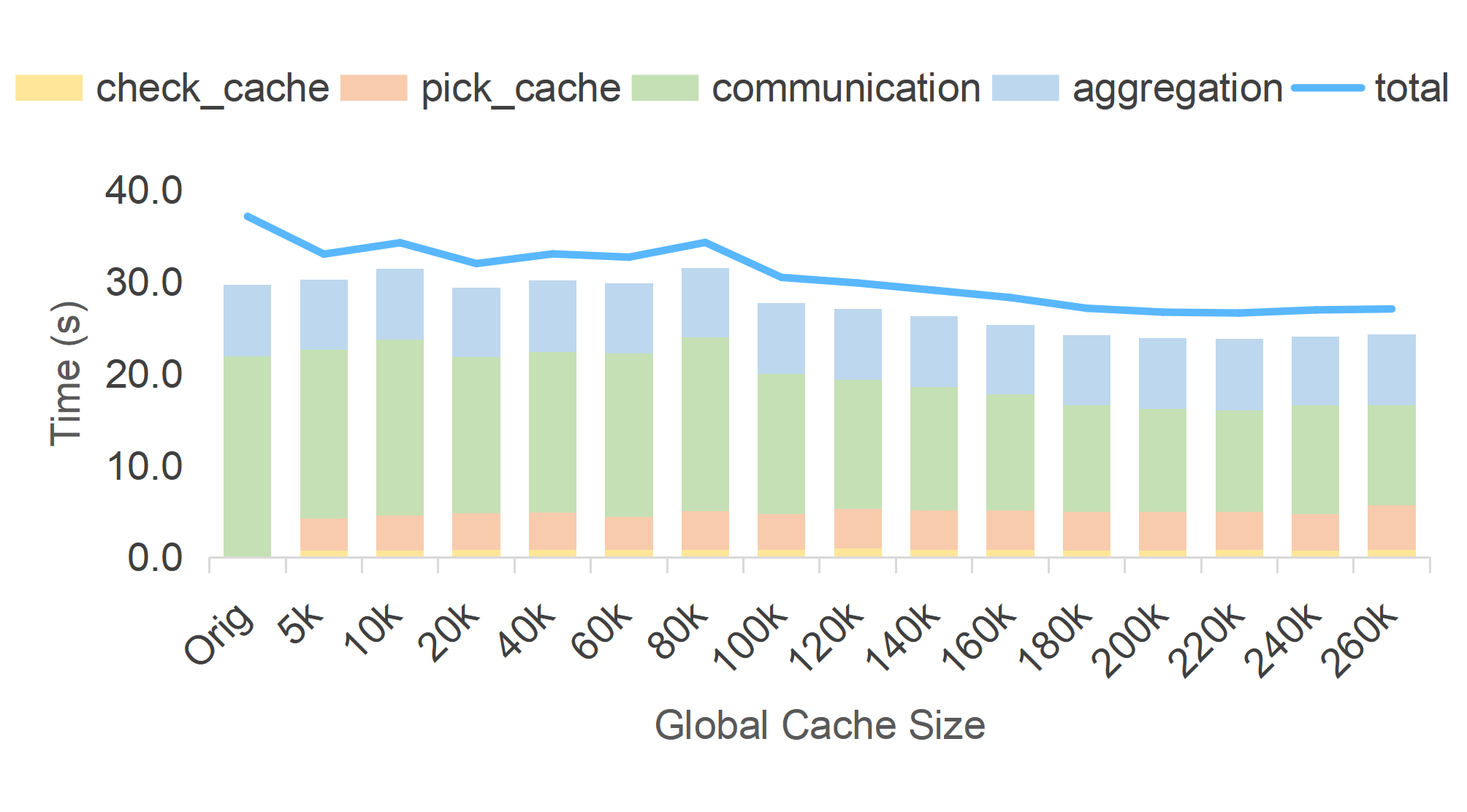}
        \caption{Fix local, 3p.}
    \end{subfigure}
    \hfill
    \begin{subfigure}[b]{0.32\textwidth}
        \setlength{\abovecaptionskip}{0pt}
        \setlength{\belowcaptionskip}{0pt}
        \includegraphics[width=\textwidth]{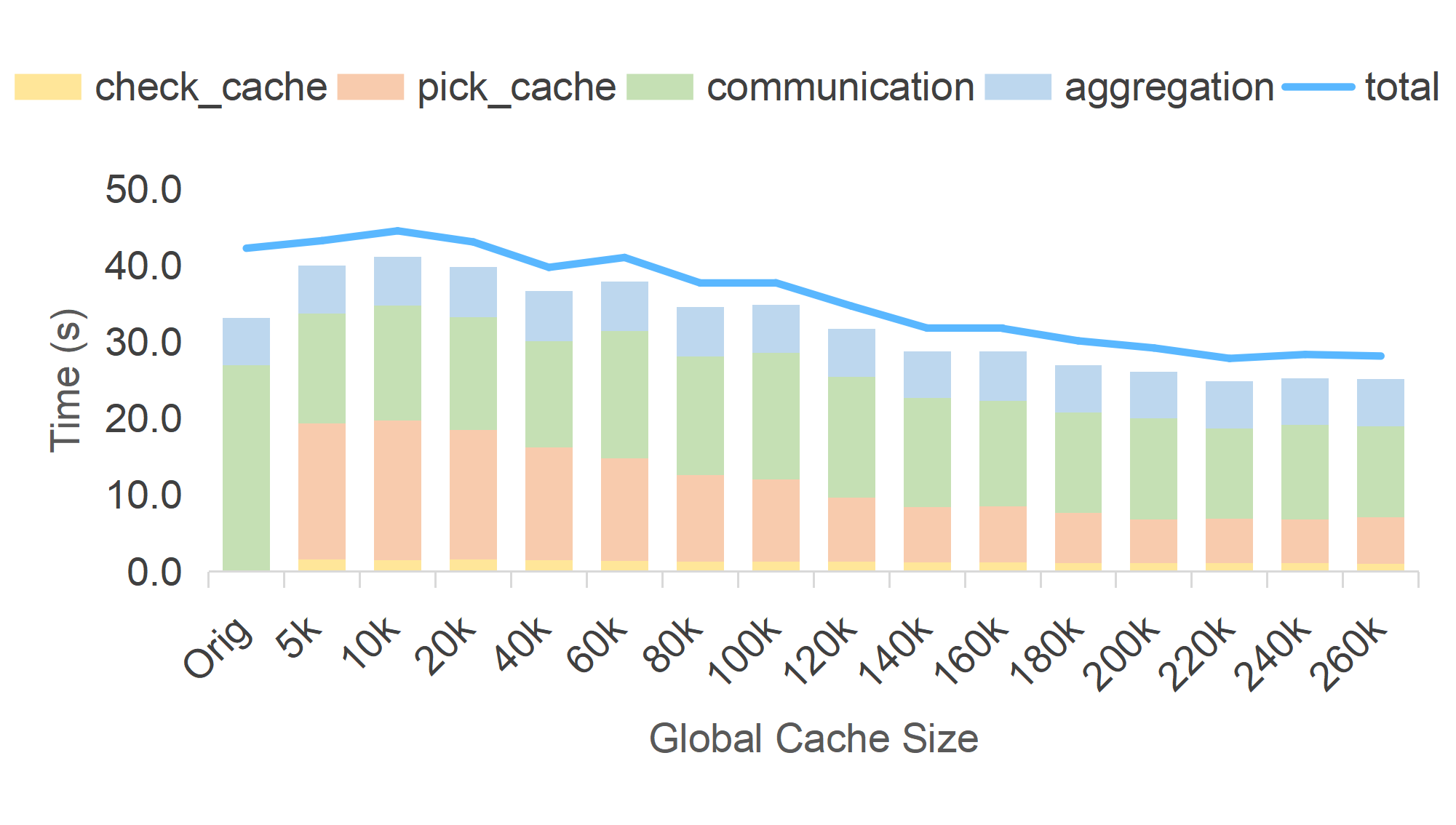}
        \caption{Fix local, 4p.}
    \end{subfigure}
    
    \begin{subfigure}[b]{0.32\textwidth}
        \setlength{\abovecaptionskip}{0pt}
        \setlength{\belowcaptionskip}{0pt}
        \includegraphics[width=\textwidth]{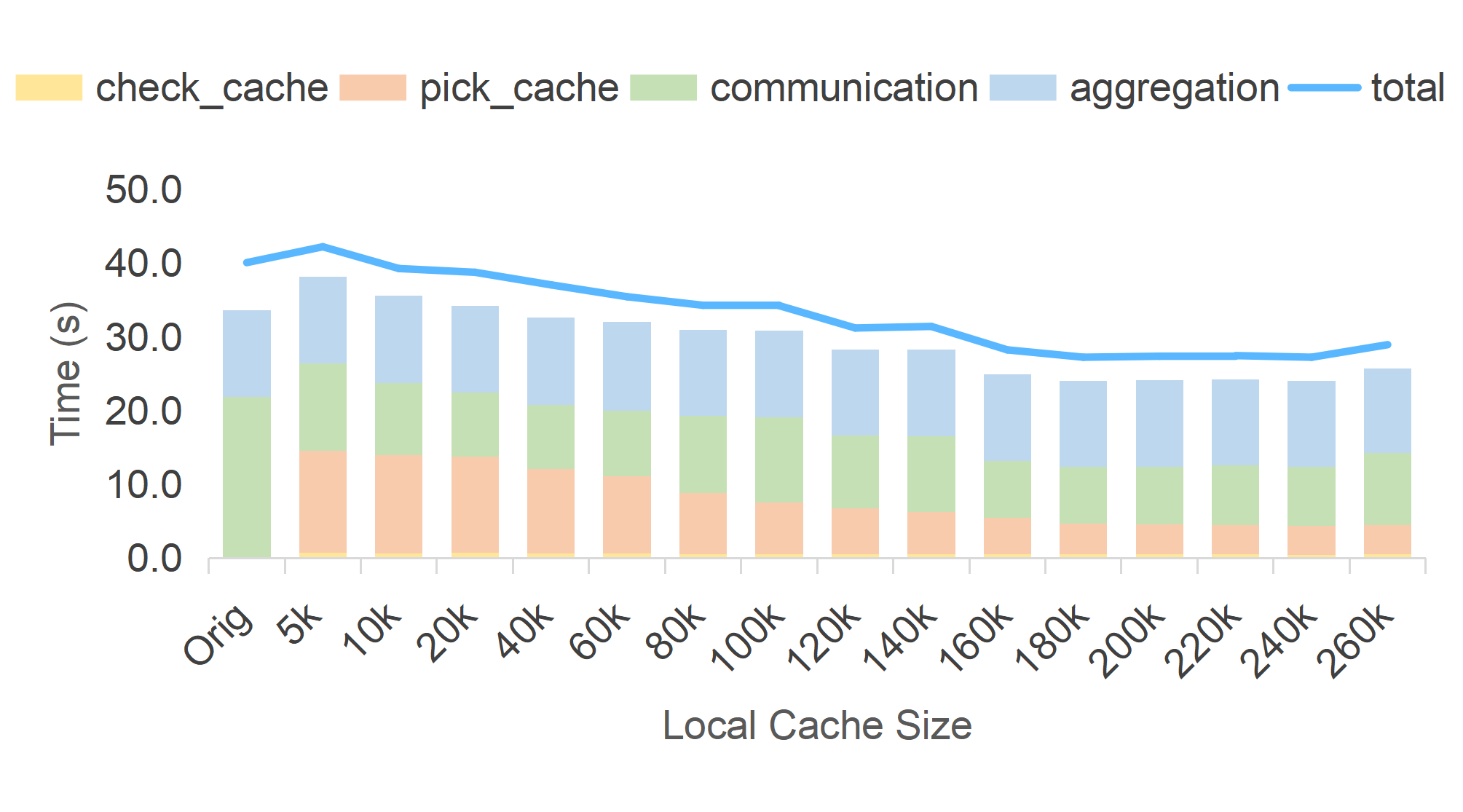}
        \caption{Fix global, 2p.}
    \end{subfigure}
    \hfill
    \begin{subfigure}[b]{0.32\textwidth}
        \setlength{\abovecaptionskip}{0pt}
        \setlength{\belowcaptionskip}{0pt}
        \includegraphics[width=\textwidth]{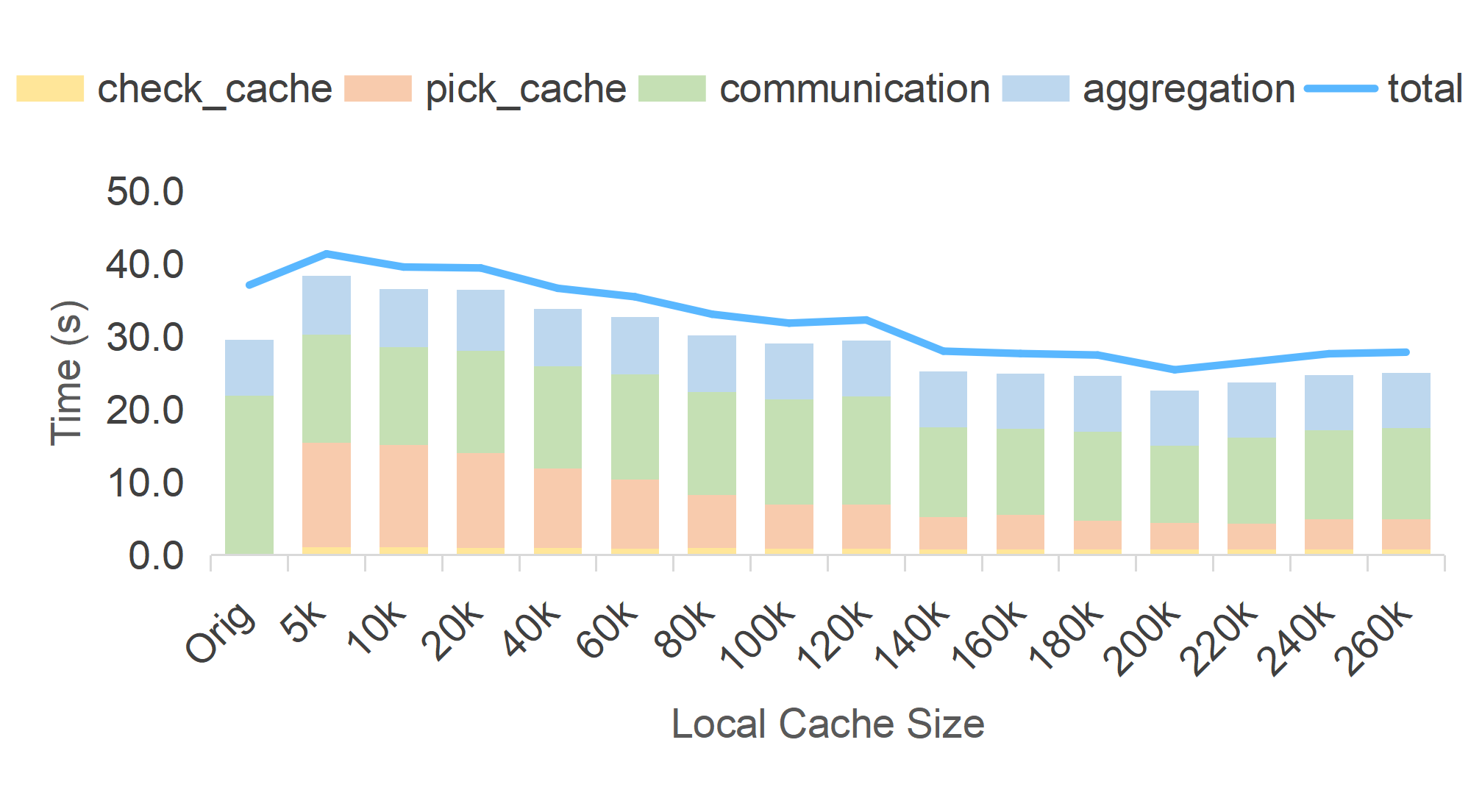}
        \caption{Fix global, 3p.}
    \end{subfigure}
    \hfill
    \begin{subfigure}[b]{0.32\textwidth}
        \setlength{\abovecaptionskip}{0pt}
        \setlength{\belowcaptionskip}{0pt}
        \includegraphics[width=\textwidth]{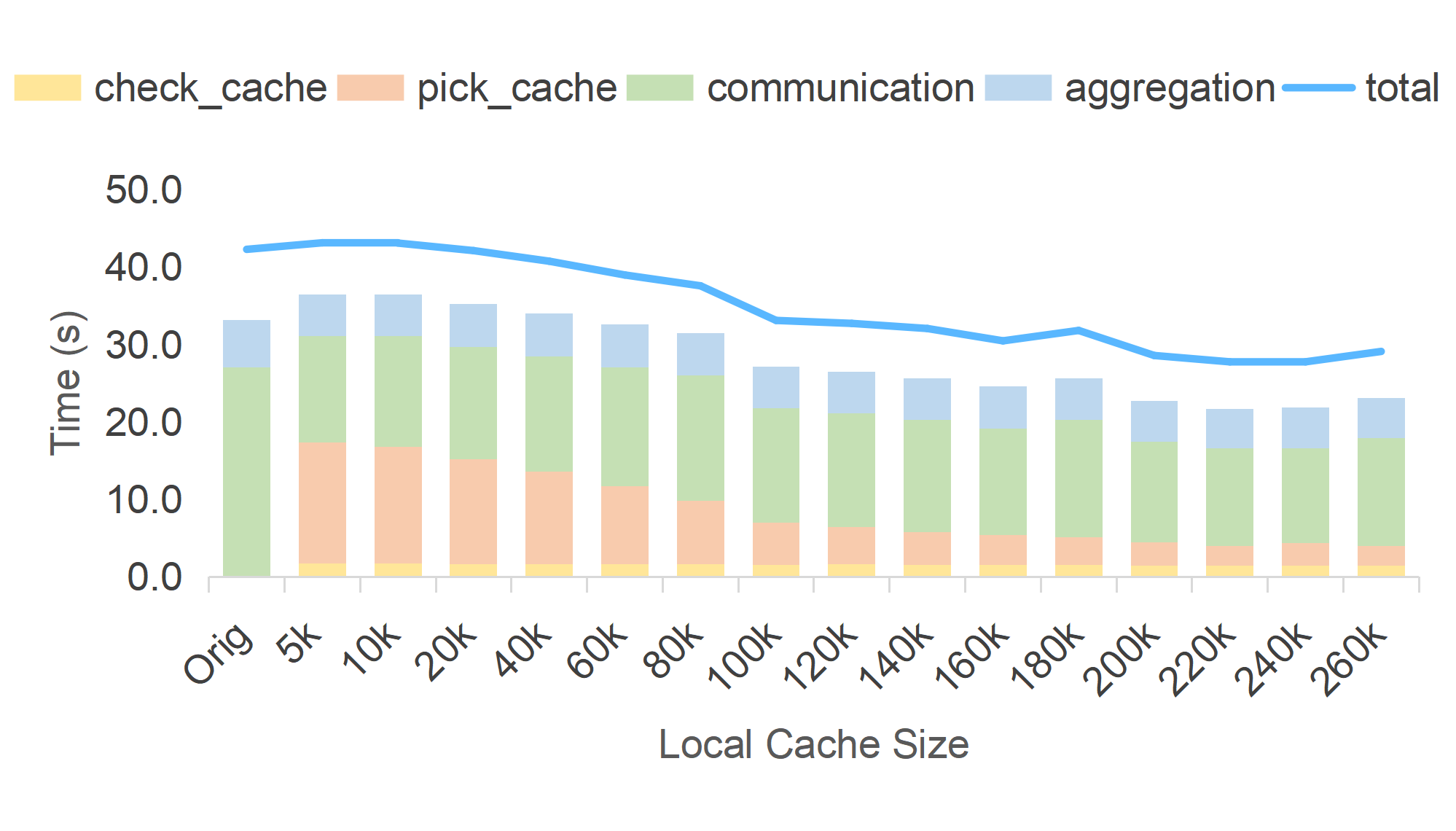}
        \caption{Fix global, 4p.}
    \end{subfigure}
    \caption{Impact of cache capacity with one fixed. 
        (a)-(c) fix local cache capacity with 2, 3, and 4 partitions (2p, 3p, 4p), respectively; 
        (d)-(f) fix global cache capacity with 2, 3, and 4 partitions (2p, 3p, 4p), respectively.
    }
    \label{fig:impact_cache_capacity_fix}
    \vspace{-1em}
\end{figure*}

\begin{figure*}[!htbp]
    \centering
    \setlength{\abovecaptionskip}{0pt}
    \setlength{\belowcaptionskip}{0pt}
    \begin{subfigure}[b]{0.32\textwidth}
        \setlength{\abovecaptionskip}{0pt}
        \setlength{\belowcaptionskip}{0pt}
        \includegraphics[width=\textwidth]{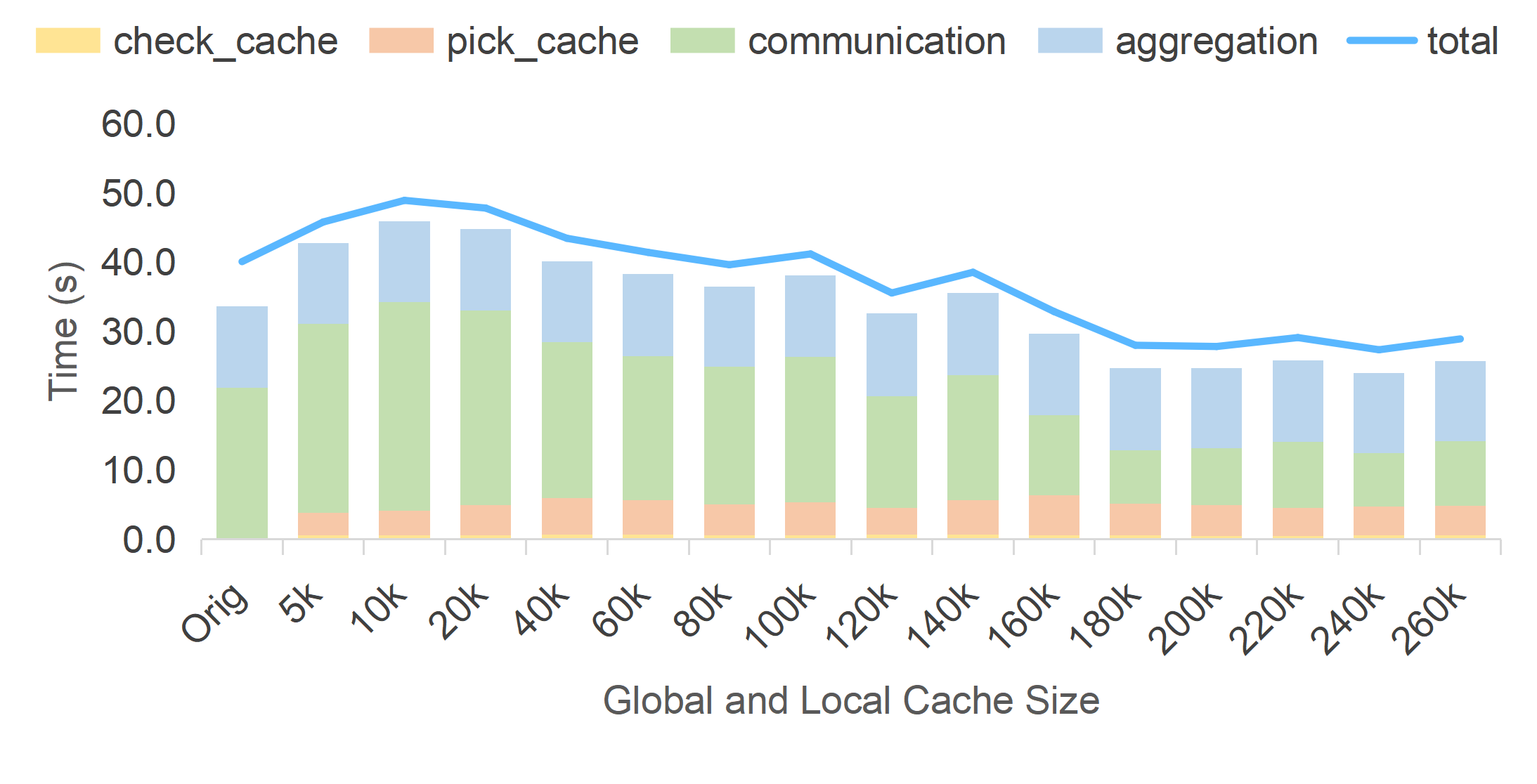}
        \caption{Both cache, 2p.}
    \end{subfigure}
    \hfill
    \begin{subfigure}[b]{0.32\textwidth}
        \setlength{\abovecaptionskip}{0pt}
        \setlength{\belowcaptionskip}{0pt}
        \includegraphics[width=\textwidth]{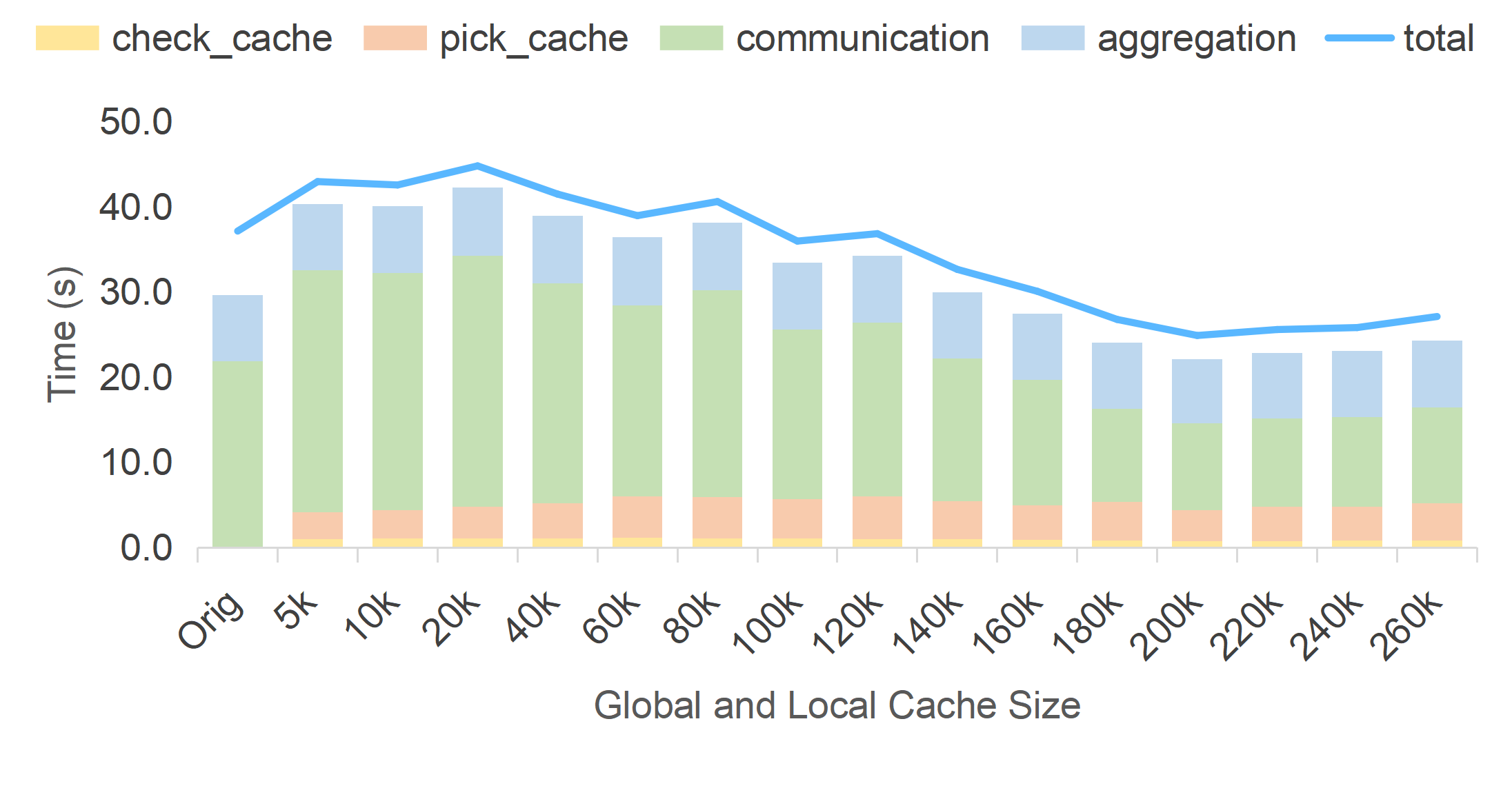}
        \caption{Both cache, 3p.}
    \end{subfigure}
    \hfill
    \begin{subfigure}[b]{0.32\textwidth}
        \setlength{\abovecaptionskip}{0pt}
        \setlength{\belowcaptionskip}{0pt}
        \includegraphics[width=\textwidth]{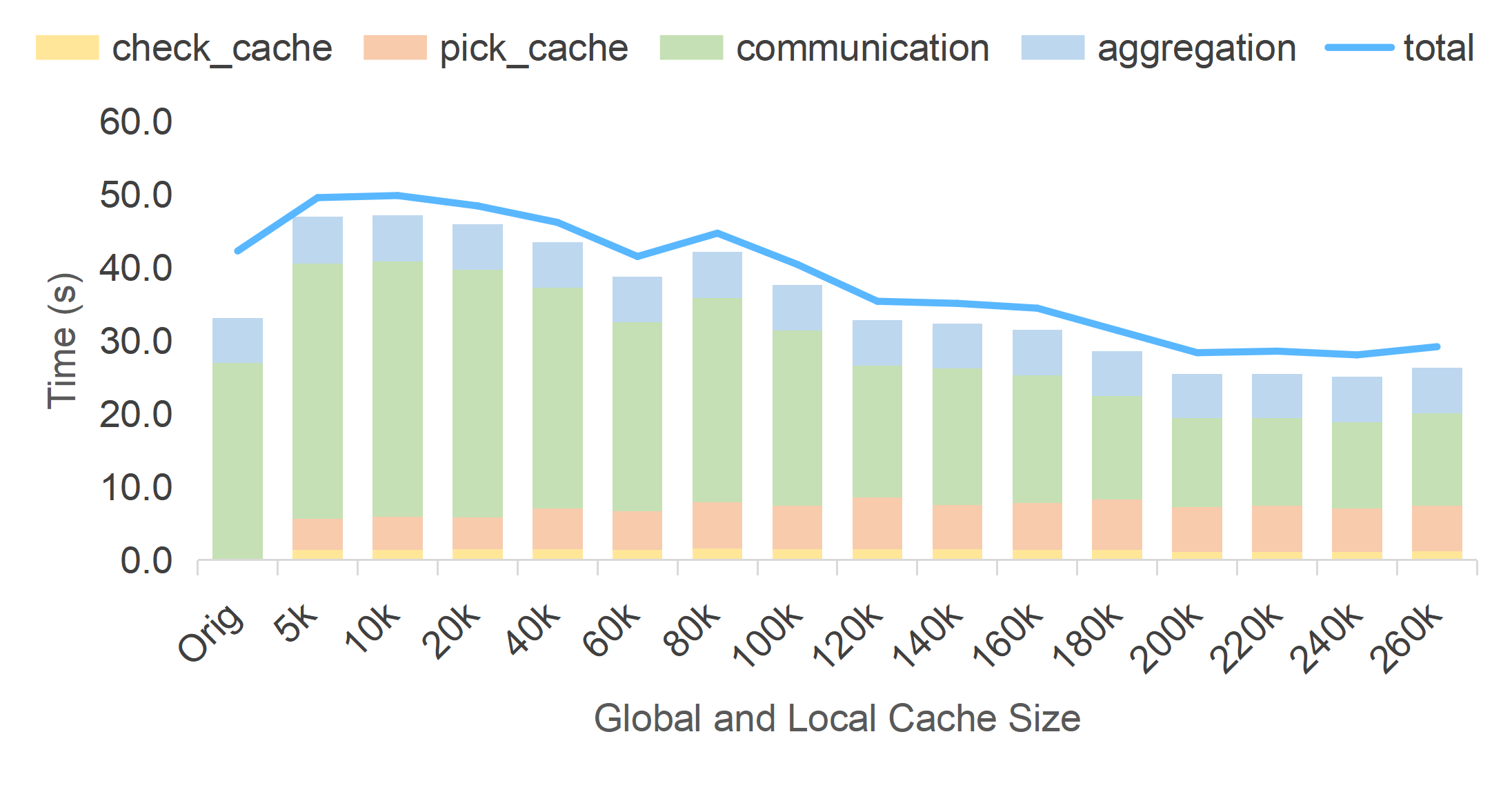}
        \caption{Both cache, 4p.}
    \end{subfigure}
    \caption{Impact of simultaneously varying both cache capacities. Experiments with 2, 3, and 4 partitions (2p, 3p, 4p), where both local and global cache capacities are adjusted together.
    }
    \label{fig:impact_cache_capacity_same}
\end{figure*}

\subsection{Impact of Cache Capacity and Partition on Epoch Time}
\label{ssec:cc_cache_etime}
To evaluate how cache capacity and the number of partitions affect GNN training efficiency, we analyze the average epoch time under different cache configurations. As shown in Fig.~\ref{fig:etime_cache_partition}, we compare three caching strategies: our proposed JACA method and two classic baselines FIFO and LRU, under varying cache capacities (5K to 260K) and partition counts (2 to 4), using both GCN and GraphSAGE models. The epoch time is decomposed into Total Time and Communication Time to capture the runtime cost breakdown. Importantly, all three strategies are implemented under the same system-level optimization settings to ensure a fair comparison. Specifically, all methods use a joint cache structure, hash-based feature indexing, pinned memory for asynchronous CPU-GPU communication, per-GPU dedicated pinned memory regions, and a shared memory buffer for global caching across GPUs. Thus, the performance differences can be attributed primarily to the caching policy rather than to differences in memory management or communication pipelines. To isolate the effects of caching alone, both RAPA and Pipeline techniques are excluded.

JACA consistently achieves the lowest epoch time across all configurations, and its communication time remains minimal and stable, highlighting the effectiveness of our joint cache design. This advantage is attributed to two main factors: (i) a high cache hit ratio ensured by adaptive vertex prioritization and (ii) a highly optimized memory access and data transfer mechanism. In contrast, FIFO and LRU exhibit significantly higher communication overhead, exhibiting a clear ``peak-and-drop'' pattern, especially at intermediate cache capacities (e.g., 40K to 100K), where cache inefficiency leads to frequent CPU-GPU data transfers. As cache capacity increases further, this overhead is mitigated, leading to a gradual decline in total epoch time. These results demonstrate that JACA improves cache utilization and achieves optimization in communication efficiency and training throughput, showing superior system adaptability and scalability.

\subsection{Impact of Cache Capacity on System Performance}
\label{ssec:cc_cache_system}
To further investigate the impact of global and local cache capacities on system performance, we adopt two experimental approaches. The training process is divided into four key stages, \textit{check\_cache}, \textit{pick\_cache}, \textit{communication}, and \textit{aggregation}, to facilitate fine-grained performance analysis of the caching mechanism. In the first approach, we fix the capacity of one cache (either global or local), while varying the other to assess the individual impact of each cache. In the second approach, we vary both caches simultaneously to explore their combined effect on performance. Experiments are conducted on the Reddit dataset using a 3-layer GCN model trained for 50 epochs, with the number of partitions ranging from 2 to 4. To minimize external variability, all experiments are run on four NVIDIA RTX 3090 GPUs. To isolate the effects of caching alone, both RAPA and Pipeline techniques are excluded. Results are shown in Fig.~\ref{fig:impact_cache_capacity_fix} and Fig.~\ref{fig:impact_cache_capacity_same}, where the first column represents the baseline without caching.

From the experimental results, we observe that as cache capacity increases, both communication time and total training time decrease significantly. Once the cache capacity is sufficient to cover all halo vertices, the time overhead stabilizes and no longer changes with cache capacity. The reported communication time includes halo feature exchanges during aggregation, necessary synchronization and gradient-related communication in the backward and update stages. Therefore, even when the cache fully covers halo vertices, the overall communication time does not drop to zero. Figs.~\ref{fig:impact_cache_capacity_fix}(a)-(c) show that the overhead of the caching strategy (i.e., \textit{check\_cache} and \textit{pick\_cache}) is relatively small and stable. Although the local cache capacity in the experiments is sufficient to store all halo vertices, the communication time for vertex updates is still affected by the global cache capacity. In Figs.~\ref{fig:impact_cache_capacity_fix}(d)-(f), the overhead of the caching strategy decreases rapidly as the local cache capacity increases. This occurs because, when local cache capacity is small, many vertices need to fetch features from the global cache, resulting in higher overhead for cache retrieval.

Additionally, as shown in Fig.~\ref{fig:impact_cache_capacity_same}, the total time overhead may even exceed that of the case without caching when the cache capacity is small, due to the inherent overhead of the caching strategy itself. This is expected that with very small caches, only a few thousand vertices, the cache hit ratio is too low to cover the high-overlap halo vertices, so the communication saving is limited. Meanwhile, JACA still incurs relatively fixed bookkeeping costs for cache lookup and selection. When the hit ratio is near zero, these fixed costs cannot be amortized, leading to negative net gain. Therefore, selecting the appropriate cache capacity is crucial. Furthermore, it can be observed that while higher partition counts reduce computation time, they significantly increase communication costs. Fortunately, the caching strategy effectively mitigates the increase of communication overhead to some extent. In future work, we will further optimize the caching algorithm to minimize the overhead introduced by the caching strategy.

\begin{figure}[!htbp]
    \centering
    \setlength{\abovecaptionskip}{0pt}
    \setlength{\belowcaptionskip}{0pt}
    \includegraphics[width=\linewidth]{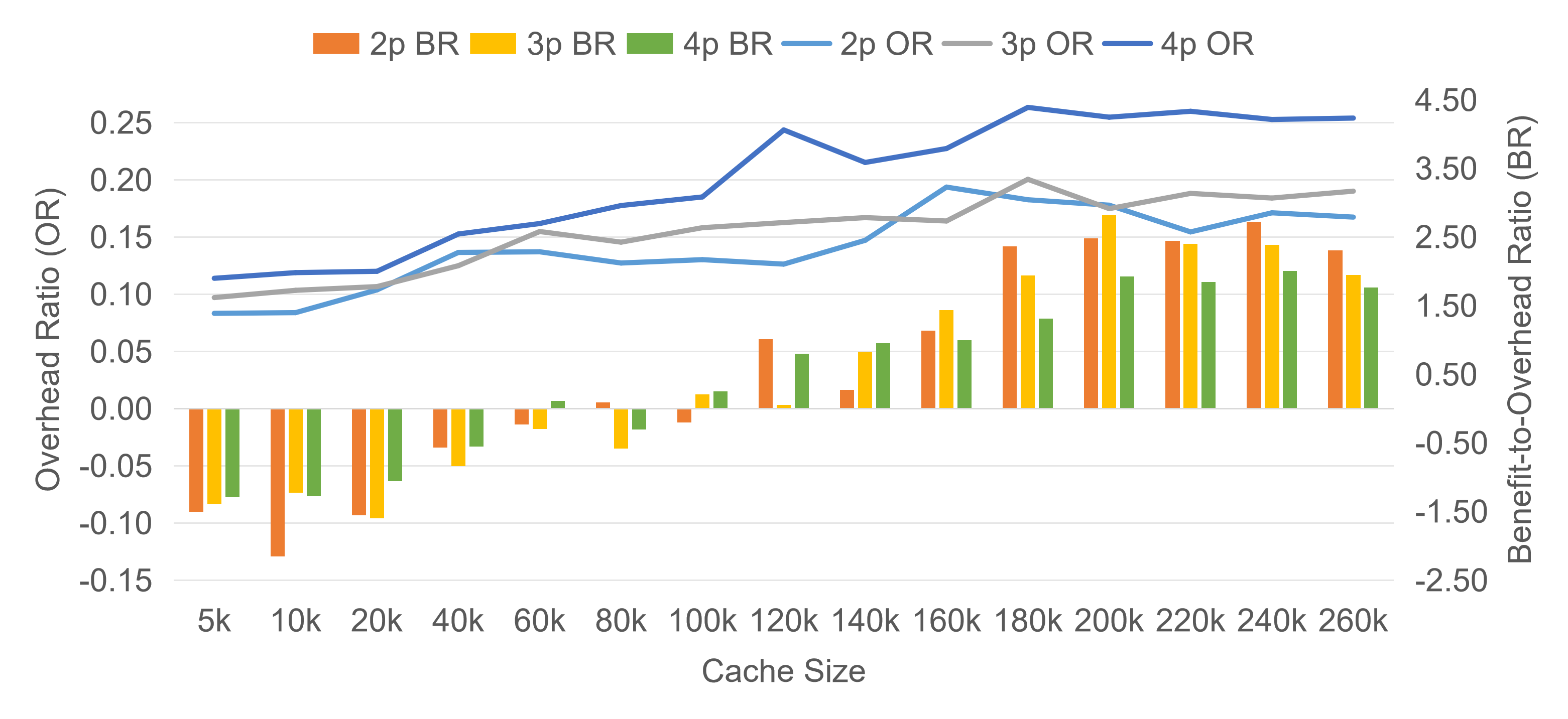}
    \caption{Overhead ratio and benefit-to-overhead ratio of CaPGNN.}
    \label{fig:net_benefit}
\end{figure}

To quantify the net overhead, we report the overhead ratio $r_{overhead} = \frac{T_{check}+T_{pick}}{T_{total}}$ and the benefit-to-overhead ratio $r_{benefit}=\frac{T^{base}_{total} - T^{JACA}_{total}}{T_{check}+T_{pick}}$ under different cache capacities, as shown in Fig.~\ref{fig:net_benefit}. Our measurements show that the bookkeeping overhead remains stable and small across a wide range of capacities, while the benefit of JACA increases with cache capacity as more high-overlap halo vertices are covered and communication is reduced. In practice, we use the adaptive cache-capacity mechanism to avoid extremely small caches. To further improve robustness under low-hit regimes, we discuss two practical optimizations.
First, the system can monitor the moving-average hit ratio; when it stays below a threshold for several epochs, it can reduce cache-update frequency or temporarily bypass the caching path to avoid negative net gain. Second, the constant-factor overhead of metadata access can be reduced via vertex ID reordering and batched cache operations.

\subsection{Partitioning Results of RAPA}
\label{ssec:pr_rapa}
\begin{figure*}[!htbp]
    \vspace{-2em}
    \centering
    \setlength{\abovecaptionskip}{0pt}
    \setlength{\belowcaptionskip}{0pt}
    \begin{subfigure}[b]{0.24\textwidth}
        \setlength{\abovecaptionskip}{0pt}
        \setlength{\belowcaptionskip}{0pt}
        \includegraphics[width=\textwidth]{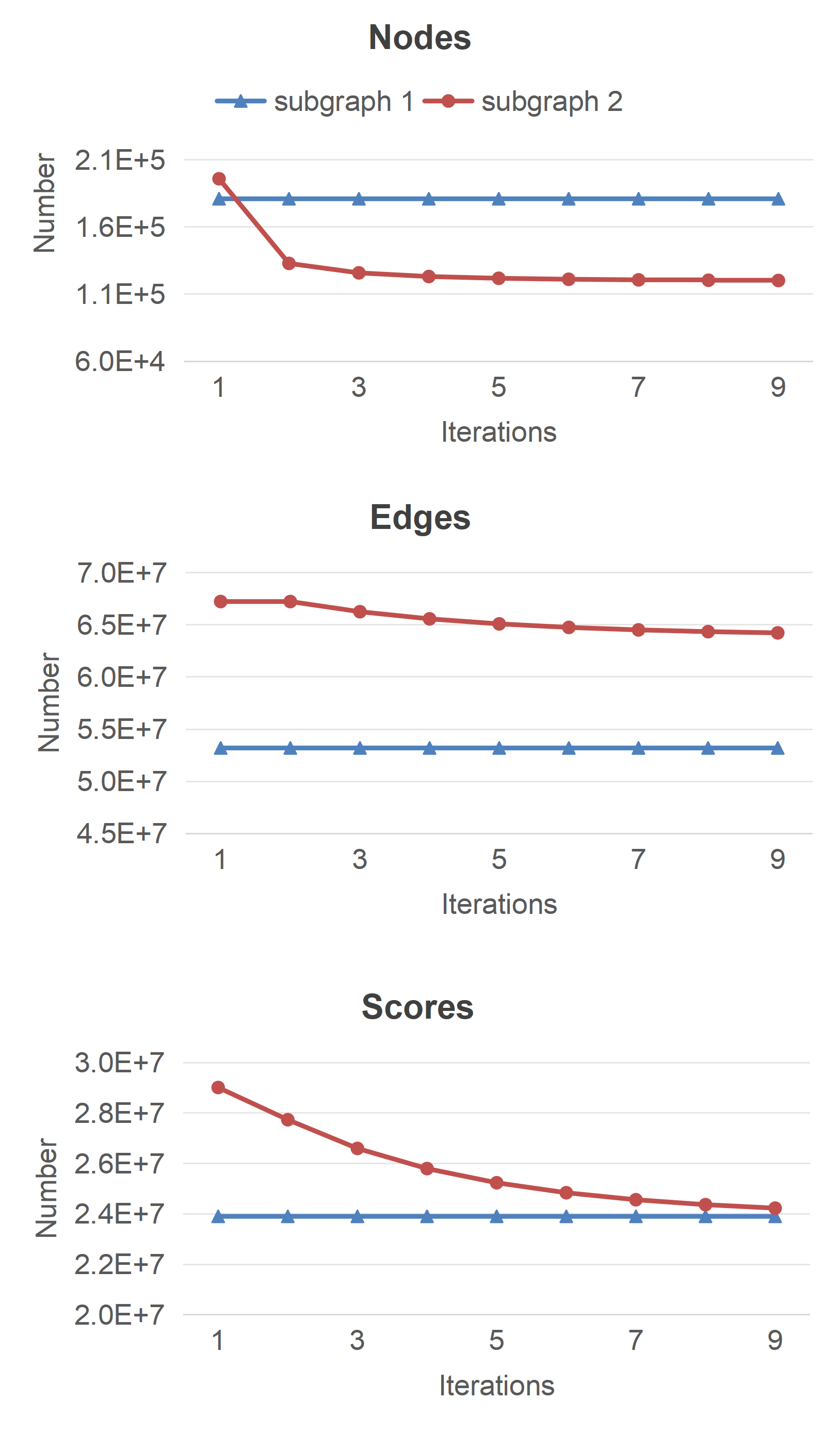}
        \caption{x2.}
    \end{subfigure}
    \hfill
    \begin{subfigure}[b]{0.24\textwidth}
        \setlength{\abovecaptionskip}{0pt}
        \setlength{\belowcaptionskip}{0pt}
        \includegraphics[width=\textwidth]{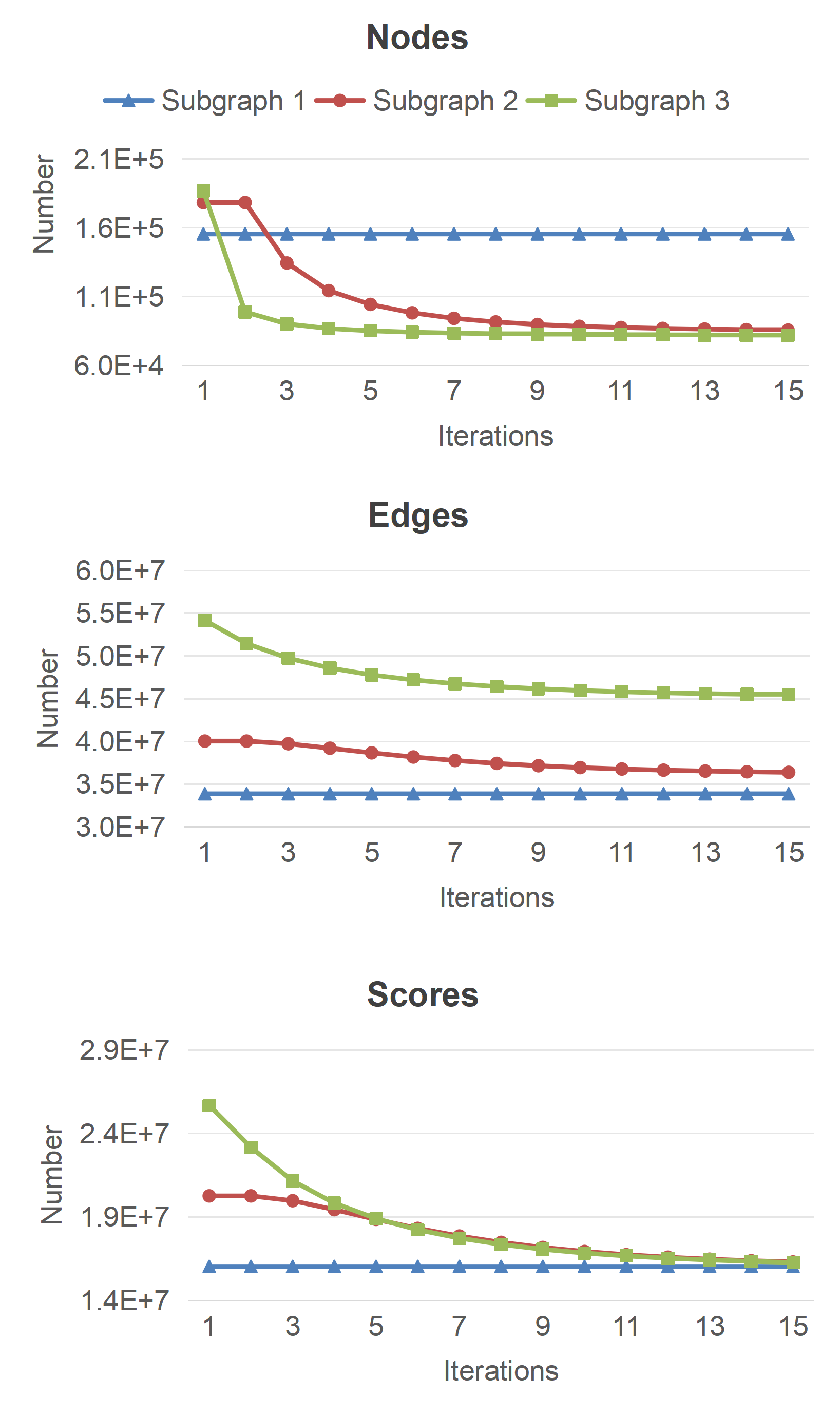}
        \caption{x3.}
    \end{subfigure}
    \hfill
    \begin{subfigure}[b]{0.24\textwidth}
        \setlength{\abovecaptionskip}{0pt}
        \setlength{\belowcaptionskip}{0pt}
        \includegraphics[width=\textwidth]{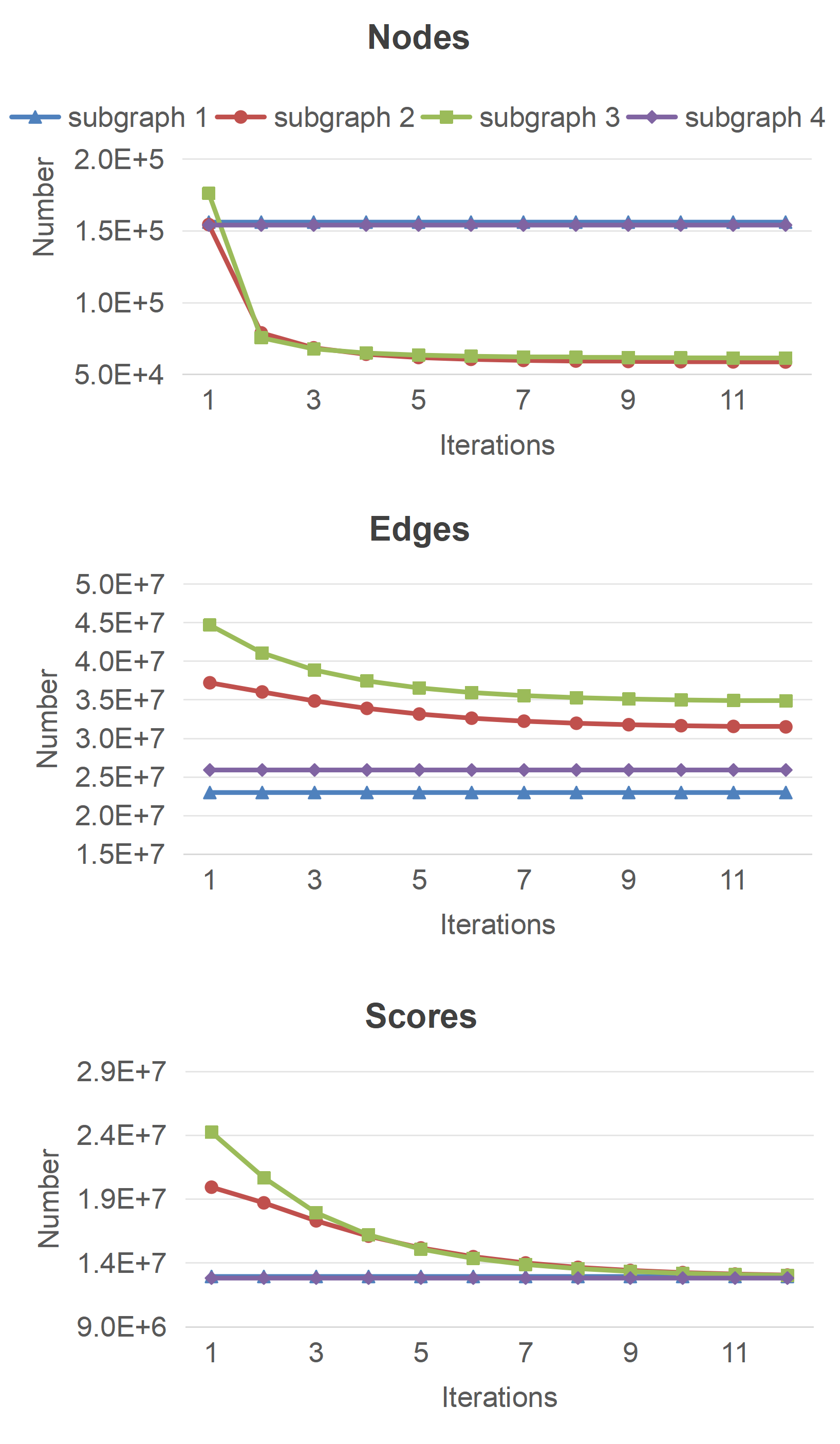}
        \caption{x4.}
    \end{subfigure}
    \hfill
    \begin{subfigure}[b]{0.24\textwidth}
        \setlength{\abovecaptionskip}{0pt}
        \setlength{\belowcaptionskip}{0pt}
        \includegraphics[width=\textwidth]{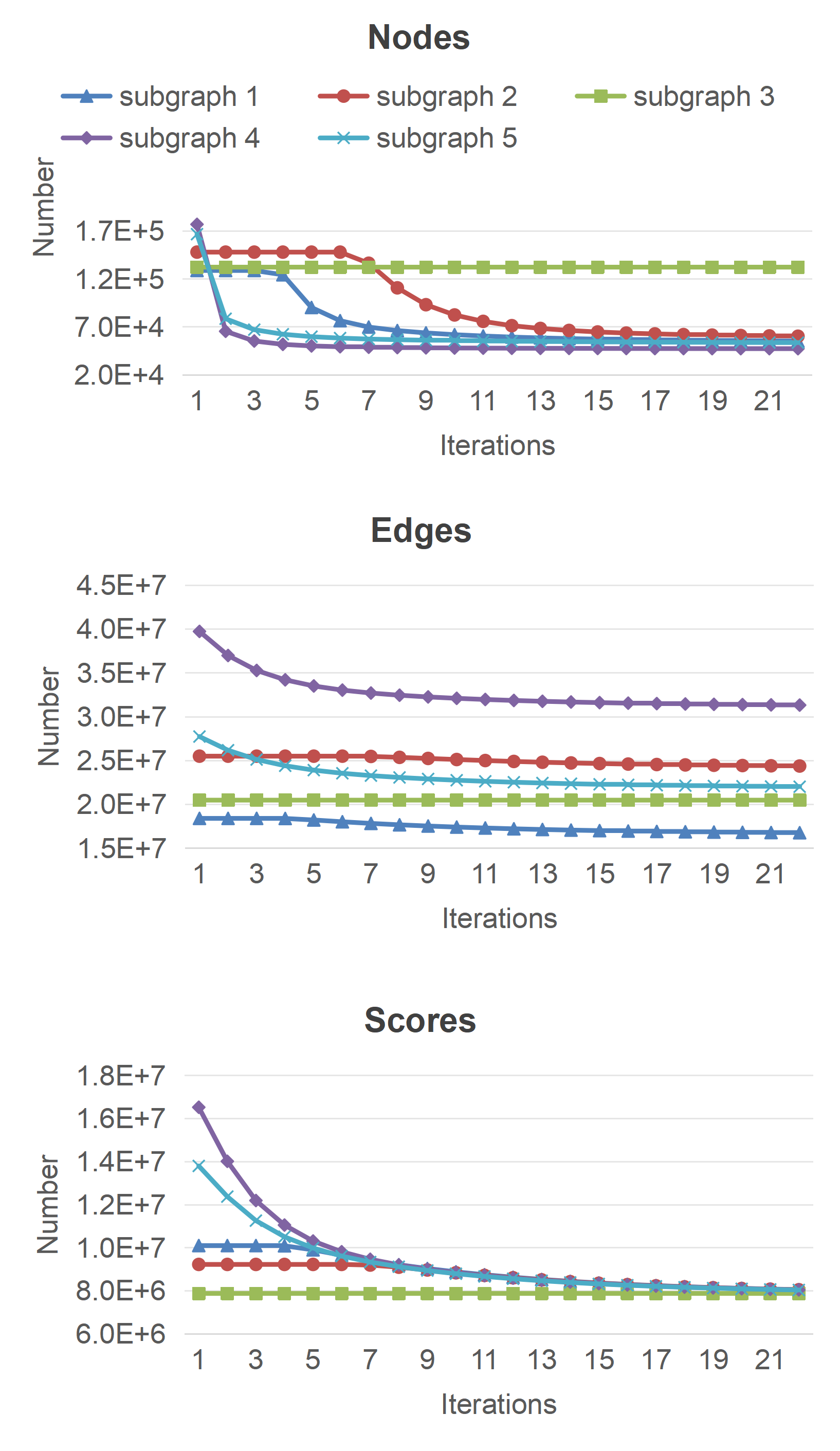}
        \caption{x5.}
    \end{subfigure}
    \caption{Changes in the number of nodes, edges, and scores for each subgraph during RAPA. RAPA progressively balances subgraph cost, with convergence observed across different partition granularities. x2-x5 are defined in Table~\ref{tab:gpu_grouping}.}
    \label{fig:score_rapa} 
    \vspace{-1em}
\end{figure*}

To evaluate the effectiveness of RAPA, we visually examine the evolution of subgraph characteristics across multiple iterations under different partition counts ranging from 2 to 5. As shown in Fig.~\ref{fig:score_rapa}, we track the changes in the number of nodes, the number of edges, and the heuristic scores for each subgraph with the number of iterations. RAPA quickly reduces the cost imbalance across subgraphs, which aligns well with the optimization objective defined in Eq.~\ref{eq:heuristic_partition}. The observed convergence of subgraph scores toward a common range confirms that RAPA effectively balances costs across partitions. Moreover, as the number of partitions increases, the initial imbalance becomes more pronounced, yet RAPA consistently guides subgraphs toward a tightly clustered cost distribution. This demonstrates the strong scalability and stability of RAPA under varying partition granularities. These results empirically validate the ability of RAPA to generate well-balanced partitions.

\begin{figure*}[!htbp]
    \centering
    \setlength{\abovecaptionskip}{0pt}
    \setlength{\belowcaptionskip}{0pt}
    \begin{subfigure}[b]{0.30\textwidth}
        \setlength{\abovecaptionskip}{0pt}
        \setlength{\belowcaptionskip}{0pt}
        \includegraphics[width=\textwidth]{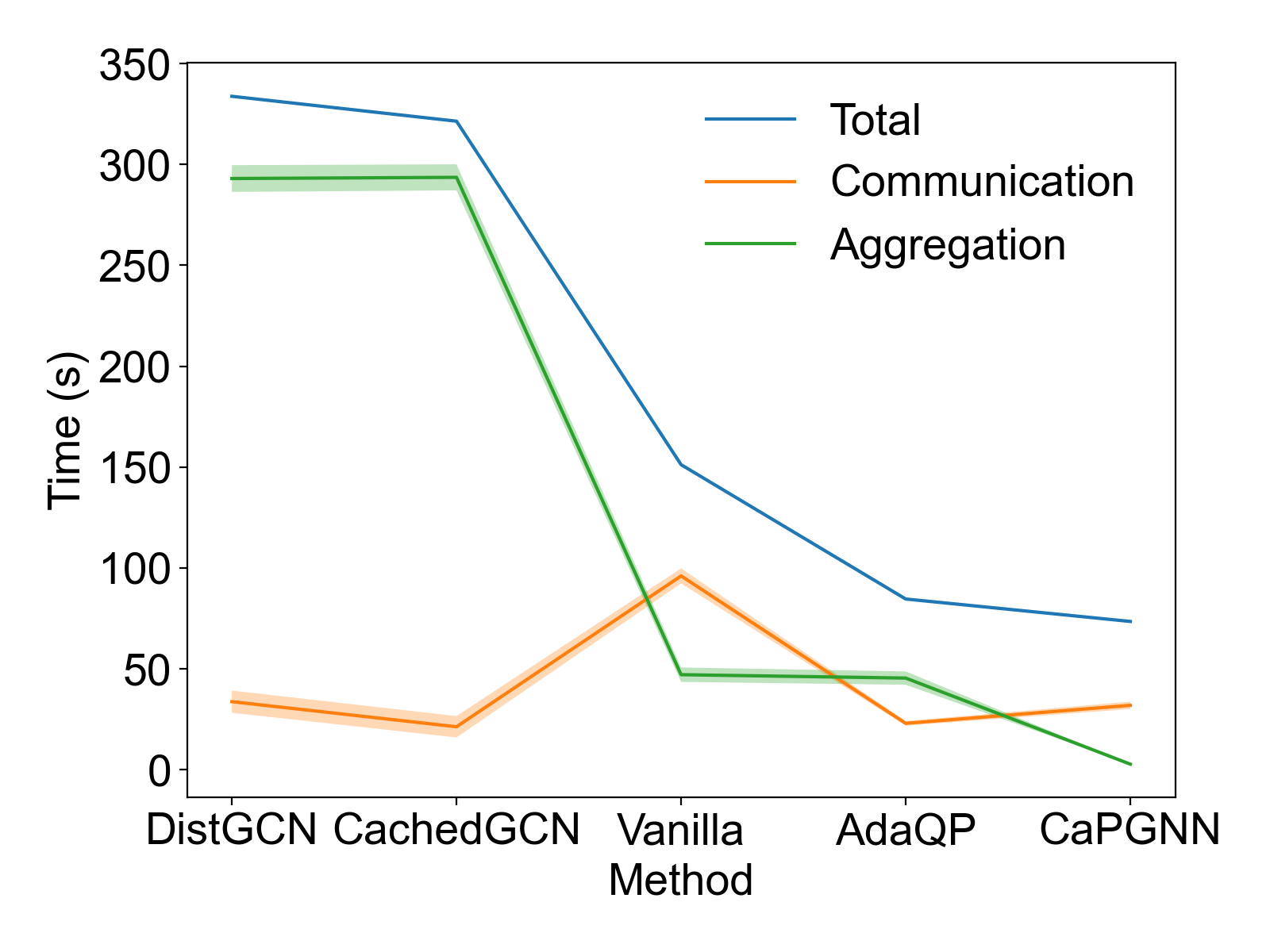}
        \caption{R9 + R9.}
    \end{subfigure}
    \hfill
    \begin{subfigure}[b]{0.30\textwidth}
        \setlength{\abovecaptionskip}{0pt}
        \setlength{\belowcaptionskip}{0pt}
        \includegraphics[width=\textwidth]{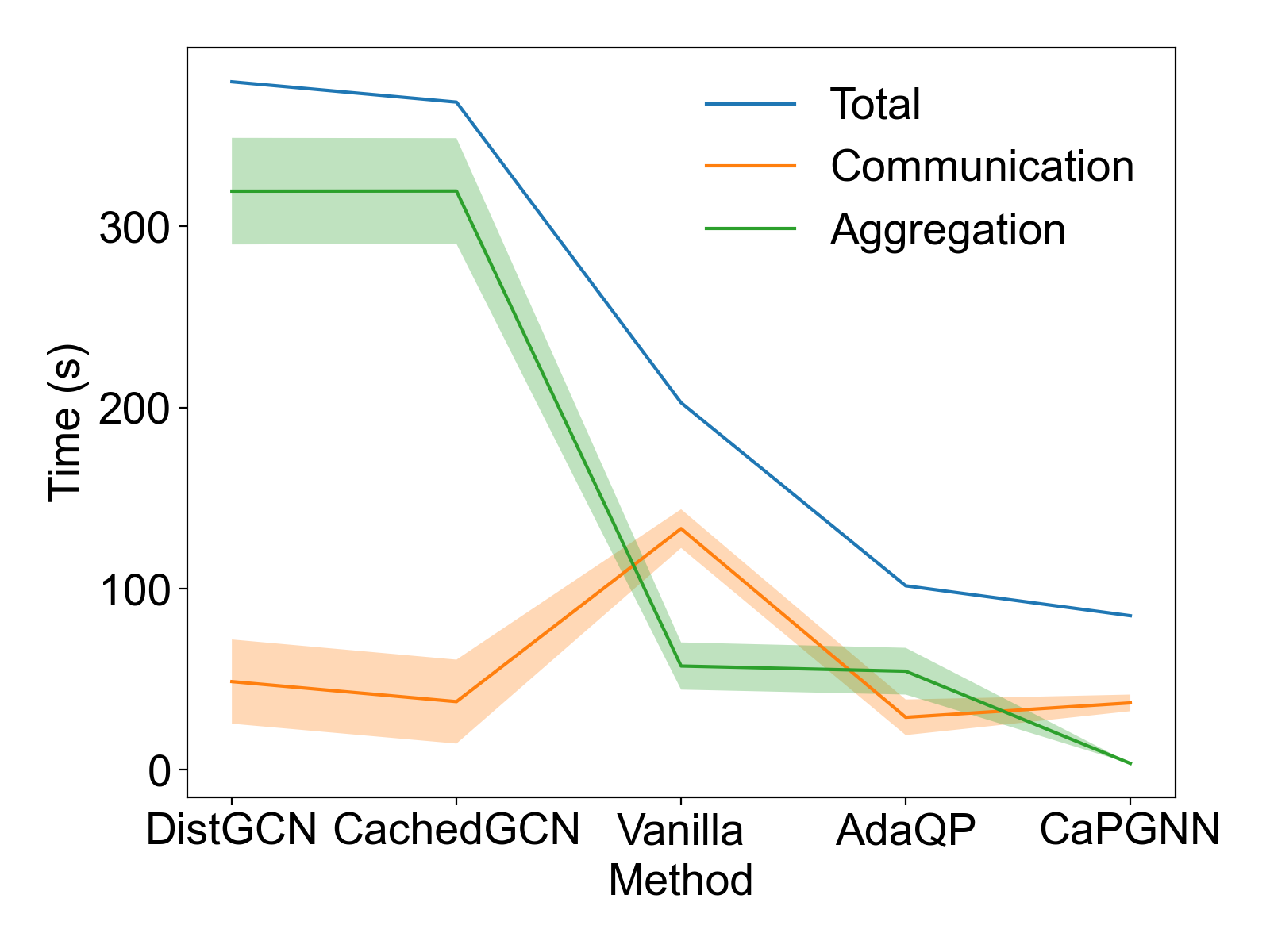}
        \caption{T4 + R9.}
    \end{subfigure}
    \hfill
    \begin{subfigure}[b]{0.30\textwidth}
        \setlength{\abovecaptionskip}{0pt}
        \setlength{\belowcaptionskip}{0pt}
        \includegraphics[width=\textwidth]{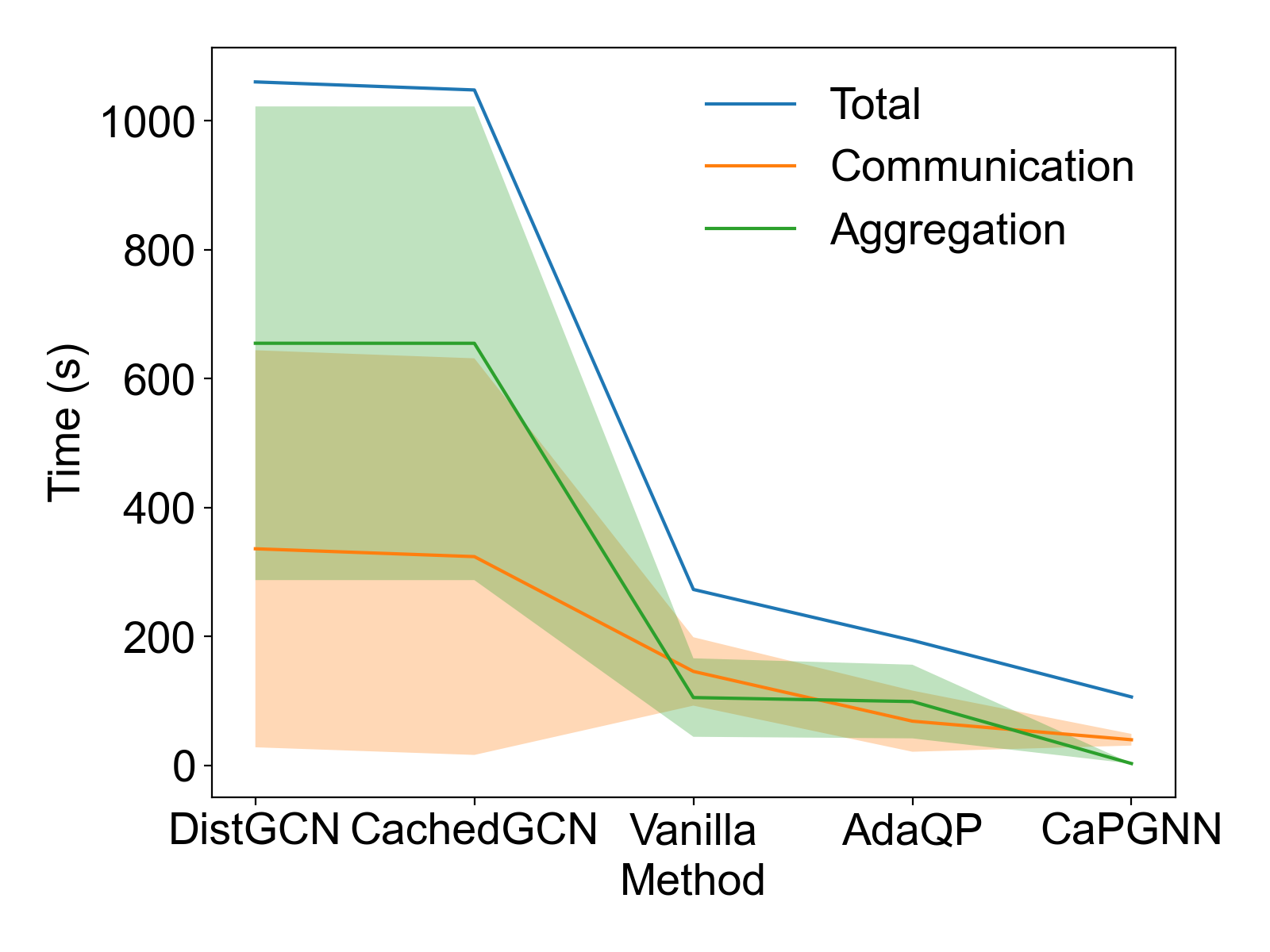}
        \caption{G6 + R9.}
    \end{subfigure}
    \begin{subfigure}[b]{0.30\textwidth}
        \setlength{\abovecaptionskip}{0pt}
        \setlength{\belowcaptionskip}{0pt}
        \includegraphics[width=\textwidth]{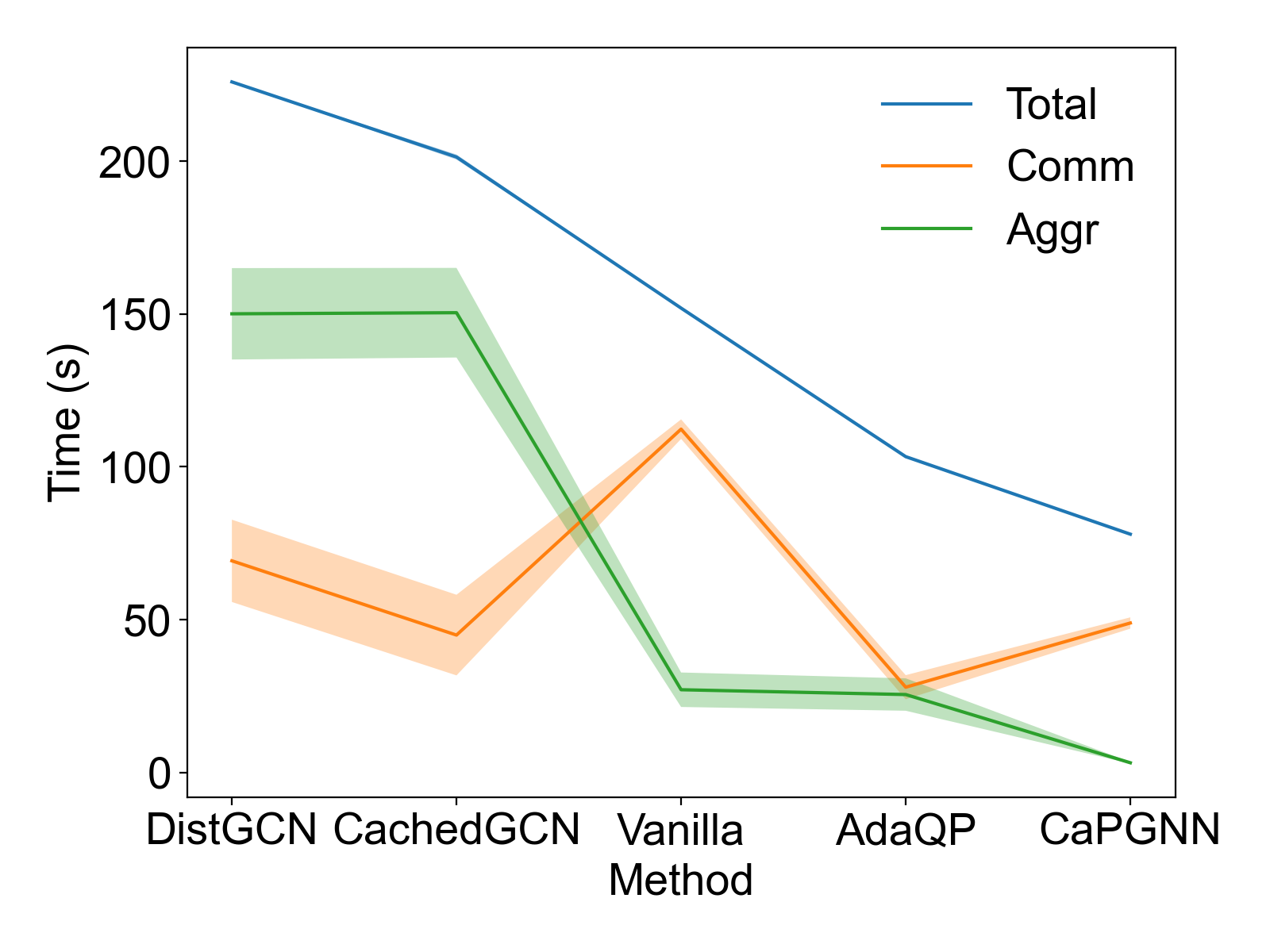}
        \caption{two T4s + two R9s.}
    \end{subfigure}
    \hfill
    \begin{subfigure}[b]{0.30\textwidth}
        \setlength{\abovecaptionskip}{0pt}
        \setlength{\belowcaptionskip}{0pt}
        \includegraphics[width=\textwidth]{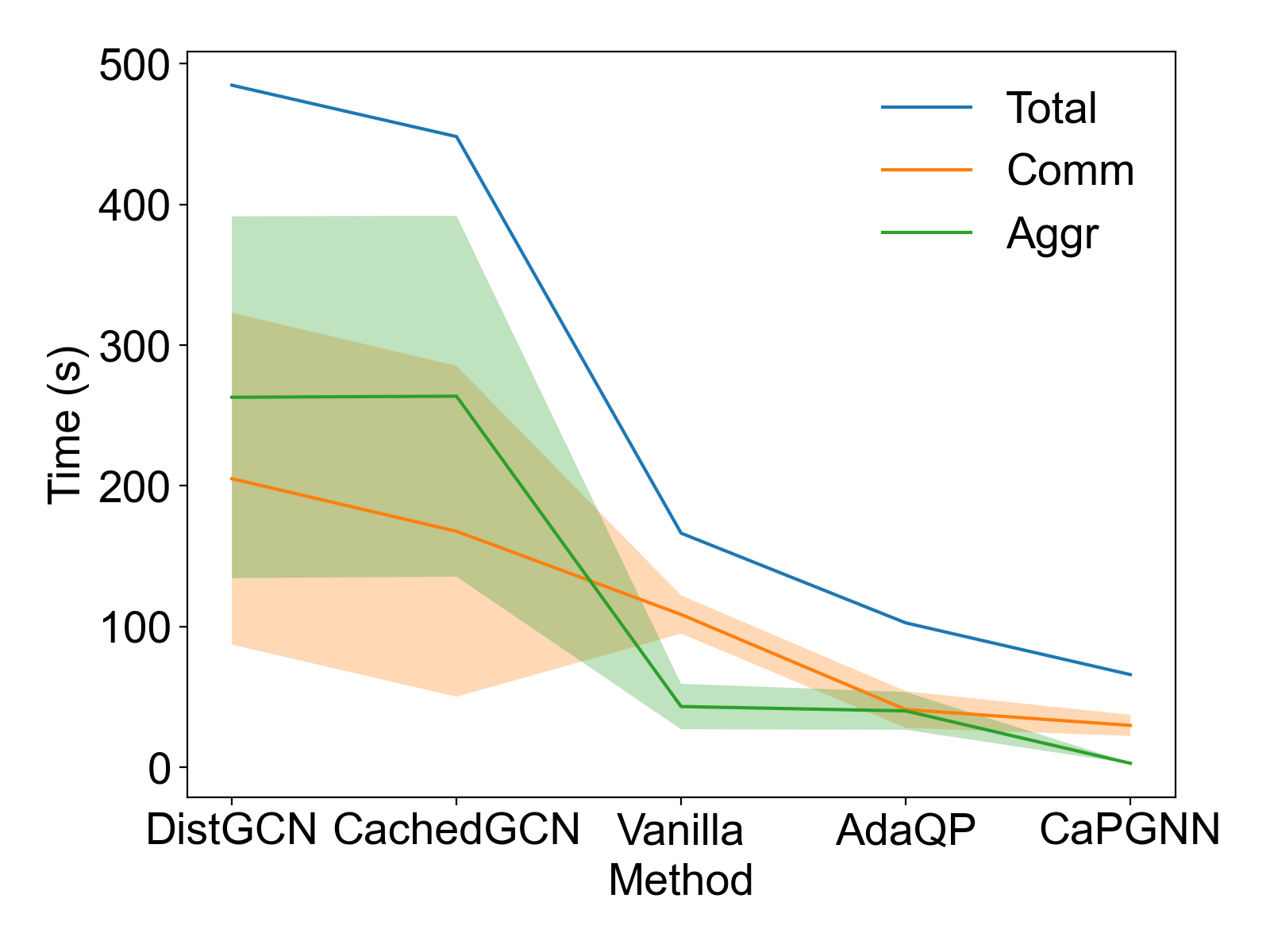}
        \caption{two G6s + two R9s.}
    \end{subfigure}
    \hfill
    \begin{subfigure}[b]{0.32\textwidth}
        \setlength{\abovecaptionskip}{0pt}
        \setlength{\belowcaptionskip}{0pt}
        \includegraphics[width=\textwidth]{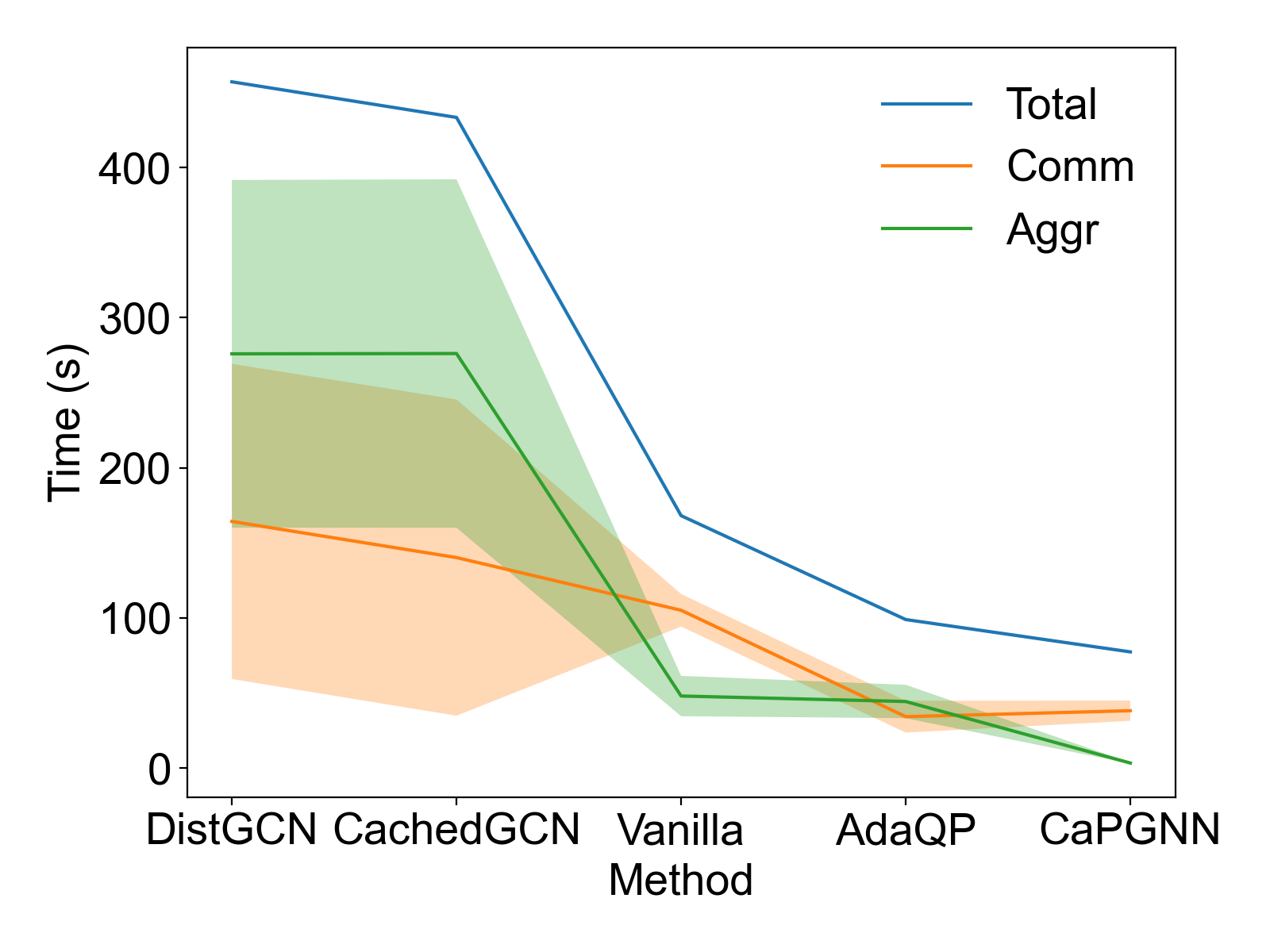}
        \caption{two T4s + two G6s.}
    \end{subfigure}
    \caption{Training performance on Reddit with GCN under heterogeneous GPU settings. Communication indicates inter-GPU data transfer, and aggregation reflects computation during message passing. CaPGNN achieves the lowest total and communication time, while maintaining stable aggregation, demonstrating effective load balancing and reduced cross-device overhead.}
    \label{fig:hetero_variance}
     \vspace{-1em}
\end{figure*}

\begin{figure*}[!htbp]
    \centering
    \setlength{\abovecaptionskip}{0pt}
    \setlength{\belowcaptionskip}{0pt}
    \begin{subfigure}[b]{0.24\textwidth}
        \setlength{\abovecaptionskip}{0pt}
        \setlength{\belowcaptionskip}{0pt}
        \includegraphics[width=\textwidth]{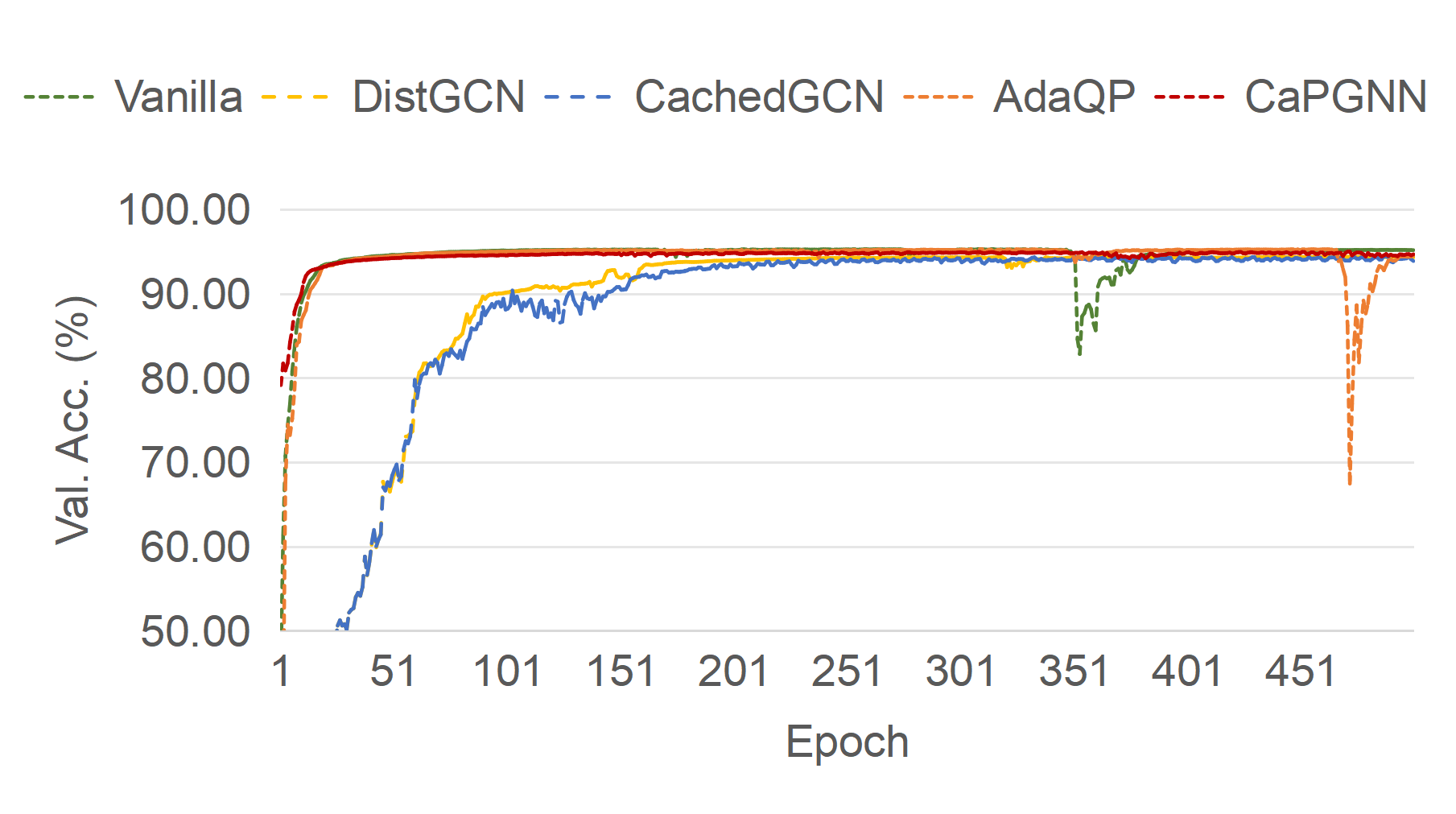}
        \caption{GCN, Rt, 2p.}
    \end{subfigure}
    \hfill
    \begin{subfigure}[b]{0.24\textwidth}
        \setlength{\abovecaptionskip}{0pt}
        \setlength{\belowcaptionskip}{0pt}
        \includegraphics[width=\textwidth]{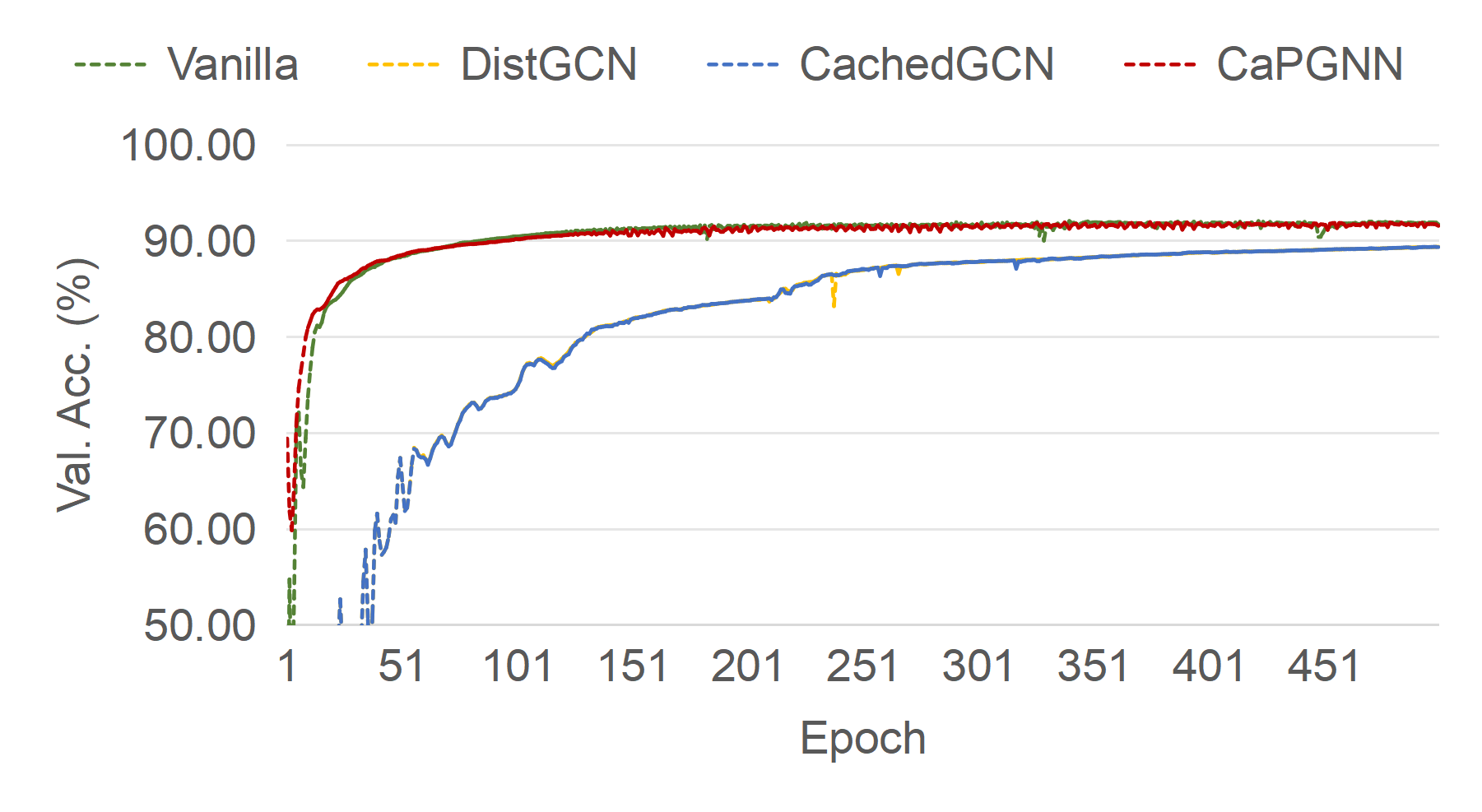}
        \caption{GCN, Os, 2p.}
    \end{subfigure}
    \hfill
    \begin{subfigure}[b]{0.24\textwidth}
        \setlength{\abovecaptionskip}{0pt}
        \setlength{\belowcaptionskip}{0pt}
        \includegraphics[width=\textwidth]{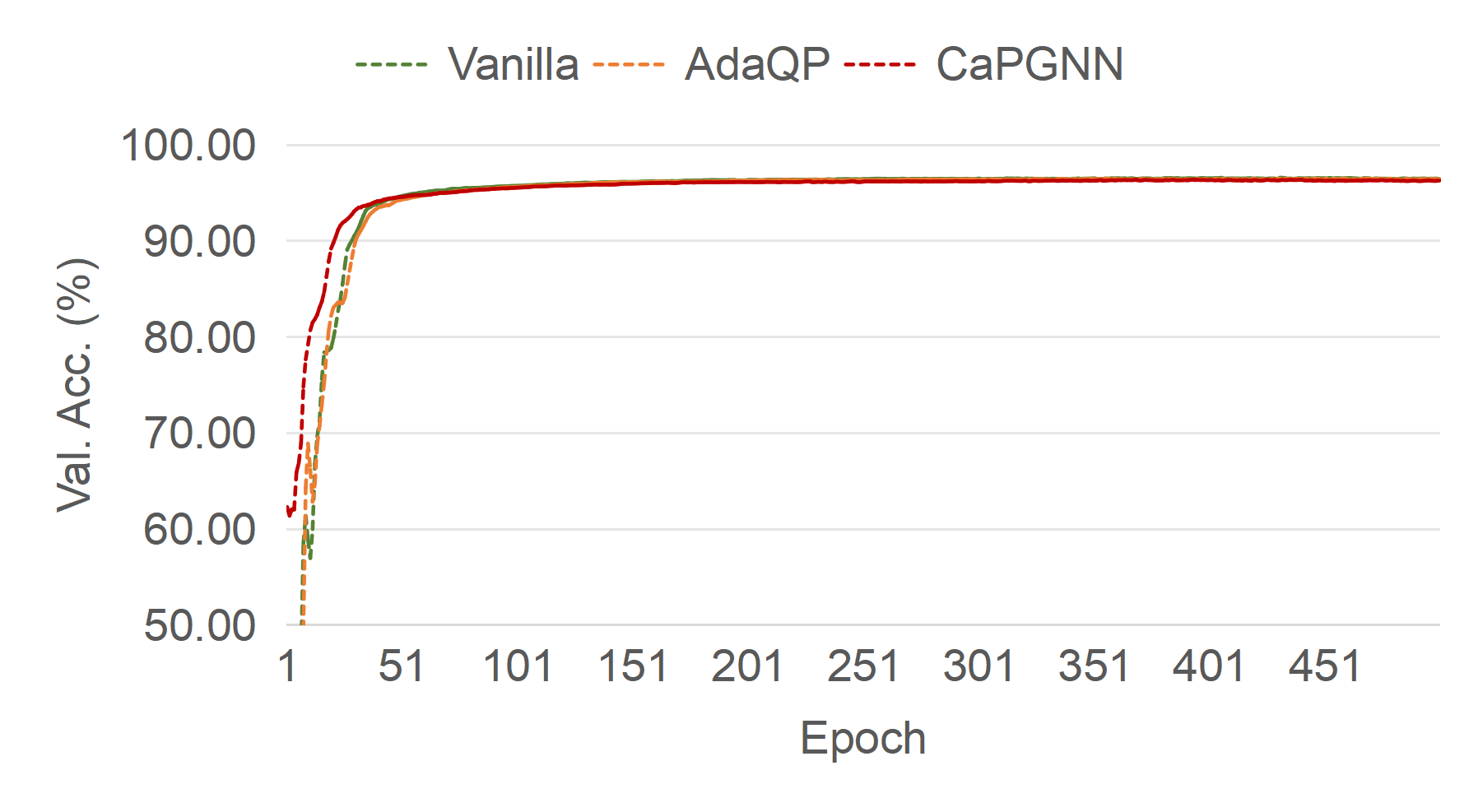}
        \caption{GraphSAGE, Rt, 2p.}
    \end{subfigure}
    \hfill
    \begin{subfigure}[b]{0.24\textwidth}
        \setlength{\abovecaptionskip}{0pt}
        \setlength{\belowcaptionskip}{0pt}
        \includegraphics[width=\textwidth]{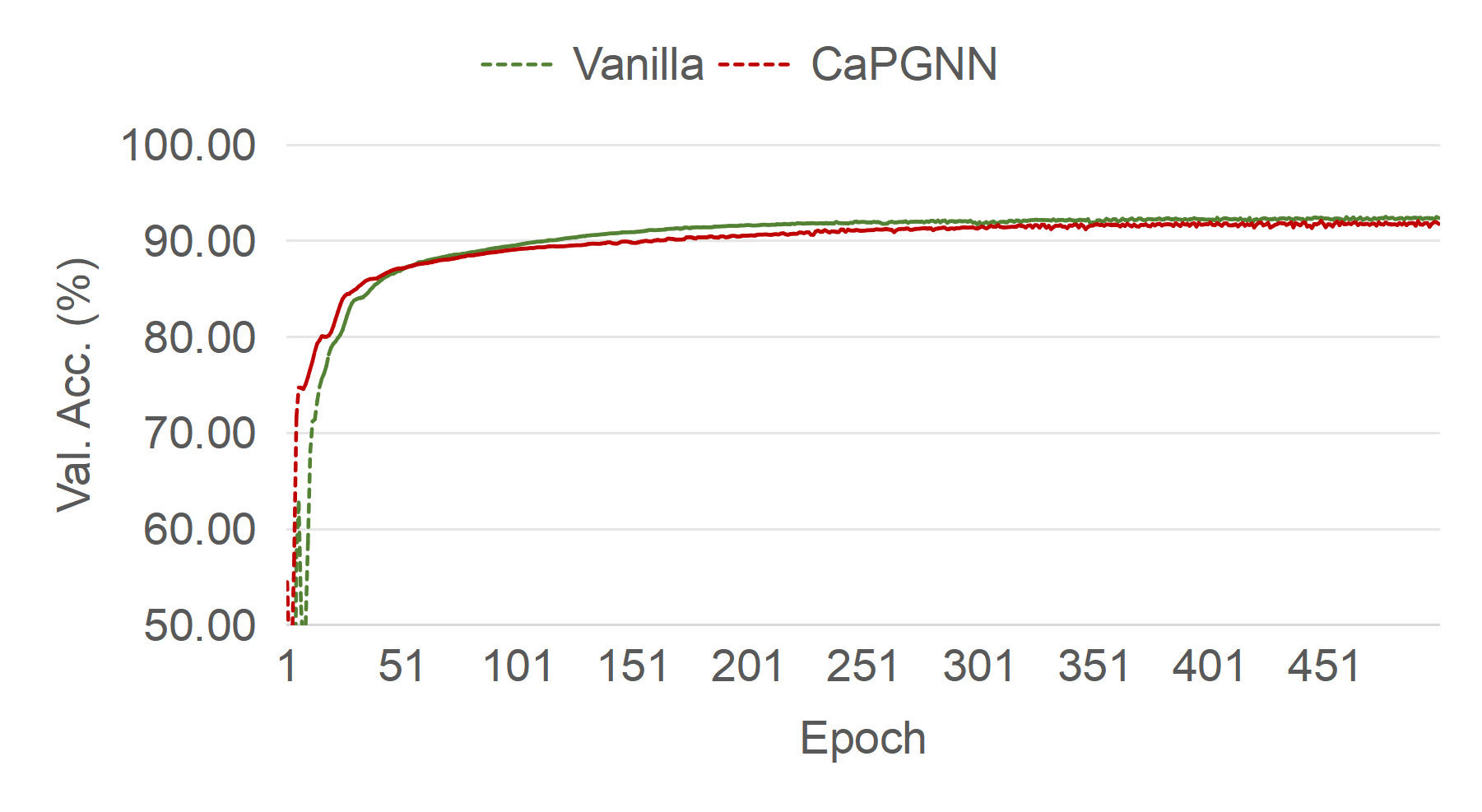}
        \caption{GraphSAGE, Os, 2p.}
    \end{subfigure}
    \hfill
    \begin{subfigure}[b]{0.24\textwidth}
        \setlength{\abovecaptionskip}{0pt}
        \setlength{\belowcaptionskip}{0pt}
        \includegraphics[width=\textwidth]{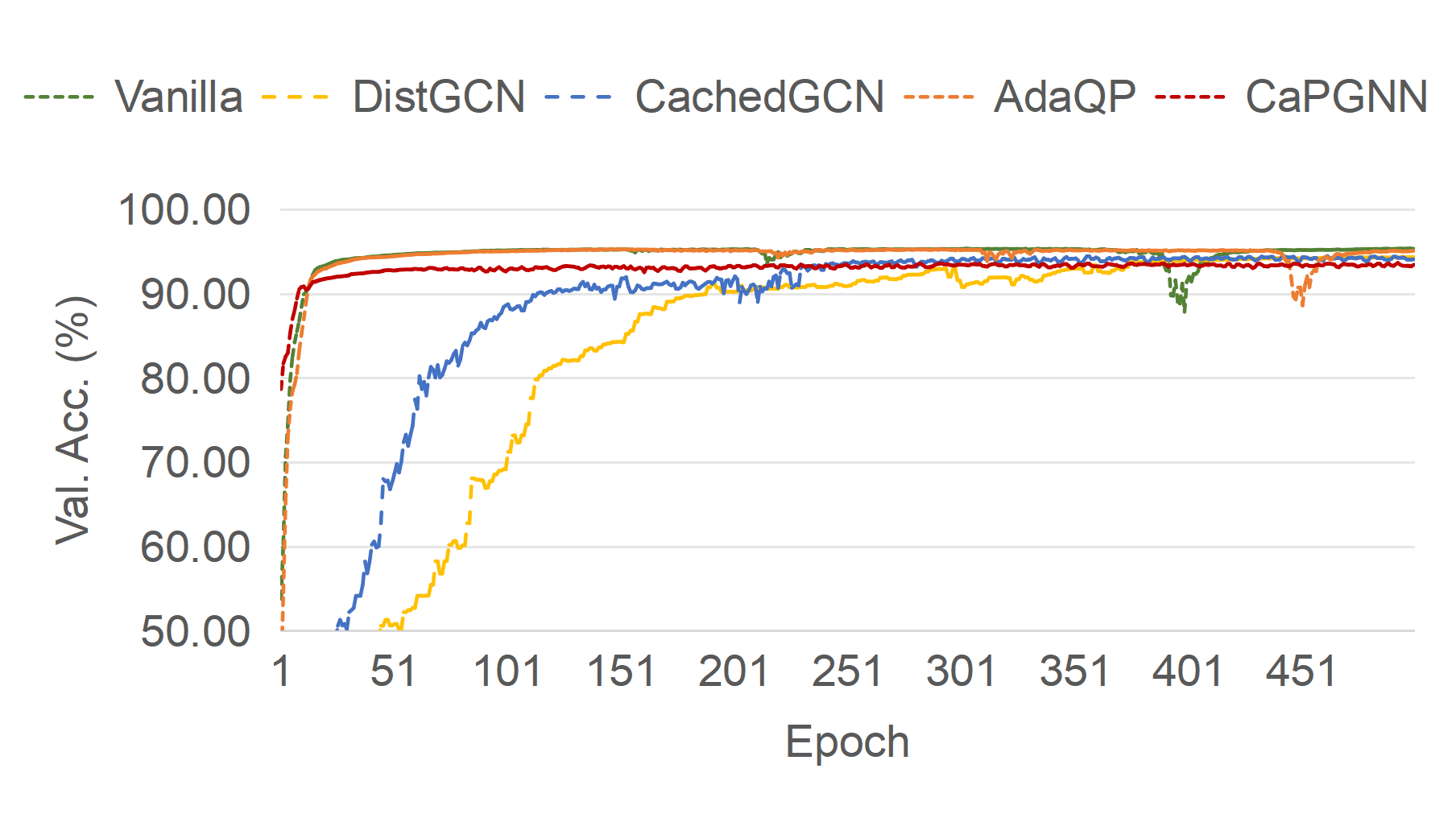}
        \caption{GCN, Rt, 4p.}
    \end{subfigure}
    \hfill
    \begin{subfigure}[b]{0.24\textwidth}
        \setlength{\abovecaptionskip}{0pt}
        \setlength{\belowcaptionskip}{0pt}
        \includegraphics[width=\textwidth]{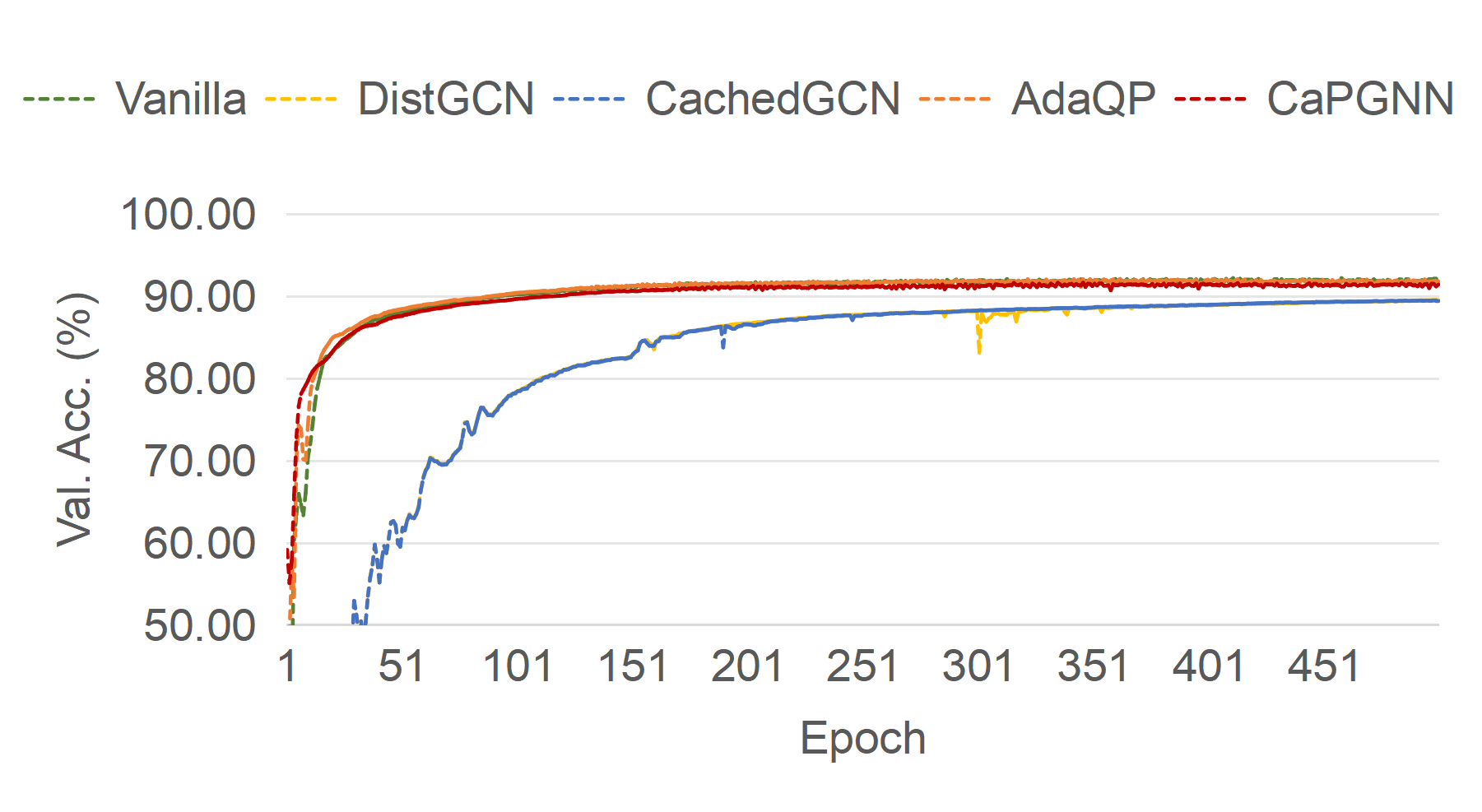}
        \caption{GCN, Os, 4p.}
    \end{subfigure}
    \hfill
    \begin{subfigure}[b]{0.24\textwidth}
        \setlength{\abovecaptionskip}{0pt}
        \setlength{\belowcaptionskip}{0pt}
        \includegraphics[width=\textwidth]{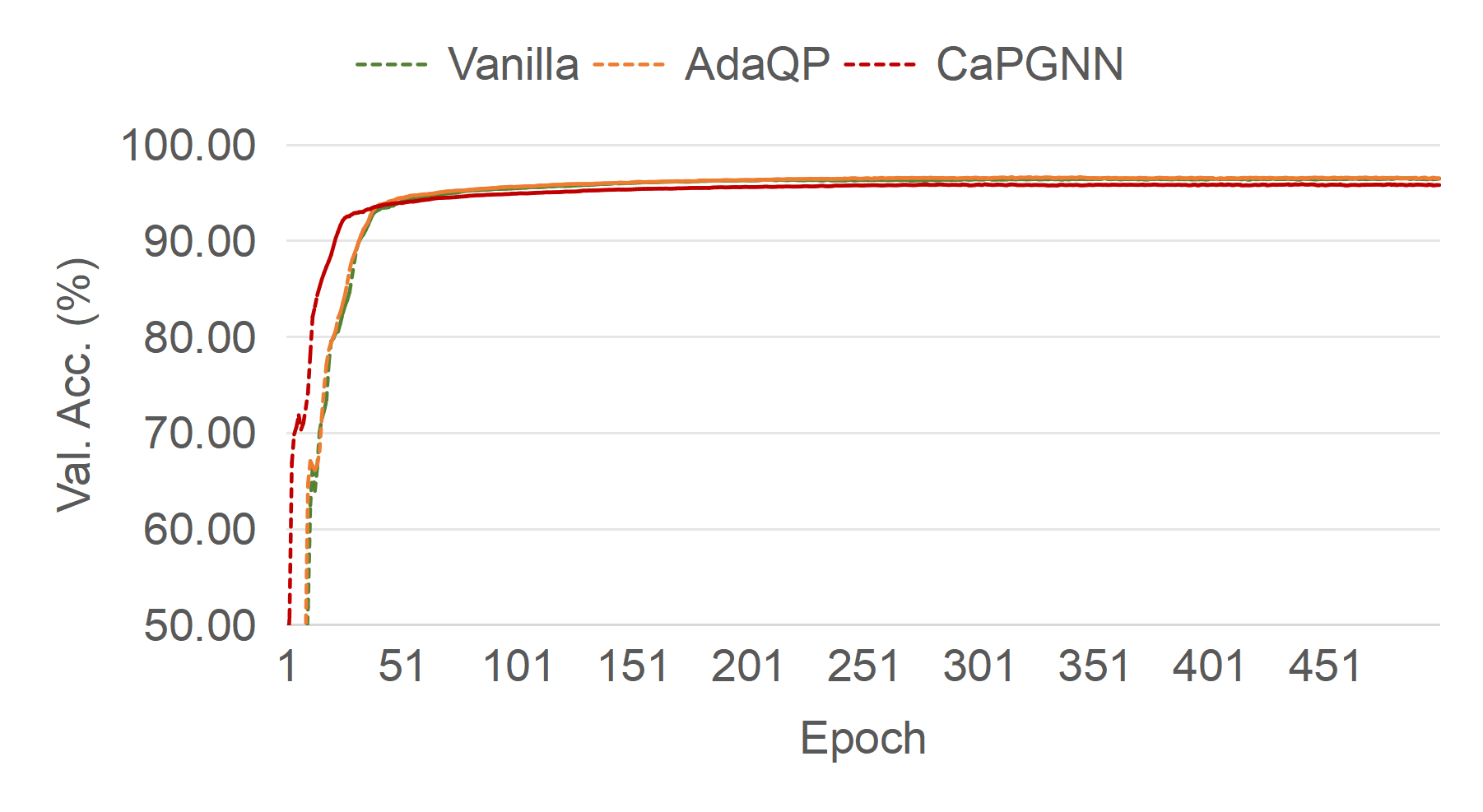}
        \caption{GraphSAGE, Rt, 4p.}
    \end{subfigure}
    \hfill
    \begin{subfigure}[b]{0.24\textwidth}
        \setlength{\abovecaptionskip}{0pt}
        \setlength{\belowcaptionskip}{0pt}
        \includegraphics[width=\textwidth]{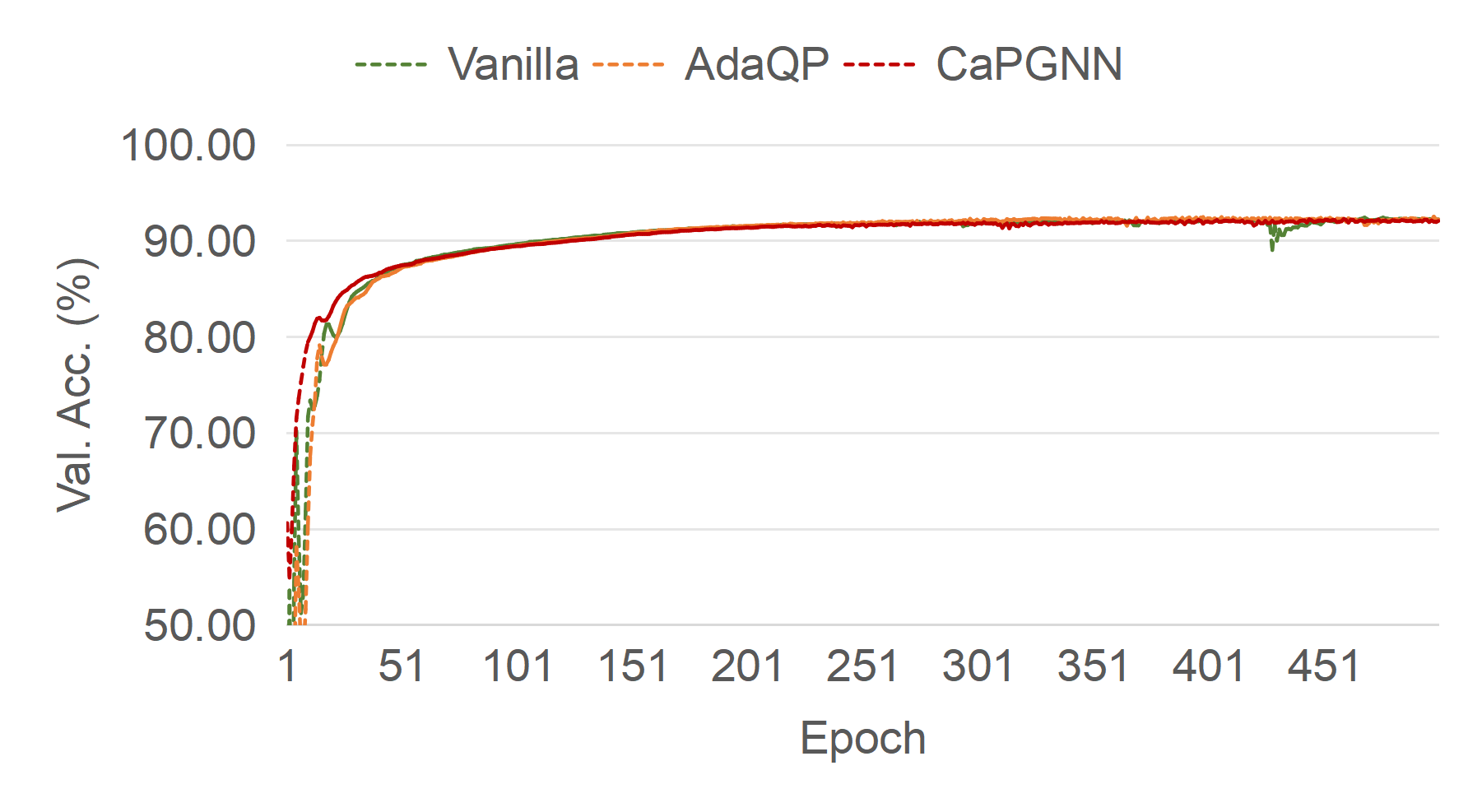}
        \caption{GraphSAGE, Os, 4p.}
    \end{subfigure}
    \caption{Epoch to validation accuracy comparison.}
    \label{fig:convergence}
\end{figure*}

\subsection{Robustness and Effectiveness of CaPGNN}
\label{ssec:eff_rapa}
To evaluate the robustness and efficiency of CaPGNN in heterogeneous GPU environments, we conduct experiments on the Reddit dataset using the GCN model. The training is performed over 200 epochs under various GPU combinations, including R9, T4 and G6, with two-partition and four-partition settings. We report three time components: total training time, communication time reflecting inter-GPU data transfer, and aggregation time reflecting computation during message passing.

As illustrated in Fig.~\ref{fig:hetero_variance}, when GPU devices have similar performance, such as two RTX 3090s or two A40s, all methods exhibit relatively balanced training behavior, reflected by low variance in both communication and aggregation time. This suggests that with uniform hardware, even hardware-agnostic strategies can maintain acceptable load balance. However, as the degree of GPU heterogeneity increases, for example, in the G6 + R9 configuration, the variance in training time becomes significantly larger for DistGCN and CachedGCN. These methods assume equal partitioning and uniform device capability, leading to load imbalance once the computational power between devices diverges. In contrast, CaPGNN remains efficient and stable in all configurations. Its communication time stays consistently low and the aggregation time exhibits minimal variance. This is attributed to RAPA's hardware-aware partitioning, which adjusts subgraphs according to the relative performance of each GPU, effectively balancing both data movement and computational load.

\subsection{Preserving Convergence Rate}
\label{ssec:convergence_rate}
We have theoretically proven the convergence of parallel training in Section~\ref{sec:joint_cache}; here, we provide empirical validation. Fig.~\ref{fig:convergence} illustrates the epoch-to-validation accuracy curves for various methods on the Reddit and ogbn-products datasets, using both GCN and GraphSAGE models under 2- and 4-partition settings. Note that AdaQP encountered an out-of-memory (OOM) issue on ogbn-products with two partitions in our environment and is therefore omitted from the figure. As shown, CaPGNN matches Vanilla in convergence speed, with nearly identical training curves. In contrast, DistGCN and CachedGCN exhibit noticeably slower or more unstable convergence, particularly in the GCN setting with two partitions, as shown in Figs.~\ref{fig:convergence}(a) and (e). These empirical findings align well with our theoretical convergence analysis and demonstrate the reliability of our method in parallel GNN training.

\subsection{Comparison of Overall Performance}
\label{ssec:overall_performance}
To evaluate overall performance, we evaluate the training time of the three methods on different datasets, with varying partition numbers and GPU combinations. Note that we only report the average training time, excluding the computation time for the validation set, as it does not affect the training results. We use GCN and GraphSAGE, setting the dimensions of the hidden layer to 256. The global and local cache capacities are identical, calculated using the proposed adaptive method described in Section~\ref{sec:joint_cache}. Evaluation metrics include the total training time for 200 epochs and the accuracy of the test set. Due to space limitations, we present only the results using METIS as the pre-partitioning method and omit detailed results for all GPU combinations. However, it is important to emphasize that the proposed CaPGNN remains effective under these conditions.

As shown in Table~\ref{tab:overall_performance}, CaPGNN achieves the shortest total training time in most cases and significantly reduces communication overhead compared to other methods. It also maintains comparable or even higher accuracy than the baseline in the vast majority of cases, with only a few instances of accuracy degradation of no more than 5\%. On large-scale datasets such as Reddit, Yelp, ogbn-products, and AmazonProducts, CaPGNN reduces communication time by 18\% to 99\%, and achieves speedups ranging from 0.17x to 18.98x relative to the baselines. Specifically, both DistGCN and CachedGCN rely on equally-sized partitions and whole-subgraph feature caching. Their performance degrades significantly when applied to larger graphs or heterogeneous GPU environments. Although their caching strategies reduce some computation, they introduce additional overhead from full feature replication and memory copy operations, which become substantial under memory pressure. These issues are particularly severe when the number of partitions increases (e.g., 5 or more), leading to exacerbated communication bottlenecks. Furthermore, AdaQP employs Gurobi to solve the bit-width optimization problem. In configurations involving a large number of partitions or high-dimensional features, the solver often encounters timeout issues. In contrast, CaPGNN is effective in reducing training time and communication cost and is more stable and scalable across various hardware settings and large graph workloads. Even in environments with heterogeneous GPU performance, CaPGNN maintains low overhead and avoids timeouts or OOM errors observed in competing methods, and achieves competitive accuracy.

\begin{table*}[htb]
    \centering
    \belowrulesep=0pt
    \aboverulesep=0pt
    \setlength{\abovecaptionskip}{0pt}
    \setlength{\belowcaptionskip}{0pt}
    \caption{Overall performance. The results report the total time in seconds for 200 epochs. \textbf{Epoch} is the epoch time; \textbf{Comm} is the communication time; \textbf{Acc} is the validation accuracy. x2-x8 are defined in Table~\ref{tab:gpu_grouping}, datasets are defined in Table~\ref{tab:dataset_info}.}
    \label{tab:overall_performance}
    \resizebox{\textwidth}{!}{%
    \begin{tabular}{cl|ccccccccccccccccccccc}
        \Xhline{1.5pt}
        \multirow{3}{*}{\textbf{Dataset}} & \multirow{3}{*}{\textbf{Alg}} & \multicolumn{21}{c}{\textbf{GCN}} \\
        \cmidrule(lr){3-23}
         &  & \multicolumn{3}{c}{\textbf{x2}} & \multicolumn{3}{c}{\textbf{x3}} & \multicolumn{3}{c}{\textbf{x4}} & \multicolumn{3}{c}{\textbf{x5}} & \multicolumn{3}{c}{\textbf{x6}} & \multicolumn{3}{c}{\textbf{x7}} & \multicolumn{3}{c}{\textbf{x8}} \\
         \cmidrule(lr){3-5} \cmidrule(lr){6-8} \cmidrule(lr){9-11} \cmidrule(lr){12-14} \cmidrule(lr){15-17} \cmidrule(lr){18-20} \cmidrule(lr){21-23}
         &  & \textbf{Epoch} & \textbf{Comm} & \textbf{Acc} & \textbf{Epoch} & \textbf{Comm} & \textbf{Acc} & \textbf{Epoch} & \textbf{Comm} & \textbf{Acc} & \textbf{Epoch} & \textbf{Comm} & \textbf{Acc} & \textbf{Epoch} & \textbf{Comm} & \textbf{Acc} & \textbf{Epoch} & \textbf{Comm} & \textbf{Acc} & \textbf{Epoch} & \textbf{Comm} & \textbf{Acc} \\
         \Xhline{1pt}
        \multirow{5}{*}{\textbf{Cl}} & \textbf{DistGCN} & 23.87 & 15.15 & 4.57 & 31.46 & 23.43 & 5.20 & 39.37 & 31.36 & 5.81 & 50.53 & 41.15 & 5.41 & 64.11 & 54.65 & 5.61 & 70.07 & 58.64 & 4.24 & 79.39 & 67.57 & 7.10 \\
         & \textbf{CachedGCN} & 10.37 & 2.08 & 4.95 & \textbf{11.54} & 3.66 & 5.18 & \textbf{13.16} & 5.20 & 5.51 & \textbf{17.08} & 7.74 & 5.56 & \textbf{18.66} & 9.48 & 5.56 & \textbf{22.14} & 11.19 & 4.22 & 26.07 & 14.51 & 7.60 \\
         & \textbf{Vanilla} & 15.86 & 7.21 & \textbf{67.28} & 22.29 & 12.97 & \textbf{66.75} & 22.30 & 11.99 & \textbf{67.46} & 21.52 & 10.97 & \textbf{67.48} & 22.74 & 10.73 & \textbf{66.95} & 25.19 & 11.05 & \textbf{66.68} & \textbf{25.47} & \textbf{11.24} & \textbf{66.85} \\
         & \textbf{AdaQP} & \multicolumn{3}{c}{\cellcolor[HTML]{EFEFEF}Timeout} & \multicolumn{3}{c}{\cellcolor[HTML]{EFEFEF}Timeout} & \multicolumn{3}{c}{\cellcolor[HTML]{EFEFEF}Timeout} & \multicolumn{3}{c}{\cellcolor[HTML]{EFEFEF}Timeout} & \multicolumn{3}{c}{\cellcolor[HTML]{EFEFEF}Timeout} & \multicolumn{3}{c}{\cellcolor[HTML]{EFEFEF}Timeout} & \multicolumn{3}{c}{\cellcolor[HTML]{EFEFEF}Timeout} \\
         & \textbf{CaPGNN} & \textbf{8.85} & \textbf{0.27} & 67.00 & 12.25 & \textbf{1.58} & 65.68 & 15.72 & \textbf{3.11} & 65.66 & 19.85 & \textbf{5.96} & 65.31 & 22.22 & \textbf{7.88} & 65.34 & 27.44 & \textbf{10.41} & 64.98 & 29.94 & 12.39 & 64.83 \\
         \Xhline{1pt}
        \multirow{5}{*}{\textbf{Fr}} & \textbf{DistGCN} & 19.23 & 10.15 & 46.77 & 24.95 & 16.85 & 46.82 & 30.36 & 22.66 & 46.98 & 39.36 & 29.89 & 47.23 & 47.50 & 38.26 & 47.64 & 54.41 & 42.76 & \textbf{47.67} & 62.00 & 49.53 & \textbf{47.67} \\
         & \textbf{CachedGCN} & \textbf{14.81} & 6.70 & 42.40 & \textbf{18.02} & 10.45 & 25.48 & 21.99 & 14.23 & 46.41 & 28.95 & 19.57 & 25.48 & 33.91 & 24.48 & 42.65 & 41.12 & 29.13 & 42.39 & 45.76 & 32.92 & 46.43 \\
         & \textbf{Vanilla} & 48.65 & 39.69 & 47.45 & 54.49 & 44.64 & 42.34 & 47.40 & 36.91 & 46.88 & 47.52 & 36.68 & 43.92 & 42.89 & 31.59 & 47.74 & 42.79 & 30.01 & 47.39 & 41.40 & 28.41 & 43.47 \\
         & \textbf{AdaQP} & 22.53 & 8.78 & \textbf{48.74} & 25.97 & 10.81 & \textbf{48.27} & 26.05 & 8.62 & \textbf{47.56} & 26.10 & 8.97 & \textbf{48.69} & 30.84 & 9.33 & \textbf{48.31} & 29.93 & 8.90 & 42.39 & \textbf{35.88} & \textbf{9.34} & 46.74 \\
         & \textbf{CaPGNN} & 19.13 & \textbf{1.42} & 47.34 & 22.58 & \textbf{4.62} & 42.86 & \textbf{22.29} & \textbf{7.10} & 42.61 & \textbf{26.09} & \textbf{8.85} & 42.42 & \textbf{21.00} & \textbf{8.54} & 42.43 & \textbf{28.08} & 12.92 & 42.44 & 36.81 & 18.52 & 42.40 \\
         \Xhline{1pt}
        \multirow{5}{*}{\textbf{Cs}} & \textbf{DistGCN} & 40.96 & 26.19 & 96.12 & 52.52 & 39.13 & 96.07 & 63.81 & 51.23 & 96.01 & 82.22 & 67.53 & 95.98 & 105.04 & 89.90 & 96.01 & 111.67 & 94.40 & 96.03 & 126.66 & 108.93 & 96.03 \\
         & \textbf{CachedGCN} & 19.05 & 4.32 & 96.12 & 18.19 & 4.95 & 96.07 & 19.87 & 7.32 & 96.01 & 27.03 & 11.57 & 95.98 & 28.87 & 14.31 & 96.00 & 33.83 & 16.90 & 96.03 & 37.68 & 20.12 & 96.01 \\
         & \textbf{Vanilla} & 41.91 & 30.80 & \textbf{96.48} & 63.07 & 51.35 & \textbf{96.54} & 58.00 & 46.20 & \textbf{96.54} & 62.33 & 49.08 & \textbf{96.58} & 67.18 & 52.59 & \textbf{96.62} & 63.50 & 47.72 & \textbf{96.52} & 62.89 & 46.42 & \textbf{96.52} \\
         & \textbf{AdaQP} & \multicolumn{3}{c}{\cellcolor[HTML]{EFEFEF}Timeout} & \multicolumn{3}{c}{\cellcolor[HTML]{EFEFEF}Timeout} & \multicolumn{3}{c}{\cellcolor[HTML]{EFEFEF}Timeout} & \multicolumn{3}{c}{\cellcolor[HTML]{EFEFEF}Timeout} & \multicolumn{3}{c}{\cellcolor[HTML]{EFEFEF}Timeout} & \multicolumn{3}{c}{\cellcolor[HTML]{EFEFEF}Timeout} & \multicolumn{3}{c}{\cellcolor[HTML]{EFEFEF}Timeout} \\
         & \textbf{CaPGNN} & \textbf{10.91} & \textbf{0.20} & 96.00 & \textbf{14.58} & \textbf{2.14} & 95.97 & \textbf{17.58} & \textbf{3.89} & 95.93 & \textbf{19.51} & \textbf{5.53} & 95.98 & \textbf{24.21} & \textbf{7.64} & 95.88 & \textbf{33.03} & \textbf{11.78} & 95.68 & \textbf{37.42} & \textbf{15.43} & 95.68 \\
         \Xhline{1pt}
        \multirow{5}{*}{\textbf{Rt}} & \textbf{DistGCN} & 333.73 & 33.76 & 94.02 & 270.05 & 64.45 & 93.93 & 225.90 & 69.23 & 93.93 & 380.56 & 207.20 & 94.04 & 350.98 & 191.17 & 94.06 & 532.15 & 339.41 & 94.01 & 482.47 & 292.28 & 93.99 \\
         & \textbf{CachedGCN} & 321.40 & 21.33 & 92.30 & 251.73 & 45.96 & 90.29 & 201.33 & 44.94 & 91.88 & 349.36 & 175.91 & 93.01 & 299.91 & 143.87 & 92.10 & 482.68 & 292.49 & 92.62 & 438.05 & 248.49 & 90.81 \\
         & \textbf{Vanilla} & 151.19 & 96.07 & \textbf{95.33} & 150.89 & 107.00 & \textbf{95.28} & 151.90 & 112.31 & 95.30 & 153.80 & 113.34 & 95.26 & 175.85 & 131.17 & \textbf{95.28} & 162.23 & 115.17 & \textbf{95.27} & 154.28 & 111.37 & \textbf{95.33} \\
         & \textbf{AdaQP} & 84.63 & 23.02 & 95.21 & 83.45 & 29.93 & 95.22 & 103.31 & 27.89 & \textbf{95.35} & 104.87 & 34.79 & \textbf{95.28} & \multicolumn{3}{c}{\cellcolor[HTML]{EFEFEF}Timeout} & \multicolumn{3}{c}{\cellcolor[HTML]{EFEFEF}Timeout} & \multicolumn{3}{c}{\cellcolor[HTML]{EFEFEF}Timeout} \\
         & \textbf{CaPGNN} & \textbf{44.83} & \textbf{2.36} & 95.09 & \textbf{44.01} & \textbf{7.88} & 94.13 & \textbf{53.98} & \textbf{12.15} & 93.68 & \textbf{45.61} & \textbf{12.35} & 93.45 & \textbf{66.16} & \textbf{24.92} & 93.05 & \textbf{62.28} & \textbf{25.44} & 93.66 & \textbf{62.06} & \textbf{24.85} & 93.66 \\
         \Xhline{1pt}
        \multirow{5}{*}{\textbf{Yp}} & \textbf{DistGCN} & 202.17 & 76.82 & 20.75 & 196.57 & 116.23 & 20.75 & 227.22 & 155.55 & 20.75 & 405.91 & 310.25 & 20.75 & 375.77 & 297.55 & 20.75 & 426.99 & 328.07 & 20.75 & 481.74 & 374.90 & 20.75 \\
         & \textbf{CachedGCN} & 152.33 & 54.41 & 20.75 & 162.54 & 82.21 & 20.75 & 182.17 & 110.55 & 20.75 & 241.40 & 162.61 & 20.75 & 298.84 & 220.65 & 20.75 & 347.45 & 249.61 & 20.75 & 394.97 & 288.13 & 20.75 \\
         & \textbf{Vanilla} & 137.36 & 107.84 & 41.92 & 161.16 & 133.32 & 41.71 & 140.54 & 114.53 & 41.55 & 158.62 & 128.20 & \textbf{42.53} & 158.22 & 128.19 & \textbf{43.21} & 151.01 & 120.01 & 41.79 & 144.36 & 113.27 & 41.84 \\
         & \textbf{AdaQP} & 60.97 & 24.74 & \textbf{43.66} & 70.31 & 32.23 & \textbf{42.95} & 96.27 & 28.99 & 41.50 & 79.39 & 31.37 & 42.46 & 83.37 & 33.97 & 42.32 & \multicolumn{3}{c}{\cellcolor[HTML]{EFEFEF}Timeout} & \multicolumn{3}{c}{\cellcolor[HTML]{EFEFEF}Timeout} \\
         & \textbf{CaPGNN} & \textbf{46.02} & \textbf{6.46} & 43.28 & \textbf{37.06} & \textbf{6.08} & 38.41 & \textbf{49.74} & \textbf{12.51} & \textbf{43.53} & \textbf{39.76} & \textbf{11.63} & 41.70 & \textbf{45.73} & \textbf{18.91} & 41.26 & \textbf{81.26} & \textbf{38.68} & \textbf{42.45} & \textbf{84.91} & \textbf{40.89} & \textbf{43.14} \\
         \Xhline{1pt}
        \multirow{5}{*}{\textbf{As}} & \textbf{DistGCN} & 943.64 & 193.89 & 11.02 & 801.19 & 282.41 & 11.02 & 799.19 & 377.00 & 11.02 & 1117.69 & 671.00 & 11.02 & 1148.38 & 731.20 & 11.02 & 1570.68 & 1070.75 & 11.02 & 1564.86 & 1053.38 & 11.02 \\
         & \textbf{CachedGCN} & 821.68 & 116.81 & 11.01 & 744.16 & 224.84 & 11.01 & 671.10 & 261.45 & 11.01 & 1018.29 & 572.27 & 11.01 & 1070.85 & 620.14 & 11.01 & \multicolumn{3}{c}{\cellcolor[HTML]{EFEFEF}OOM} & \multicolumn{3}{c}{\cellcolor[HTML]{EFEFEF}OOM} \\
         & \textbf{Vanilla} & 363.51 & 223.27 & 37.98 & 417.01 & 299.81 & 38.85 & 374.93 & 263.00 & 37.65 & 361.98 & 272.63 & 37.51 & 464.24 & 326.92 & 38.05 & 428.67 & 311.09 & 37.96 & \multicolumn{3}{c}{\cellcolor[HTML]{EFEFEF}OOM} \\
         & \textbf{AdaQP} & \multicolumn{3}{c}{\cellcolor[HTML]{EFEFEF}OOM} & 226.42 & 109.35 & 38.56 & 225.94 & 111.96 & 39.09 & 185.03 & 91.75 & 39.76 & 294.18 & 158.07 & 23.47 & 259.10 & 143.20 & 38.82 & \multicolumn{3}{c}{\cellcolor[HTML]{EFEFEF}Timeout} \\
         & \textbf{CaPGNN} & \textbf{124.99} & \textbf{22.52} & \textbf{49.50} & \textbf{79.10} & \textbf{13.47} & \textbf{50.00} & \textbf{56.37} & \textbf{8.35} & \textbf{50.55} & \textbf{116.31} & \textbf{40.55} & \textbf{50.20} & \textbf{99.23} & \textbf{35.66} & \textbf{50.97} & \textbf{96.38} & \textbf{30.24} & \textbf{52.06} & \textbf{183.83} & \textbf{46.14} & \textbf{50.86} \\
         \Xhline{1pt}
        \multirow{5}{*}{\textbf{Os}} & \textbf{DistGCN} & 931.56 & 294.40 & 83.83 & 1022.72 & 514.16 & 85.75 & 1028.06 & 607.62 & 86.72 & 1611.85 & 1114.16 & 86.97 & 1847.03 & 1341.05 & 86.88 & 2638.53 & 2005.49 & 86.77 & 2647.99 & 2009.64 & 86.49 \\
         & \textbf{CachedGCN} & 889.51 & 253.78 & 83.80 & 962.86 & 452.15 & 86.67 & 947.17 & 526.09 & 86.51 & 1504.53 & 1006.94 & 86.95 & 1705.32 & 1199.54 & 86.84 & \multicolumn{3}{c}{\cellcolor[HTML]{EFEFEF}OOM} & \multicolumn{3}{c}{\cellcolor[HTML]{EFEFEF}OOM} \\
         & \textbf{Vanilla} & 328.92 & 209.76 & 91.79 & 331.79 & 233.68 & 91.85 & 282.71 & 199.66 & 91.77 & 272.99 & 191.31 & 91.70 & 280.27 & 192.78 & \textbf{91.62} & \multicolumn{3}{c}{\cellcolor[HTML]{EFEFEF}OOM} & \multicolumn{3}{c}{\cellcolor[HTML]{EFEFEF}OOM} \\
         & \textbf{AdaQP} & \multicolumn{3}{c}{\cellcolor[HTML]{EFEFEF}OOM} & 141.60 & 45.36 & \textbf{91.86} & 155.59 & 42.15 & \textbf{91.85} & 189.71 & 49.92 & \textbf{91.71} & \multicolumn{3}{c}{\cellcolor[HTML]{EFEFEF}OOM} & \multicolumn{3}{c}{\cellcolor[HTML]{EFEFEF}OOM} & \multicolumn{3}{c}{\cellcolor[HTML]{EFEFEF}OOM} \\
         & \textbf{CaPGNN} & \textbf{92.83} & \textbf{5.22} & \textbf{92.09} & \textbf{83.16} & \textbf{14.58} & 91.74 & \textbf{80.12} & \textbf{17.42} & 91.75 & \textbf{94.26} & \textbf{28.50} & 91.67 & \textbf{150.50} & \textbf{47.84} & 91.18 & \textbf{146.38} & \textbf{56.89} & \textbf{91.72} & \textbf{139.48} & \textbf{47.55} & \textbf{91.61} \\
         \Xhline{2 pt}
     &  & \multicolumn{21}{c}{\textbf{GraphSAGE}} \\
    \cmidrule(lr){3-23}
     &  & \multicolumn{3}{c}{\textbf{x2}} & \multicolumn{3}{c}{\textbf{x3}} & \multicolumn{3}{c}{\textbf{x4}} & \multicolumn{3}{c}{\textbf{x5}} & \multicolumn{3}{c}{\textbf{x6}} & \multicolumn{3}{c}{\textbf{x7}} & \multicolumn{3}{c}{\textbf{x8}} \\
     \cmidrule(lr){3-5} \cmidrule(lr){6-8} \cmidrule(lr){9-11} \cmidrule(lr){12-14} \cmidrule(lr){15-17} \cmidrule(lr){18-20} \cmidrule(lr){21-23}
    \multirow{-3}{*}{\textbf{Dataset}} & \multirow{-3}{*}{\textbf{Alg}} & \textbf{Epoch} & \textbf{Comm} & \textbf{Acc} & \textbf{Epoch} & \textbf{Comm} & \textbf{Acc} & \textbf{Epoch} & \textbf{Comm} & \textbf{Acc} & \textbf{Epoch} & \textbf{Comm} & \textbf{Acc} & \textbf{Epoch} & \textbf{Comm} & \textbf{Acc} & \textbf{Epoch} & \textbf{Comm} & \textbf{Acc} & \textbf{Epoch} & \textbf{Comm} & \textbf{Acc} \\
    \Xhline{1pt}
     & \textbf{Vanilla} & 19.00 & 7.21 & 67.84 & 25.09 & 12.74 & \textbf{67.33} & 23.97 & 10.45 & \textbf{68.01} & 26.77 & 10.92 & \textbf{68.01} & 27.39 & 10.40 & \textbf{68.37} & \textbf{31.51} & \textbf{11.72} & \textbf{68.34} & \textbf{32.59} & \textbf{11.44} & \textbf{68.37} \\
     & \textbf{AdaQP} & \multicolumn{3}{c}{\cellcolor[HTML]{EFEFEF}Timeout} & \multicolumn{3}{c}{\cellcolor[HTML]{EFEFEF}Timeout} & \multicolumn{3}{c}{\cellcolor[HTML]{EFEFEF}Timeout} & \multicolumn{3}{c}{\cellcolor[HTML]{EFEFEF}Timeout} & \multicolumn{3}{c}{\cellcolor[HTML]{EFEFEF}Timeout} & \multicolumn{3}{c}{\cellcolor[HTML]{EFEFEF}Timeout} & \multicolumn{3}{c}{\cellcolor[HTML]{EFEFEF}Timeout} \\
    \multirow{-3}{*}{\textbf{Cl}} & \textbf{CaPGNN} & \textbf{10.77} & \textbf{0.49} & \textbf{67.94} & \textbf{16.13} & \textbf{3.23} & 66.90 & \textbf{17.60} & \textbf{3.36} & 67.74 & \textbf{22.83} & \textbf{6.05} & 66.50 & \textbf{26.21} & \textbf{8.18} & 67.21 & 42.15 & 15.89 & 66.55 & 41.63 & 14.89 & 66.93 \\
    \Xhline{1pt}
     & \textbf{Vanilla} & 48.29 & 38.96 & 49.18 & 53.30 & 43.00 & 50.95 & 48.61 & 37.71 & 42.37 & 50.04 & 38.37 & \textbf{50.95} & 45.51 & 33.04 & \textbf{50.56} & 43.81 & 29.71 & \textbf{50.09} & 42.50 & 28.08 & 51.44 \\
     & \textbf{AdaQP} & 20.59 & 7.40 & 47.92 & 24.76 & 8.52 & \textbf{49.90} & 26.69 & 8.55 & \textbf{51.51} & 32.84 & \textbf{9.26} & 50.68 & 34.27 & \textbf{9.61} & 48.03 & 40.82 & \textbf{9.11} & 49.74 & 44.26 & \textbf{9.67} & \textbf{51.49} \\
    \multirow{-3}{*}{\textbf{Fr}} & \textbf{CaPGNN} & \textbf{20.51} & \textbf{2.34} & \textbf{49.83} & \textbf{24.41} & \textbf{5.82} & 49.16 & \textbf{22.45} & \textbf{7.45} & 49.11 & \textbf{29.06} & 11.22 & 48.08 & \textbf{25.37} & 12.56 & 49.23 & \textbf{32.97} & 16.15 & 48.58 & \textbf{42.62} & 22.97 & 48.62 \\
    \Xhline{1pt}
     & \textbf{Vanilla} & 45.91 & 31.27 & \textbf{97.01} & 60.05 & 44.75 & 97.01 & 65.93 & 49.64 & 97.03 & 67.40 & 48.99 & 97.09 & 69.75 & 49.67 & 97.09 & 70.78 & 49.05 & 97.06 & 70.72 & 47.89 & 97.20 \\
     & \textbf{AdaQP} & \multicolumn{3}{c}{\cellcolor[HTML]{EFEFEF}Timeout} & \multicolumn{3}{c}{\cellcolor[HTML]{EFEFEF}Timeout} & \multicolumn{3}{c}{\cellcolor[HTML]{EFEFEF}Timeout} & \multicolumn{3}{c}{\cellcolor[HTML]{EFEFEF}Timeout} & \multicolumn{3}{c}{\cellcolor[HTML]{EFEFEF}Timeout} & \multicolumn{3}{c}{\cellcolor[HTML]{EFEFEF}Timeout} & \multicolumn{3}{c}{\cellcolor[HTML]{EFEFEF}Timeout} \\
    \multirow{-3}{*}{\textbf{Cs}} & \textbf{CaPGNN} & \textbf{15.14} & \textbf{3.19} & 97.00 & \textbf{18.85} & \textbf{4.03} & \textbf{97.17} & \textbf{19.67} & \textbf{4.35} & \textbf{97.09} & \textbf{24.17} & \textbf{7.10} & \textbf{97.14} & \textbf{28.38} & \textbf{8.93} & \textbf{97.15} & \textbf{52.97} & \textbf{18.90} & \textbf{96.79} & \textbf{54.29} & \textbf{19.85} & \textbf{96.89} \\
    \Xhline{1pt}
     & \textbf{Vanilla} & 179.69 & 123.87 & 96.37 & 148.65 & 104.41 & 96.27 & 146.56 & 107.33 & 96.35 & 156.22 & 114.86 & \textbf{96.35} & 182.81 & 136.98 & \textbf{96.37} & 159.59 & 112.46 & \textbf{96.41} & 156.95 & 113.13 & \textbf{96.27} \\
     & \textbf{AdaQP} & 84.39 & 21.37 & 96.38 & 95.84 & 32.44 & \textbf{96.29} & 114.24 & 34.16 & 95.08 & \multicolumn{3}{c}{\cellcolor[HTML]{EFEFEF}Timeout} & \multicolumn{3}{c}{\cellcolor[HTML]{EFEFEF}Timeout} & \multicolumn{3}{c}{\cellcolor[HTML]{EFEFEF}Timeout} & \multicolumn{3}{c}{\cellcolor[HTML]{EFEFEF}Timeout} \\
    \multirow{-3}{*}{\textbf{Rt}} & \textbf{CaPGNN} & \textbf{43.93} & \textbf{2.56} & \textbf{96.41} & \textbf{42.91} & \textbf{8.14} & 95.90 & \textbf{52.93} & \textbf{11.97} & \textbf{95.85} & \textbf{47.02} & \textbf{13.87} & 93.34 & \textbf{69.42} & \textbf{27.20} & 95.22 & \textbf{76.02} & \textbf{34.14} & 95.52 & \textbf{72.75} & \textbf{28.84} & 93.96 \\
    \Xhline{1pt}
     & \textbf{Vanilla} & 142.50 & 111.45 & 58.14 & 156.42 & 127.79 & 58.21 & 127.67 & 28.30 & 57.57 & 152.77 & 121.68 & 57.97 & 164.70 & 131.97 & 57.90 & 153.03 & 119.57 & 57.22 & 143.54 & 111.06 & 57.55 \\
     & \textbf{AdaQP} & 67.06 & 27.71 & 57.70 & 74.76 & 30.91 & 58.38 & 127.67 & 28.30 & 57.57 & 90.49 & 31.29 & 57.62 & 97.59 & 34.93 & 58.18 & \multicolumn{3}{c}{\cellcolor[HTML]{EFEFEF}Timeout} & \multicolumn{3}{c}{\cellcolor[HTML]{EFEFEF}Timeout} \\
    \multirow{-3}{*}{\textbf{Yp}} & \textbf{CaPGNN} & \textbf{47.69} & \textbf{5.60} & \textbf{59.47} & \textbf{43.56} & \textbf{9.42} & \textbf{59.36} & \textbf{45.70} & \textbf{12.69} & \textbf{59.28} & \textbf{42.58} & \textbf{15.92} & \textbf{58.97} & \textbf{49.15} & \textbf{20.93} & \textbf{60.14} & \textbf{106.44} & \textbf{53.09} & \textbf{60.79} & \textbf{106.58} & \textbf{54.43} & \textbf{60.81} \\
    \Xhline{1pt}
     & \textbf{Vanilla} & 388.49 & 245.20 & 63.43 & 434.06 & 313.61 & \textbf{64.50} & 359.77 & 247.83 & 60.60 & 366.00 & 274.26 & 63.35 & 469.97 & 330.98 & \textbf{63.87} & 407.81 & 290.41 & 62.28 & \multicolumn{3}{c}{\cellcolor[HTML]{EFEFEF}OOM} \\
     & \textbf{AdaQP} & \multicolumn{3}{c}{\cellcolor[HTML]{EFEFEF}OOM} & 234.89 & 114.65 & 62.24 & 223.84 & 108.27 & 60.64 & 189.41 & 92.85 & 61.54 & 298.78 & 163.01 & 61.18 & 261.04 & 141.20 & 62.57 & \multicolumn{3}{c}{\cellcolor[HTML]{EFEFEF}Timeout} \\
    \multirow{-3}{*}{\textbf{As}} & \textbf{CaPGNN} & \textbf{115.60} & \textbf{33.20} & \textbf{63.54} & \textbf{78.41} & \textbf{17.10} & 64.49 & \textbf{57.62} & \textbf{10.98} & \textbf{64.89} & \textbf{129.89} & \textbf{48.08} & \textbf{64.23} & \textbf{101.85} & \textbf{37.77} & 63.78 & \textbf{157.49} & 57.21 & \textbf{64.57} & \textbf{236.39} & \textbf{79.95} & \textbf{62.48} \\
    \Xhline{1pt}
     & \textbf{Vanilla} & 277.92 & 155.75 & 91.02 & 307.38 & 210.19 & 91.55 & 281.09 & 196.53 & 91.51 & 267.94 & 184.14 & 91.34 & 275.60 & 187.09 & 91.60 & \multicolumn{3}{c}{\cellcolor[HTML]{EFEFEF}OOM} & \multicolumn{3}{c}{\cellcolor[HTML]{EFEFEF}OOM} \\
     & \textbf{AdaQP} & \multicolumn{3}{c}{\cellcolor[HTML]{EFEFEF}OOM} & 143.15 & 43.76 & 91.58 & 148.02 & 42.09 & 91.45 & 163.56 & 50.29 & 91.46 & 294.32 & 71.35 & 91.55 & \multicolumn{3}{c}{\cellcolor[HTML]{EFEFEF}OOM} & \multicolumn{3}{c}{\cellcolor[HTML]{EFEFEF}OOM} \\
    \multirow{-3}{*}{\textbf{Os}} & \textbf{CaPGNN} & \textbf{90.96} & \textbf{22.23} & \textbf{92.24} & \textbf{81.22} & \textbf{12.72} & \textbf{92.22} & \textbf{83.01} & \textbf{18.48} & \textbf{92.19} & \textbf{99.53} & \textbf{32.51} & \textbf{92.03} & \textbf{138.41} & \textbf{45.95} & \textbf{91.96} & \textbf{201.70} & \textbf{72.32} & \textbf{92.16} & \textbf{199.11} & \textbf{67.98} & \textbf{92.23} \\
    \Xhline{1.5pt}
    \end{tabular}%
    }
\end{table*}

\subsection{Ablation Study}
\label{ssec:ablation_study}
To further evaluate the individual and combined contributions of caching, partitioning, and pipelining, we conduct comprehensive ablation studies on both GCN and GraphSAGE models. As shown in Table~\ref{tab:ablation_study}, we compare five configurations: a baseline version without caching or partitioning (Vanilla), caching only (+JACA), partitioning only (+RAPA), both caching and partitioning (+JACA+RAPA), and the full optimization with pipelining (+JACA+RAPA+Pipe.). All experiments are conducted using four partitions, two NVIDIA RTX 3090 GPUs and two NVIDIA A40 GPUs, with each model trained for 200 epochs.

The results reveal several important findings. Introducing JACA alone significantly reduces communication overhead across all datasets while maintaining comparable accuracy. For instance, on the Reddit dataset, the communication time for GCN drops from 112.31s to 41.86s, with accuracy only slightly decreasing from 95.30\% to 93.76\%. Similarly, on ogbn-products, GraphSAGE sees communication reduced from 196.53s to 97.78s, and the accuracy even improved. These results demonstrate that our two-level caching mechanism can effectively eliminate redundant data movement, thanks to its prioritization of high-importance vertices. Applying RAPA alone also leads to a considerable reduction in communication. For example, on AmazonProducts, GCN's communication time decreases from 263.00s to 26.17s, and on Reddit from 112.31s to 54.57s. However, compared to JACA, RAPA tends to introduce more accuracy degradation on certain datasets. For instance, on Yelp, GCN’s accuracy drops from 41.55\% to 39.57\% under RAPA alone, whereas JACA maintains a higher accuracy of 41.21\%. This may be due to the fact that the influence score used by RAPA is a heuristic structural proxy rather than an explicit measure of task importance. Although high-degree or frequently reused vertices often correlate with being more informative for message passing, the correlation is not guaranteed and can vary with graph properties, e.g., weaker homophily or higher noise. Therefore, in a few cases RAPA alone may introduce a larger accuracy drop.

The combination of JACA and RAPA yields the most balanced result between training efficiency and accuracy. On Flickr, GCN's total execution time drops by more than 50\% (from 47.40s to 23.33s), while preserving accuracy at 47.89\%, which is higher than the baseline. Likewise, for GraphSAGE on ogbn-products, combining JACA and RAPA reduces total time from 281.09s to 113.31s, while accuracy improves from 91.51\% to 92.35\%. Furthermore, integration of pipelining brings additional performance gains, especially on large-scale datasets. By overlapping data transfer and computation, pipelining further reduces communication time. On ogbn-products, GCN's communication time is reduced to 17.42s, and GraphSAGE achieves 18.48s, both with maintained or improved accuracy. These results demonstrate the scalability of our full system design.

Overall, the ablation study confirms that both caching and partitioning independently contribute to performance improvement, but their combination yields the best results in terms of both efficiency and accuracy. When further enhanced with pipelining, the system achieves maximum speedup without compromising predictive performance. These benefits are consistently observed across both GCN and GraphSAGE, validating the generality of our approach.

\begin{table*}[htb]
    \centering
    \belowrulesep=0pt
    \aboverulesep=0pt
    \renewcommand{\arraystretch}{1.3}
    \caption{Ablation analysis of CaPGNN. Vanilla denotes baseline.}
    \label{tab:ablation_study}
    \resizebox{\linewidth}{!}{%
        \begin{tabular}{cl|ccccccccccccccccccccc}
            \Xhline{1.5pt}
            \multirow{2}{*}{\textbf{Model}} &\multirow{2}{*}{\textbf{Alg}}  & \multicolumn{3}{c}{\textbf{Cl}} & \multicolumn{3}{c}{\textbf{Fr}} & \multicolumn{3}{c}{\textbf{Cs}} & \multicolumn{3}{c}{\textbf{Rt}} & \multicolumn{3}{c}{\textbf{Yp}} & \multicolumn{3}{c}{\textbf{As}} & \multicolumn{3}{c}{\textbf{Os}} \\
            \cmidrule(lr){3-5} \cmidrule(lr){6-8} \cmidrule(lr){9-11} \cmidrule(lr){12-14} \cmidrule(lr){15-17} \cmidrule(lr){18-20} \cmidrule(lr){21-23}
             & & \textbf{Epoch} & \textbf{Comm} & \textbf{Acc} &  \textbf{Epoch} & \textbf{Comm} & \textbf{Acc} & \textbf{Epoch} & \textbf{Comm} & \textbf{Acc} & \textbf{Epoch} & \textbf{Comm} & \textbf{Acc} & \textbf{Epoch} & \textbf{Comm} & \textbf{Acc} & \textbf{Epoch} & \textbf{Comm} & \textbf{Acc} & \textbf{Epoch} & \textbf{Comm} & \textbf{Acc} \\
             \Xhline{1pt}
            \multirow{7}{*}{\textbf{GCN}} & \textbf{Vanilla} & 22.30 & 11.99 & 67.46 & 47.40 & 36.91 & 46.88 & 58.00 & 46.20 & 96.54 & 151.90 & 112.31 & 95.30 & 140.54 & 114.53 & 41.55 & 374.93 & 263.00 & 37.65 & 282.71 & 199.66 & 91.77 \\
            \cmidrule(lr){2-23}
            & \textbf{+JACA} & 18.03 & 1.51 & 64.88 & 32.34 & 9.55 & 47.62 & 30.02 & 5.89 & 95.68 & 103.94 & 41.86 & 93.76 & 108.16 & 49.29 & 41.21 & 276.70 & 133.40 & 47.5 & 236.02 & 102.20 & 91.46 \\
            \cmidrule(lr){2-23}
            & \textbf{+RAPA} & 16.05 & 4.71 & 65.84 & 26.92 & 12.43 & 49.71 & 20.02 & 7.70 & 95.87 & 77.95 & 48.89 & 93.83 & 69.40 & 45.13 & 39.57 & 77.14 & 26.17 & 51.39 & 117.66 & 61.49 & 91.51 \\
            \cmidrule(lr){2-23}
            & \textbf{\makecell[l]{+JACA\\+RAPA}} & 16.30 & \textbf{1.32} & 65.46 & 23.33 & \textbf{5.28} & 47.89 & 20.12 & 3.83 & 95.91 & 64.00 & 22.92 & 93.93 & 54.62 & 22.25 & 43.57 & 68.59 & 14.29 & 50.92 & 109.51 & 42.75 & 91.63 \\
            \cmidrule(lr){2-23}
            & \textbf{\makecell[l]{+JACA\\+RAPA\\+Pipe.}} & \textbf{15.72} & 3.11 & 65.66 & \textbf{22.29} & 7.10 & 42.61 & \textbf{17.58} & \textbf{3.39} & 95.93 & \textbf{53.98} & \textbf{12.15} & 93.68 & \textbf{49.74} & \textbf{12.51} & 43.53 & \textbf{56.37} & \textbf{8.35} & 50.55 & \textbf{80.12} & \textbf{17.42} & 91.75 \\
            \Xhline{1pt}
            \multirow{7}{*}{\textbf{GraphSAGE}} & \textbf{Vanilla} & 23.97 & 10.45 & 68.01 & 48.61 & 37.71 & 42.37 & 65.93 & 49.64 & 97.03 & 146.56 & 107.33 & 96.35 & 127.67 & 28.30 & 57.57 & 359.77 & 247.83 & 60.6 & 281.09 & 196.53 & 91.51 \\
            \cmidrule(lr){2-23}
            & \textbf{+JACA} & 22.14 & 2.18 & 67.96 & 29.94 & 7.96 & 50.64 & 35.63 & 6.38 & 97.10 & 104.92 & 44.11 & 95.96 & 110.27 & 50.64 & 59.72 & 280.52 & 132.07 & 67.28 & 229.90 & 97.78 & 92.31 \\
            \cmidrule(lr){2-23}
            & \textbf{+RAPA} & 18.74 & 4.24 & 67.71 & 25.30 & 11.30 & 51.42 & 22.42 & 7.59 & 96.77 & 80.19 & 47.77 & 96.11 & 65.97 & 42.14 & 57.11 & 76.03 & 25.28 & 74.29 & 109.36 & 52.27 & 92.24 \\
            \cmidrule(lr){2-23}
            & \textbf{\makecell[l]{+JACA\\+RAPA}} & 18.59 & \textbf{1.09} & 67.71 & 23.48 & \textbf{5.22} & 50.83 & 22.30 & 4.08 & 96.98 & 63.77 & 21.64 & 95.98 & 59.62 & 25.06 & 58.58 & 70.85 & 15.58 & 67.48 & 113.31 & 44.10 & 92.35 \\
            \cmidrule(lr){2-23}
            & \textbf{\makecell[l]{+JACA\\+RAPA\\+Pipe.}} & \textbf{17.60} & 3.36 & 67.74 & \textbf{22.45} & 7.45 & 49.11 & \textbf{19.67} & \textbf{3.85} & 97.09 & \textbf{52.93} & \textbf{11.97} & 95.85 & \textbf{45.70} & \textbf{12.69} & 59.28 & \textbf{57.62} & \textbf{10.98} & 64.89 & \textbf{83.01} & \textbf{18.48} & 92.19 \\
            \Xhline{1.5pt}
        \end{tabular}%
    }
\end{table*}

\subsection{Extension to Distributed Systems}
\label{ssec:distributed_systems}
\begin{figure}[htbp]
    \centering
    \setlength{\abovecaptionskip}{0em}
    \setlength{\belowcaptionskip}{0em}
    \includegraphics[width=\linewidth]{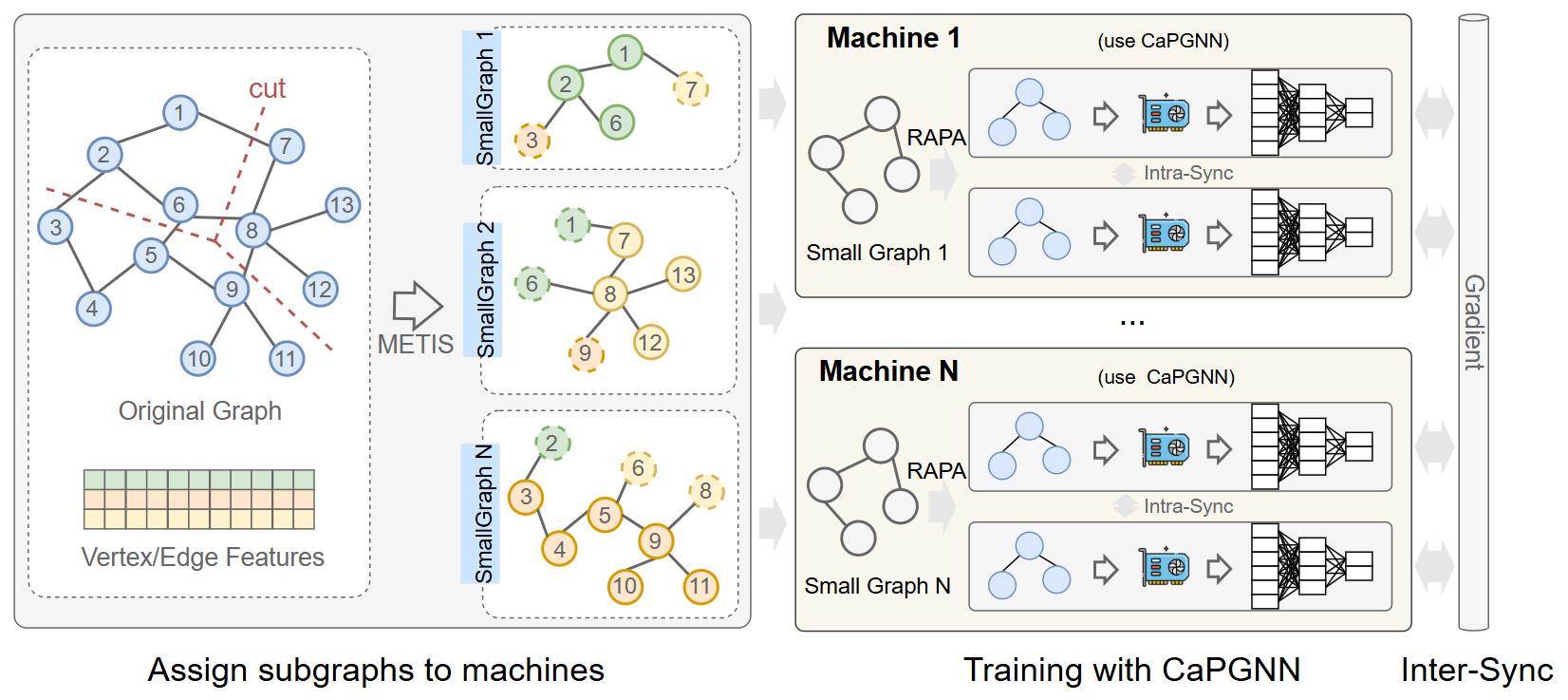}
    \caption{A prototype implementation of CaPGNN in a distributed environment. Each machine runs CaPGNN independently and then synchronizes gradients across machines.}
    \label{fig:distributed_version}
\end{figure}

We provide a prototype implementation of CaPGNN in a distributed environment\footnote{https://github.com/songxf1024/CaPGNN/tree/dist}, as shown in Fig.~\ref{fig:distributed_version}. In our implementation, the original graph is partitioned into multiple small graphs, and each small graph is assigned to a separate server equipped with multiple GPUs. Within each server, CaPGNN executes the proposed joint caching and resource-aware partitioning strategy as described in Section~IV, while inter-server communication is handled through standard message passing over Ethernet or InfiniBand. In this way, the framework scales to distributed training without modifying the core algorithms.

We conduct preliminary experiments on a GPU cluster consisting of two machines, each with four GPUs. Table~\ref{tab:distributed_capgnn} shows that CaPGNN can smoothly scale from single-machine multi-GPU to multi-machine multi-GPU environments while maintaining training throughput and accuracy.

\begin{table}[]
    \centering
    \belowrulesep=0pt
    \aboverulesep=0pt
    \renewcommand{\arraystretch}{1.2}
    \caption{Training performance of distributed CaPGNN.}
    \label{tab:distributed_capgnn}
    \resizebox{\linewidth}{!}{%
        \begin{tabular}{c|cccccc}
            \Xhline{1.5pt}
            \textbf{Dataset} & \textbf{Partitions} & \textbf{Machines} & \textbf{GPUs} & \textbf{Model} & \textbf{\makecell[c]{Epoch/s}} & \textbf{Accuracy (\%)} \\
            \Xhline{1pt}
             \multirow{6}{*}{As} & \multirow{2}{*}{\makecell[c]{1M-4D\\(4 workers)}} & \multirow{2}{*}{M1} & \multirow{2}{*}{\makecell[c]{R9(x2)\\T4(x2)}} & GCN & 3.54 & 50.55 \\
             &  &  &  & GraphSAGE & 3.47 & 64.89 \\
             \cmidrule(lr){2-7}
             & \multirow{2}{*}{\makecell[c]{2M-2D\\(4 workers)}} & M1 & R9(x2) & GCN & 3.07 & 51.21 \\
             &  & M2 & R9(x2) & GraphSAGE & 2.93 & 65.60 \\
             \cmidrule(lr){2-7}
             & \multirow{2}{*}{\makecell[c]{2M-4D\\(8 workers)}} & M1 & \makecell[c]{R9(x2)\\T4(x2)} & GCN & 3.27 & 47.92 \\
             &  & M2 & R9(x4) & GraphSAGE & 3.23 & 65.32 \\
             \Xhline{1.2pt}
             
             \multirow{6}{*}{Os} & \multirow{2}{*}{\makecell[c]{1M-4D\\(4 workers)}} & \multirow{2}{*}{M1} & \multirow{2}{*}{\makecell[c]{R9(x2)\\T4(x2)}} & GCN & 2.49 & 91.75 \\
             &  &  &  & GraphSAGE & 2.41 & 92.19 \\
             \cmidrule(lr){2-7}
             & \multirow{2}{*}{\makecell[c]{2M-2D\\(4 workers)}} & M1 & R9(x2) & GCN & 3.50 & 91.20 \\
             &  & M2 & R9(x2) & GraphSAGE & 3.75 & 91.91 \\
             \cmidrule(lr){2-7}
             & \multirow{2}{*}{\makecell[c]{2M-4D\\(8 workers)}} & M1 & \makecell[c]{R9(x2)\\T4(x2)} & GCN & 3.70 & 91.15 \\
             &  & M2 & R9(x4) & GraphSAGE & 3.55 & 92.18 \\
             \Xhline{1.5pt}
        \end{tabular}%
    }
\end{table}

These results demonstrate the applicability of CaPGNN to distributed training. However, we also observe that the \textit{As} dataset, which contains a significantly larger number of edges than \textit{Os}, introduces higher communication overhead and leads to reduced throughput. Therefore, more sophisticated optimizations, such as topology-aware graph partitioning across servers and advanced communication scheduling for large-scale deployments, remain as directions for future work.

\section{Discussion}
\label{sec:discussion}
The proposed CaPGNN demonstrates excellent performance. However, we have identified several areas for improvement, which are outlined below for further discussion.

\textbf{Partitioning.} Currently, CaPGNN adjusts only halo vertices during partitioning, leaving inner vertices unaffected. As a result, when significant performance differences between GPUs, removing all halo vertices might still fail to achieve load balance. In future work, we plan to optimize the partitioning process by adjusting both halo and inner vertices to achieve a more comprehensive balance.

\textbf{Distributed Training.} CaPGNN is primarily designed for single-machine multi-GPU parallel training. We have also conducted preliminary experiments in a multi-machine multi-GPU distributed setting, showing that the idea naturally generalizes beyond a single node. In future work, we plan to further optimize CaPGNN for high-performance distributed environments and large-scale maps of hundreds of billions and even larger.

\textbf{Adaptive Staleness Control.} CaPGNN uses a fixed staleness refresh policy, a promising direction is to design an adaptive staleness control mechanism that dynamically adjusts the refresh interval or update frequency to keep the effective staleness within a target bound.

These improvements aim to further enhance the performance, scalability, and flexibility of CaPGNN in more diverse and challenging scenarios.

\section{Conclusion}
\label{sec:conclusion}
In this paper, we propose CaPGNN, an optimization framework that integrates joint adaptive caching with resource-aware graph partitioning to improve the efficiency of parallel full-batch GNN training. By leveraging its two key innovations, i.e., JACA for efficient joint adaptive caching across CPU and GPU and RAPA for effective resource-aware partitioning, CaPGNN significantly reduces communication overhead. Both theoretical analysis and experimental results confirm the convergence of CaPGNN. Experiments across multiple graph datasets and GPU configurations show that CaPGNN outperforms existing methods, reducing communication overhead by up to 99\% and achieving speedups of up to 18.98x. This work introduces a novel solution to address key bottlenecks in GNN training. Future research can build on this foundation to explore advanced resource scheduling strategies and further optimize distributed training at scale.

\section*{Acknowledgments}
\label{sec:acknowledgments}
This research is supported partly by National Natural Science Foundation of China Fund No. U24B20151, as well as partly by the SCUT Research Startup Fund No. K3200890. Any opinions, findings, and conclusions or recommendations expressed in this publication are those of the authors and do not necessarily reflect the views of the sponsoring agencies.

\bibliographystyle{elsarticle-harv} 
\bibliography{reference}

@String{Computing = "Computing" }

@String{Computer = "{IEEE} Computer" }

@String{Springer = "Springer-Verlag" }

@article{mirhoseini2021graph,
  title={A graph placement methodology for fast chip design},
  author={Mirhoseini, Azalia and Goldie, Anna and Yazgan, Mustafa and Jiang, Joe Wenjie and Songhori, Ebrahim and Wang, Shen and Lee, Young-Joon and Johnson, Eric and Pathak, Omkar and Nazi, Azade and others},
  journal={Nature},
  volume={594},
  number={7862},
  pages={207--212},
  year={2021},
  publisher={Nature Publishing Group}
}

@inproceedings{sarlin2020superglue,
  title={Superglue: Learning feature matching with graph neural networks},
  author={Sarlin, Paul-Edouard and DeTone, Daniel and Malisiewicz, Tomasz and Rabinovich, Andrew},
  booktitle={Proceedings of the IEEE/CVF conference on computer vision and pattern recognition},
  pages={4938--4947},
  year={2020}
}

@inproceedings{yang2022gnnlab,
  title={GNNLab: a factored system for sample-based GNN training over GPUs},
  author={Yang, Jianbang and Tang, Dahai and Song, Xiaoniu and Wang, Lei and Yin, Qiang and Chen, Rong and Yu, Wenyuan and Zhou, Jingren},
  booktitle={Proceedings of the Seventeenth European Conference on Computer Systems},
  pages={417--434},
  year={2022}
}

@article{shao2024distributed,
  title={Distributed graph neural network training: A survey},
  author={Shao, Yingxia and Li, Hongzheng and Gu, Xizhi and Yin, Hongbo and Li, Yawen and Miao, Xupeng and Zhang, Wentao and Cui, Bin and Chen, Lei},
  journal={ACM Computing Surveys},
  volume={56},
  number={8},
  pages={1--39},
  year={2024},
  publisher={ACM New York, NY}
}

@inproceedings{gong2022graphite,
  title={Graphite: optimizing graph neural networks on CPUs through cooperative software-hardware techniques},
  author={Gong, Zhangxiaowen and Ji, Houxiang and Yao, Yao and Fletcher, Christopher W and Hughes, Christopher J and Torrellas, Josep},
  booktitle={Proceedings of the 49th Annual International Symposium on Computer Architecture},
  pages={916--931},
  year={2022}
}

@article{grattarola2021graph,
  title={Graph neural networks in tensorflow and keras with spektral [application notes]},
  author={Grattarola, Daniele and Alippi, Cesare},
  journal={IEEE Computational Intelligence Magazine},
  volume={16},
  number={1},
  pages={99--106},
  year={2021},
  publisher={IEEE}
}

@inproceedings{huang2020ge,
  title={Ge-spmm: General-purpose sparse matrix-matrix multiplication on gpus for graph neural networks},
  author={Huang, Guyue and Dai, Guohao and Wang, Yu and Yang, Huazhong},
  booktitle={SC20: International Conference for High Performance Computing, Networking, Storage and Analysis},
  pages={1--12},
  year={2020},
  organization={IEEE}
}

@article{CDFGNN,
  author       = {Shuai Zhang and
                  Zite Jiang and
                  Haihang You},
  title        = {{CDFGNN:} a Systematic Design of Cache-based Distributed Full-Batch
                  Graph Neural Network Training with Communication Reduction},
  journal      = {CoRR},
  volume       = {abs/2408.00232},
  year         = {2024},
}

@inproceedings{schlichtkrull2018modeling,
  title={Modeling relational data with graph convolutional networks},
  author={Schlichtkrull, Michael and Kipf, Thomas N and Bloem, Peter and Van Den Berg, Rianne and Titov, Ivan and Welling, Max},
  booktitle={The semantic web: 15th international conference, ESWC 2018, Heraklion, Crete, Greece, June 3--7, 2018, proceedings 15},
  pages={593--607},
  year={2018},
  organization={Springer}
}

@article{windgp,
  author       = {Li Zeng and
                  Haohan Huang and
                  Binfan Zheng and
                  Kang Yang and
                  Sheng Cheng Shao and
                  Jinhua Zhou and
                  Jun Xie and
                  Rongqian Zhao and
                  Xin Chen},
  title        = {WindGP: Efficient Graph Partitioning on Heterogenous Machines},
  journal      = {CoRR},
  volume       = {abs/2403.00331},
  year         = {2024},
}

@article{karypis1998metis,
  title={METIS: A software package for partitioning unstructured graphs},
  author={Karypis, George and Kumar, Vipin},
  journal={Partitioning Meshes, and Computing Fill-Reducing Orderings of Sparse Matrices, Version},
  volume={4},
  number={0},
  pages={0--80},
  year={1998}
}

@article{wan2023adaqp,
  title={Adaptive message quantization and parallelization for distributed full-graph gnn training},
  author={Wan, Borui and Zhao, Juntao and Wu, Chuan},
  journal={Proceedings of Machine Learning and Systems},
  volume={5},
  year={2023}
}

@article{verma2017an,
  title={An Experimental Comparison of Partitioning Strategies in Distributed Graph Processing},
  author={Verma, Shiv and Leslie, Luke M. and Shin, Yosub and Gupta, Indranil},
  journal={Proceedings of the VLDB Endowment},
  volume={10},
  number={5},
  pages={493--504},
  year={2017},
  publisher={VLDB Endowment}
}

@ARTICLE{9761891Huang,
  author={Huang, Linyong and Zhang, Zhe and Li, Shuangchen and Niu, Dimin and Guan, Yijin and Zheng, Hongzhong and Xie, Yuan},
  journal={IEEE Access}, 
  title={Practical Near-Data-Processing Architecture for Large-Scale Distributed Graph Neural Network}, 
  year={2022},
  volume={10},
  number={},
  pages={46796-46807}
}

@ARTICLE{8763922,
    author={Li, Ang and Song, Shuaiwen Leon and Chen, Jieyang and Li, Jiajia and Liu, Xu and Tallent, Nathan R. and Barker, Kevin J.},
    journal={ IEEE Transactions on Parallel \& Distributed Systems },
    title={{ Evaluating Modern GPU Interconnect: PCIe, NVLink, NV-SLI, NVSwitch and GPUDirect }},
    year={2020},
    volume={31},
    number={01},
    ISSN={1558-2183},
    pages={94-110},
    publisher={IEEE Computer Society},
    address={Los Alamitos, CA, USA},
    month=jan
}

@inproceedings{zhao2024neutroncache,
  title={NeutronCache: An Efficient Cache-Enhanced Distributed Graph Neural Network Training System},
  author={Zhao, Chu and Dong, Shengjie and Zhao, Yuhai and Li, Yuan},
  booktitle={Proceedings of the 33rd ACM International Conference on Information and Knowledge Management},
  pages={3310--3319},
  year={2024}
}

@inproceedings{zhang2023two,
  title={Two-level Graph Caching for Expediting Distributed GNN Training},
  author={Zhang, Zhe and Luo, Ziyue and Wu, Chuan},
  booktitle={IEEE INFOCOM 2023-IEEE Conference on Computer Communications},
  pages={1--10},
  year={2023},
  organization={IEEE}
}

@article{sancus,
author = {Peng, Jingshu and Chen, Zhao and Shao, Yingxia and Shen, Yanyan and Chen, Lei and Cao, Jiannong},
title = {Sancus: staleness-aware communication-avoiding full-graph decentralized training in large-scale graph neural networks},
year = {2022},
issue_date = {May 2022},
publisher = {VLDB Endowment},
volume = {15},
number = {9},
issn = {2150-8097},
journal = {Proc. VLDB Endow.},
month = may,
pages = {1937–1950},
numpages = {14}
}

@article{Kipf_Welling_2016,  
    title={Semi-Supervised Classification with Graph Convolutional Networks}, 
    journal={arXiv: Learning,arXiv: Learning}, 
    author={Kipf, Thomas and Welling, Max}, 
    year={2016}, 
    month={Sep}, 
    language={en-US} 
}

@article{Hamilton_Ying_Leskovec_2017,  
    title={Inductive Representation Learning on Large Graphs}, 
    journal={Neural Information Processing Systems,Neural Information Processing Systems}, 
    author={Hamilton, WilliamL. and Ying, Zhitao and Leskovec, Jure}, 
    year={2017}, 
    month={Jun}, 
    language={en-US} 
}

@article{wan2023adaptive,
  title={Adaptive message quantization and parallelization for distributed full-graph gnn training},
  author={Wan, Borui and Zhao, Juntao and Wu, Chuan},
  journal={Proceedings of Machine Learning and Systems},
  volume={5},
  year={2023}
}

@article{merkel2024can,
  title={Can Graph Reordering Speed Up Graph Neural Network Training? An Experimental Study},
  author={Merkel, Nikolai and Toussing, Pierre and Mayer, Ruben and Jacobsen, Hans-Arno},
  journal={arXiv preprint arXiv:2409.11129},
  year={2024}
}

@article{jia2020improving,
  title={Improving the accuracy, scalability, and performance of graph neural networks with roc},
  author={Jia, Zhihao and Lin, Sina and Gao, Mingyu and Zaharia, Matei and Aiken, Alex},
  journal={Proceedings of Machine Learning and Systems},
  volume={2},
  pages={187--198},
  year={2020}
}

@article{peng2022sancus,
  title={Sancus: sta le n ess-aware c omm u nication-avoiding full-graph decentralized training in large-scale graph neural networks},
  author={Peng, Jingshu and Chen, Zhao and Shao, Yingxia and Shen, Yanyan and Chen, Lei and Cao, Jiannong},
  journal={Proceedings of the VLDB Endowment},
  volume={15},
  number={9},
  pages={1937--1950},
  year={2022},
  publisher={VLDB Endowment}
}

@inproceedings{wang2022neutronstar,
  title={Neutronstar: distributed GNN training with hybrid dependency management},
  author={Wang, Qiange and Zhang, Yanfeng and Wang, Hao and Chen, Chaoyi and Zhang, Xiaodong and Yu, Ge},
  booktitle={Proceedings of the 2022 International Conference on Management of Data},
  pages={1301--1315},
  year={2022}
}

@inproceedings{zheng2020distdgl,
  title={DistDGL: Distributed graph neural network training for billion-scale graphs},
  author={Zheng, Da and Ma, Chao and Wang, Minjie and Zhou, Jinjing and Su, Qidong and Song, Xiang and Gan, Quan and Zhang, Zheng and Karypis, George},
  booktitle={2020 IEEE/ACM 10th Workshop on Irregular Applications: Architectures and Algorithms (IA3)},
  pages={36--44},
  year={2020},
  organization={IEEE}
}

@inproceedings{yang2019aligraph,
  title={Aligraph: A comprehensive graph neural network platform},
  author={Yang, Hongxia},
  booktitle={Proceedings of the 25th ACM SIGKDD international conference on knowledge discovery \& data mining},
  pages={3165--3166},
  year={2019}
}

@inproceedings{wan2022pipegcn,
  title={{PipeGCN}: Efficient Full-Graph Training of Graph Convolutional Networks with Pipelined Feature Communication},
  author={Wan, C and Li, Y and Wolfe, Cameron R and Kyrillidis, A and Kim, Nam S and Lin, Y},
  booktitle={The Tenth International Conference on Learning Representations (ICLR 2022)},
  year={2022}
}

@inproceedings{ma2019neugraph,
  title={$\{$NeuGraph$\}$: Parallel deep neural network computation on large graphs},
  author={Ma, Lingxiao and Yang, Zhi and Miao, Youshan and Xue, Jilong and Wu, Ming and Zhou, Lidong and Dai, Yafei},
  booktitle={2019 USENIX Annual Technical Conference (USENIX ATC 19)},
  pages={443--458},
  year={2019}
}

@inproceedings{lin2020pagraph,
  title={Pagraph: Scaling gnn training on large graphs via computation-aware caching},
  author={Lin, Zhiqi and Li, Cheng and Miao, Youshan and Liu, Yunxin and Xu, Yinlong},
  booktitle={Proceedings of the 11th ACM Symposium on Cloud Computing},
  pages={401--415},
  year={2020}
}

@inproceedings{sun2023legion,
  title={Legion: Automatically Pushing the Envelope of $\{$Multi-GPU$\}$ System for $\{$Billion-Scale$\}$$\{$GNN$\}$ Training},
  author={Sun, Jie and Su, Li and Shi, Zuocheng and Shen, Wenting and Wang, Zeke and Wang, Lei and Zhang, Jie and Li, Yong and Yu, Wenyuan and Zhou, Jingren and others},
  booktitle={2023 USENIX Annual Technical Conference (USENIX ATC 23)},
  pages={165--179},
  year={2023}
}

@article{ai2023neutronorch,
  title={NeutronOrch: Rethinking Sample-based GNN Training under CPU-GPU Heterogeneous Environments},
  author={Ai, Xin and Wang, Qiange and Cao, Chunyu and Zhang, Yanfeng and Chen, Chaoyi and Yuan, Hao and Gu, Yu and Yu, Ge},
  journal={arXiv preprint arXiv:2311.13225},
  year={2023}
}

@inproceedings{malewicz2010pregel,
  title={Pregel: a system for large-scale graph processing},
  author={Malewicz, Grzegorz and Austern, Matthew H and Bik, Aart JC and Dehnert, James C and Horn, Ilan and Leiser, Naty and Czajkowski, Grzegorz},
  booktitle={Proceedings of the 2010 ACM SIGMOD International Conference on Management of data},
  pages={135--146},
  year={2010}
}

@inproceedings{gonzalez2012powergraph,
  title={$\{$PowerGraph$\}$: Distributed $\{$Graph-Parallel$\}$ computation on natural graphs},
  author={Gonzalez, Joseph E and Low, Yucheng and Gu, Haijie and Bickson, Danny and Guestrin, Carlos},
  booktitle={10th USENIX symposium on operating systems design and implementation (OSDI 12)},
  pages={17--30},
  year={2012}
}

@article{chen2019powerlyra,
  title={Powerlyra: Differentiated graph computation and partitioning on skewed graphs},
  author={Chen, Rong and Shi, Jiaxin and Chen, Yanzhe and Zang, Binyu and Guan, Haibing and Chen, Haibo},
  journal={ACM Transactions on Parallel Computing (TOPC)},
  volume={5},
  number={3},
  pages={1--39},
  year={2019},
  publisher={ACM New York, NY, USA}
}

@article{karypis1998fast,
  title={A fast and high quality multilevel scheme for partitioning irregular graphs},
  author={Karypis, George and Kumar, Vipin},
  journal={SIAM Journal on scientific Computing},
  volume={20},
  number={1},
  pages={359--392},
  year={1998},
  publisher={SIAM}
}

@inproceedings{chiang2019cluster,
  title={Cluster-gcn: An efficient algorithm for training deep and large graph convolutional networks},
  author={Chiang, Wei-Lin and Liu, Xuanqing and Si, Si and Li, Yang and Bengio, Samy and Hsieh, Cho-Jui},
  booktitle={Proceedings of the 25th ACM SIGKDD international conference on knowledge discovery \& data mining},
  pages={257--266},
  year={2019}
}

@inproceedings{cai2021dgcl,
  title={DGCL: An efficient communication library for distributed GNN training},
  author={Cai, Zhenkun and Yan, Xiao and Wu, Yidi and Ma, Kaihao and Cheng, James and Yu, Fan},
  booktitle={Proceedings of the Sixteenth European Conference on Computer Systems},
  pages={130--144},
  year={2021}
}

@inproceedings{liu2023bgl,
  title={$\{$BGL$\}$:$\{$GPU-Efficient$\}$$\{$GNN$\}$ training by optimizing graph data $\{$I/O$\}$ and preprocessing},
  author={Liu, Tianfeng and Chen, Yangrui and Li, Dan and Wu, Chuan and Zhu, Yibo and He, Jun and Peng, Yanghua and Chen, Hongzheng and Chen, Hongzhi and Guo, Chuanxiong},
  booktitle={20th USENIX Symposium on Networked Systems Design and Implementation (NSDI 23)},
  pages={103--118},
  year={2023}
}

@article{xue2023sugar,
  title={Sugar: Efficient subgraph-level training via resource-aware graph partitioning},
  author={Xue, Zihui and Yang, Yuedong and Marculescu, Radu},
  journal={IEEE Transactions on Computers},
  year={2023},
  publisher={IEEE}
}

@article{sharma2024survey,
  title={A survey of graph neural networks for social recommender systems},
  author={Sharma, Kartik and Lee, Yeon-Chang and Nambi, Sivagami and Salian, Aditya and Shah, Shlok and Kim, Sang-Wook and Kumar, Srijan},
  journal={ACM Computing Surveys},
  volume={56},
  number={10},
  pages={1--34},
  year={2024},
  publisher={ACM New York, NY}
}

@article{wu2022graph,
  title={Graph neural networks in recommender systems: a survey},
  author={Wu, Shiwen and Sun, Fei and Zhang, Wentao and Xie, Xu and Cui, Bin},
  journal={ACM Computing Surveys},
  volume={55},
  number={5},
  pages={1--37},
  year={2022},
  publisher={ACM New York, NY}
}

@article{ye2022comprehensive,
  title={A comprehensive survey of graph neural networks for knowledge graphs},
  author={Ye, Zi and Kumar, Yogan Jaya and Sing, Goh Ong and Song, Fengyan and Wang, Junsong},
  journal={IEEE Access},
  volume={10},
  pages={75729--75741},
  year={2022},
  publisher={IEEE}
}

@article{li2023task,
  title={Task placement and resource allocation for edge machine learning: A gnn-based multi-agent reinforcement learning paradigm},
  author={Li, Yihong and Zhang, Xiaoxi and Zeng, Tianyu and Duan, Jingpu and Wu, Chuan and Wu, Di and Chen, Xu},
  journal={IEEE Transactions on Parallel and Distributed Systems},
  year={2023},
  publisher={IEEE}
}

@article{yang2021graphformers,
  title={Graphformers: Gnn-nested transformers for representation learning on textual graph},
  author={Yang, Junhan and Liu, Zheng and Xiao, Shitao and Li, Chaozhuo and Lian, Defu and Agrawal, Sanjay and Singh, Amit and Sun, Guangzhong and Xie, Xing},
  journal={Advances in Neural Information Processing Systems},
  volume={34},
  pages={28798--28810},
  year={2021}
}

@inproceedings{bojchevski2018deep,
    title={Deep Gaussian Embedding of Graphs:  Unsupervised Inductive Learning via Ranking},
    author={Aleksandar Bojchevski and Stephan Günnemann},
    booktitle={International Conference on Learning Representations},
    year={2018}
}

@inproceedings{graphsainticlr20,
    title={{GraphSAINT}: Graph Sampling Based Inductive Learning Method},
    author={Hanqing Zeng and Hongkuan Zhou and Ajitesh Srivastava and Rajgopal Kannan and Viktor Prasanna},
    booktitle={International Conference on Learning Representations},
    year={2020}
}

@article{shchur2018pitfalls,
  title={Pitfalls of Graph Neural Network Evaluation},
  author={Shchur, Oleksandr and Mumme, Maximilian and Bojchevski, Aleksandar and G{\"u}nnemann, Stephan},
  journal={Relational Representation Learning Workshop, NeurIPS 2018},
  year={2018}
}

@article{hamilton2017inductive,
  title={Inductive representation learning on large graphs},
  author={Hamilton, Will and Ying, Zhitao and Leskovec, Jure},
  journal={Advances in neural information processing systems},
  volume={30},
  year={2017}
}

@article{hu2020open,
  title={Open graph benchmark: Datasets for machine learning on graphs},
  author={Hu, Weihua and Fey, Matthias and Zitnik, Marinka and Dong, Yuxiao and Ren, Hongyu and Liu, Bowen and Catasta, Michele and Leskovec, Jure},
  journal={Advances in neural information processing systems},
  volume={33},
  pages={22118--22133},
  year={2020}
}

@article{
  velickovic2018graph,
  title="{Graph Attention Networks}",
  author={Veli{\v{c}}kovi{\'{c}}, Petar and Cucurull, Guillem and Casanova, Arantxa and Romero, Adriana and Li{\`{o}}, Pietro and Bengio, Yoshua},
  journal={International Conference on Learning Representations},
  year={2018},
  note={accepted as poster},
}

@inproceedings{gilmer2017neural,
  title={Neural message passing for quantum chemistry},
  author={Gilmer, Justin and Schoenholz, Samuel S and Riley, Patrick F and Vinyals, Oriol and Dahl, George E},
  booktitle={International conference on machine learning},
  pages={1263--1272},
  year={2017},
  organization={PMLR}
}

@ARTICLE{10738209,
  author={Shi, Zheng and Zou, Yi and Song, Xianfeng and Li, Shupeng and Liu, Fangming and Xue, Quan},
  journal={IEEE Transactions on Parallel and Distributed Systems}, 
  title={DyLaClass: Dynamic Labeling Based Classification for Optimal Sparse Matrix Format Selection in Accelerating SpMV}, 
  year={2024},
  volume={35},
  number={12},
  pages={2624-2639}
}

@INPROCEEDINGS{9355302,
  author={Huang, Guyue and Dai, Guohao and Wang, Yu and Yang, Huazhong},
  booktitle={SC20: International Conference for High Performance Computing, Networking, Storage and Analysis}, 
  title={GE-SpMM: General-Purpose Sparse Matrix-Matrix Multiplication on GPUs for Graph Neural Networks}, 
  year={2020},
  volume={},
  number={},
  pages={1-12}
}

@article{wang2023hongtu,
  title={Hongtu: Scalable full-graph GNN training on multiple gpus},
  author={Wang, Qiange and Chen, Yao and Wong, Weng-Fai and He, Bingsheng},
  journal={Proceedings of the ACM on Management of Data},
  volume={1},
  number={4},
  pages={1--27},
  year={2023},
  publisher={ACM New York, NY, USA}
}

@article{wan2023scalable,
  title={Scalable and efficient full-graph gnn training for large graphs},
  author={Wan, Xinchen and Xu, Kaiqiang and Liao, Xudong and Jin, Yilun and Chen, Kai and Jin, Xin},
  journal={Proceedings of the ACM on Management of Data},
  volume={1},
  number={2},
  pages={1--23},
  year={2023},
  publisher={ACM New York, NY, USA}
}

@article{wan2022bns,
  title={Bns-gcn: Efficient full-graph training of graph convolutional networks with partition-parallelism and random boundary node sampling},
  author={Wan, Cheng and Li, Youjie and Li, Ang and Kim, Nam Sung and Lin, Yingyan},
  journal={Proceedings of Machine Learning and Systems},
  volume={4},
  pages={673--693},
  year={2022}
}

@inproceedings{tsourakakis2014fennel,
  title={Fennel: Streaming graph partitioning for massive scale graphs},
  author={Tsourakakis, Charalampos and Gkantsidis, Christos and Radunovic, Bozidar and Vojnovic, Milan},
  booktitle={Proceedings of the 7th ACM international conference on Web search and data mining},
  pages={333--342},
  year={2014}
}

\newpage
\noindent\textbf{Author biography}

\begin{wrapfigure}{l}{3cm}
  \centering
  \includegraphics[width=3.2cm,keepaspectratio]{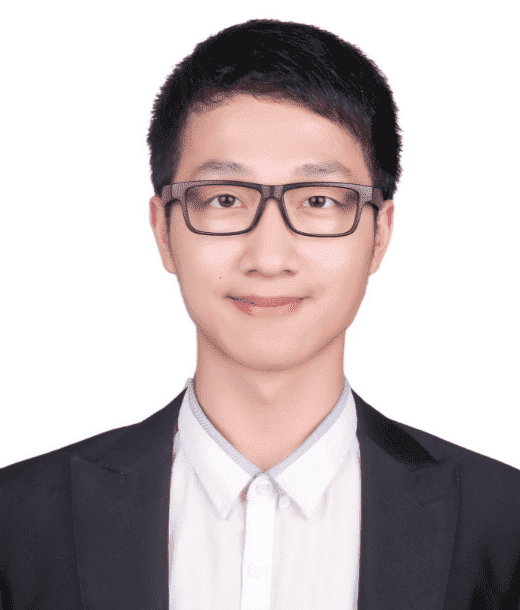}
\end{wrapfigure}
\noindent
\textbf{Xianfeng Song} was born in Huzhou, Zhejiang, China. He received the B.S. degree from Nanchang University, Jiangxi, China, in 2015. He is currently pursuing the Ph.D. degree with the College of Microelectronics, South China University of Technology, Guangdong, China. His research interest includes \textit{Distributed Computing}, \textit{Graph Neural Networks}, and \textit{Distributed Storage}.

\vspace{1em}

\begin{wrapfigure}{l}{3.2cm}
  \centering
  \includegraphics[width=3cm,keepaspectratio]{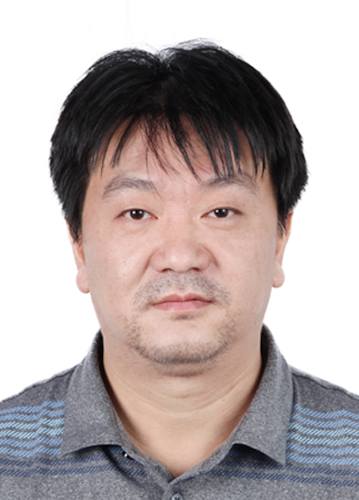}
\end{wrapfigure}
\noindent
\textbf{Yi Zou} is currently a Chair Professor at the South China University of Technology (SCUT), China. He received the M.Eng. degree from Nanyang Technological University, Singapore in 2002 and Ph. D. in Computer Engineering from Duke University, USA in 2004. Before joining SCUT, he was a postdoctoral fellow at Duke University, USA and a Senior Staff Research Scientist at Research Labs, Intel Corp., USA. He has published more than 70 publications including book chapters, technical papers, and patents. He serves as a frequent program committee member and reviewer for IEEE transactions and conferences. Most recently, he is a TPC member at IEEE NAS'21, ACM SEC'20/21, IEEE DataComp’19, etc. His current research areas of interest are intelligent compute architectures and systems, scale-out memory and storage, data and sensor fusion, edge computing and AI, etc. 

\vspace{1em}

\begin{wrapfigure}{l}{3cm}
  \centering
  \includegraphics[width=3.2cm,keepaspectratio]{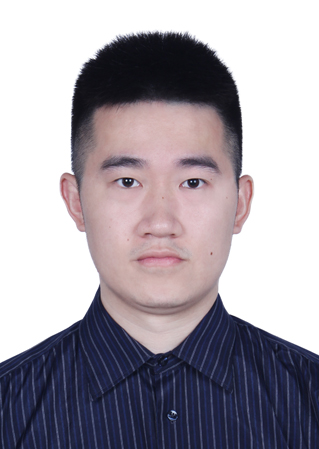}
\end{wrapfigure}
\noindent
\textbf{Zheng Shi} was born in Jiangxi, China, in 1995, He is currently pursuing the Ph.D. degree with College of Microelectronics, South China University of Technology(SCUT), Guangdong, China. His research interest includes distributed Computing, Calculate acceleration and EDA optimization.

\end{document}